\def\valpha{\vec{\alpha}}
\def\vbeta{\vec{\beta}}

\def\snr{{\rm (S/N)}}

\def\thetae{{\theta_{\rm E}}}
\def\drel{{D_{\rm rel}}}

\documentclass[graybox]{svmult}

\usepackage{type1cm}        
\usepackage{float}
\usepackage{makeidx}         
\usepackage{graphicx}        
\usepackage{multicol}        
\usepackage[bottom]{footmisc}

\usepackage{newtxtext}       %
\usepackage{newtxmath}       

\makeindex             

\begin{document}

\title*{The Demographics of Wide-Separation Planets}
\author{B. Scott Gaudi}
\institute{Department of Astronomy, The Ohio State University, Columbus, Ohio, USA\\Jet Propulsion Laboratory, California Institute of Technology, Pasadena, CA, USA\\ 
\email{gaudi.1@osu.edu}}
%
%
\maketitle

\vskip -1.7in
\abstract{\\
\noindent I begin this review by first defining what is meant by exoplanet demographics, and then motivating why we would like as broad a picture of exoplanet demographics as possible. I then outline the methodology and pitfalls to measuring exoplanet demographics in practice. I next review the methods of detecting exoplanets, focusing on the ability of these methods to detect wide separation planets. For the purposes of this review, I define wide separation as separations beyond the ``snow line'' of the protoplanetary disk, which is at $\simeq3{\rm au}$ for a sunlike star.  I note that this definition is somewhat arbitrary, and the practical boundary depends on the host star mass, planet mass and radius, and detection method.  I review the approximate scaling relations for the signal-to-noise ratio for the detectability of exoplanets as a function of the relevant physical parameters, including the host star properties. I provide a broad overview of what has already been learned from the transit, radial velocity, direct imaging, and microlensing methods. I outline the challenges to synthesizing the demographics using different methods and discuss some preliminary first steps in this direction. Finally, I describe future prospects for providing a nearly complete statistical census of exoplanets. 
}
\vskip .1in
{\em Review chapter to appear in Lecture Notes of the 3rd Advanced School on Exoplanetary Science (Editors L. Mancini, K. Biazzo, V. Bozza, A. Sozzetti)}

\section{Introduction: The Demographics of Exoplanets}
\label{sec:intro}

The demographics of exoplanets can be defined as the distribution of exoplanets as a function of a set of physical properties of the planets, their host stars, or the environment of the planetary systems. It can be distinguished from exoplanet characterization by the depth of information that is measured about the exoplanet.  Demographic surveys generally only focus on the most basic properties of the planets, host stars, or their environment. These properties are generally measured directly as parameters intrinsic to the detection technique being used, or are inferred from these parameters, sometimes requiring auxiliary information. In contrast, characterizing exoplanets generally requires more sophisticated techniques to determine the detailed properties of the exoplanets.  Of course, the distinction between exoplanet demographics and characterization is somewhat arbitrary, but nevertheless it provides a useful framework for outlining the basic goals of studies of exoplanets.

The primary motivation for measuring exoplanet demographics is to test planet formation theories. A complete, ab initio theory of planet formation must be able to describe the physical processes by which micron-sized dust grains grow by $\sim 13-14$ orders of magnitude in size and $\sim 38-41$ orders of magnitude in mass to their final raddi and masses between the radius and mass of the Earth and the radius and mass of Jupiter.  As is described in detail in the review chapter in this collection by Sean Raymond and Alessando Morbidelli and references therein, as protoplanets grow by these many orders of magnitude in size and mass, the physical mechanisms that govern their growth and migration vary significantly.  In principle, the signature of many of these physical mechanisms should be imprinted in the distribution of properties of mature planetary systems, including the properties of the planets, the nature and diversity of system architectures, and their dependencies on the host star properties and environment.  Thus, a {\it robust} and {\it unbiased} measurement of the demographics of exoplanets over a broad range of planet and host star properties provides one of the most fundamental empirical tests of planet formation theories.

This review discusses methods for characterizing the population of wide-separation planets, and constraints on the demographics of such planets from various surveys using these methods. For the purposes of this review, I will define wide-separation as semimajor axes and periods greater than the "snow line" in the protoplanetary disk
\begin{equation}
a_{\rm sl}\simeq 3~{\rm au}\left(\frac{M_*}{M_\odot}\right)^{\sim 1}, \hskip0.5cm
P_{\rm sl}\simeq5~{\rm yr}\left(\frac{M_*}{M_\odot}\right)^{\sim 5/4},
\label{eqn:snowline}
\end{equation}
where $M_*$ is the mass of the host star.  Equations \ref{eqn:snowline} are motivated by the estimated distance of the snow line in the solar protoplanetary disk (e.g., \cite{Morbidelli:2016}), and the scaling of the snow line with host star mass estimated by \cite{Kennedy:2008}, although I note that this scaling is uncertain, and the location of the snow line itself is a function of the age of the protoplanetary disk.  This definition is useful to the extent that planet formation likely proceeds differently beyond the snow line due to the larger surface density of solid material, but I note that this is also uncertain. For a more comprehensive discussion of the snowline in protoplanetary disks, its time evolution, and its possible effect on planet formation, see the discussion in the review chapter in this collection by Sean Raymond and Alessandro Morbidelli. Despite these uncertainties, I will adopt Equations \ref{eqn:snowline} as approximate boundaries between close and wide separation planets. As discussed in Section \ref{sec:methods}, the practical boundary between close- and wide-separation planets depends on the detection method and specifics of the survey, as well as on the planet mass and radius.   Thus this boundary should be not applied too rigidly.

There is a large body of literature on topics related to exoplanet demographics, which cannot be comprehensively covered in this relatively short review chapter.  For a more comprehensive (if somewhat outdated) introduction to exoplanet demographics, I refer the reader to \cite{Winn:2015}. 

The plan for this chapter is as follows.  Section \ref{sec:formalism} summarizes the mathematical formalism for constraining exoplanet demographics. Section \ref{sec:practice} discusses a few of the possible pitfalls when constraining exoplanet demographics in practice. Section \ref{sec:methods} briefly summarizes the primary methods of detecting planets, focusing on the scaling of the sensitivity of each method with the planet and host star properties.
Section \ref{sec:demo} highlights some of the main results from surveys for wide-separation planets, including results from radial velocity (Sec.~\ref{sec:rvsurveys}), transit (Sec.~\ref{sec:trsurveys}), direct imaging (Sec.~\ref{sec:disurveys}), and microlensing (Sec.~\ref{sec:mlsurveys}) surveys.  Section \ref{sec:synthesis} reviews the relatively few attempts to synthesize results from multiple methods, and Section \ref{sec:compare} discusses a few notable comparisons between the predictions of the demographics of planets from ab initio planet formation theories and results from exoplanet surveys.  Section \ref{sec:future} discusses future prospects for determining wide-orbit exoplanet demographics.  Finally, I briefly conclude in Section \ref{sec:conclusion}.

\section{Mathematical Formalism}\label{sec:formalism}

Although this review focuses on methods for measuring the demographics of wide-separation planets, and highlights some observational results to this end, it is worthwhile to present the general mathematical formalism for measuring the demographics of planets (of all separations). 
Following \cite{Clanton:2014}, in very general terms, I can mathematically define the goal of demographic surveys of exoplanets to be a measurement of the distribution function $d^n N_{\rm pl}/d\valpha$, where $\valpha$ is a vector containing the set of $n$ physical parameters upon which the planet frequency intrinsically depends.  These can include (but are not limited to): parameters of the planets (mass $M_p$, radius $R_p$), their orbits (period $P$, semimajor axis $a$, eccentricity $e$), properties of the host stars (mass $M_*$, radius $R_*$, luminosity $L_*$, effective temperature $T_{\rm eff}$, metallicity [Fe/H], multiplicity), and many others.  
The total number of planets $N_{\rm pl}$ in the domain covered by $\valpha$ is
\begin{equation}
N_{\rm pl}=\int_{\alpha_1} d\alpha_1 \int_{\alpha_2} d\alpha_2 \cdots \int_{\alpha_n}d\alpha_n
\frac{d^n N_{\rm pl}}{d\valpha}
\end{equation}
Therefore, in principle, one could simply count the number of planets as a function of the parameters $\valpha$ and then differentiate with respect to those parameters to derive $d^n N_{\rm pl}/d\valpha$. This is essentially equivalent to binning in the parameters and counting the number of planets per bin. This distribution function represents the most fundamental quantity that describes the demographics of a population of planets.  With such a distribution function in hand, one can then compare to the predictions for this distribution that are the outputs of ab initio planet formation theories (e.g., \cite{Ida:2013,Mordasini:2015}) and determine how well they match the observations.  Furthermore, one can then vary the input physics in these models, or parameters of semi-analytic parameterizations of the physics, to provide the best match to the observed distribution of physical parameters.  Thus these models can be refined, improving our understanding of planet formation and migration. 

Unfortunately, it is generally not possible to measure $d^n N_{\rm pl}/d\valpha$ directly, for several reasons.  First, all surveys are limited in the range of parameters to which they are sensitive.  Some of these are intrinsic to the detection method itself, while others are due to the survey design.  Second, all surveys suffer from inefficiencies and detection biases.  Thus the measured distribution of planet properties is not equal to the true distribution.  The effects of these first two issues can be accounted for by carefully determining the survey completeness.  Finally, each exoplanet detection method is sensitive to a different set of parameters of the planetary system. Some of these parameters belong to the set of the physical parameters of interest $\valpha$, others are a function of these parameters, others are parameters that are not directly constrained by the data under consideration and must be accounted for using external information or derived from external measurements, and others are essentially nuisance parameters that must be marginalized over.  

As a concrete example, consider radial velocity (RV) surveys for exoplanets. The RV of a star arising from the reflex motion due to a planetary companion can be described by 6 parameters: the RV semi-amplitude $K$, the orbital period $P$, eccentricity $e$, argument of periastron $\omega$, and time of periastron $T_p$ (or the time of some fiducial point in the orbit for circular orbits, such as the time of periastron), and the barycentric velocity $\gamma$.  Of these parameters, generally only $P$ and $e$ are fundamental physical parameters that (potentially) contain information about the formation and evolution of the system.  The parameter $\omega$ is a geometrical parameter that describes the orientation of the orbit with respect to the plane of the sky from the perspective of the observer\footnote{I note that in systems with multiple planets, any apsidal alignment can be inferred by the individual values of $\omega$. The existance of an apsidal alignment (or alignments) can provide constraints on the formation and/or evoluation of that system.  See, e.g., \cite{Chiang:2001}.}, and $T_P$ is an (essentially) arbitrary conventional definition for the zero point of the radial velocity time series. The parameter $\gamma$ does not contain any information about the planet or its orbit, although (if absolutely calibrated) does contain information about the Galactic orbit of the system and thus the Galactic stellar population to which the host star belongs.  Finally, $K$ depends on $P$, $e$, the mass of the planet $M_p$, the mass of the host star $M_*$, and the inclination $i$ of the orbital plane with respect the plane of the sky, with $i=90^\circ$ is edge-on). The inclination is another geometrical parameter, and thus is not of intrinsic interest.   However, $M_p$ and $M_*$ are fundamental physical parameters of interest.  The latter cannot be estimated from RV measurements of the star alone, and thus must be determined from external information.  With an estimate of $M_*$ in hand, the remaining physical quantity of interest, $M_p$, still cannot be determined directly, since the inclination $i$ is not constrained by RV measurements. Assuming $M_p\ll M_*$, one can then infer the minimum mass of the companion $M_p\sin{i}$. In order to infer $M_p$, one must use the known {\it a priori} distribution of $i$ (namely that $\cos{i}$ is uniformly distributed), {\it and} an assumed prior on the distribution of true planet masses $d N/d M_p$, to infer the posterior distribution of $M_p$ for any given detection\footnote{I note that the posterior distribution of $M_p$ given a measurement $M_p\sin{i}$ is often estimated by simply assuming that $\cos{i}$ is uniformly distributed. This leads to the familiar result that the median of the true mass is only $(\sin{[{\rm acos}(0.5)]})^{-1}\simeq 1.15$ larger than the minimum mass. Unfortunately, this is only correct if the distribution of true planet masses is such that $d N/d \log{M_p}$ is a constant \cite{Ho:2011,Stevens:2013}. For other distributions, the true mass can be larger or smaller than this naive estimate.}. However, $d N/d M_p$ is typically a distribution one would like to infer, and thus one must deconvolve a distribution of minimum masses of a sample of detections to infer the posterior distribution $d N/d M_p$ (see, e.g., \cite{Zucker:2001}).

In order to deal with the difficulties in inferring the true distribution function $d^n N_{\rm pl}/d\valpha$ of some set of physical parameters $\valpha$, one must account for not only the survey completeness, but also for the fact that the detection method (or methods) being used to survey for planets are generally not directly sensitive to (all of the) parameters of interest $\valpha$.  To deal with the latter point, several approaches can be taken.  First, one can attempt to transform the observable parameters into the physical parameters of interest. This often requires introducing external information (such as the properties of the star), or adopting priors for, and then marginalizing over, parameters that are not directly constrained. Alternatively, one can simply choose to constrain the distribution functions of the observable parameters that are most closely related to the physical parameters of interest. It is also possible to adopt both approaches. 

An important point is that the completeness of each individual target is obviously a function of the observable parameters, not the transformed physical parameters.  Therefore the completeness of each target must first be determined in terms of the observable parameters, and then transformed to the physical (or more physical) parameters of interest (if desired).

In order to determine the completeness of a given survey of a set of targets $N_{\rm tar}$, one must first specify the criteria adopted to define a detection.  An essential (but often overlooked) point is that the criteria used to determine the completeness of the entire sample of $N_{\rm tar}$ targets must {\it strictly} be the same criteria used to detect the planets in the survey data to begin with.  Failure to adhere to this requirement can lead to (and indeed, has led to) erroneous inferences about the distribution of planet properties.  This is particularly important in regions of parameter space where the number of detected planets is a strong function of the specific detection criteria.  

Typically one computes the completeness or efficiency (hereafter referred to efficiency for definiteness) of a given target as a function of the primary observables whose distributions affect the detectability of the planets.  I define this set of observables as $\vbeta$.  I note that this set can be subdivided into two subsets. I define $\vbeta_{\cal I}$ to be the subset of the $\vbeta$ observable parameters of interest whose distributions one would like to constrain (or observables which will subsequently be transformed to other, more physical parameters, whose distributions are to be constrained), and $\vbeta_{\cal N}$ to be the remainder of the set of observable `nuisance' parameters (e.g., the parameters that affect the detectability but are either considered nuisance parameters or parameters whose distributions are not specifically of interest).  The efficiency of a given target $j$ is given by
\begin{equation}
\epsilon_j(\vbeta_{\cal I}) =
\int_{\beta_{\cal N}} d\beta_{{\cal N},1} 
\int_{\beta_{\cal N}} d\beta_{{\cal N},2} 
\cdots 
\int_{\beta_{{\cal N},n}} d\beta_{{\cal N},n} 
\frac{d^n N_{\rm pl}}{d\vbeta_{\cal N}}
\cal{D}(\vbeta_{\cal N}).
\label{eqn:1efficiency}
\end{equation}
Here $d^n N_{\rm pl}/d \vbeta_{\cal N}$ is the assumed prior distribution of the observable 'nuisance' parameters $\vbeta_{\cal N}$, and ${\cal{D}}(x)$ is the set of detection criteria used to select planet candidates.  It is worth noting that the prior distribution for the nuisance parameters $\vbeta_{\cal N}$ is generally trivial, and does not contain any valuable physical information. In the simplest case of a $\chi^2$ threshold as the only detection criterion, then ${\cal{D}}={\cal{H}}[\Delta\chi^2(\vbeta_{\cal N})-\Delta\chi^2_{\rm min}]$, where ${\cal{H}}(x)$ is the Heaviside step function and $\Delta\chi^2$ is the difference between the $\chi^2$ of a fit to the data with a planet and the null hypothesis of no planet, and $\Delta\chi^2_{\rm min}$ is the minimum criterion for detection.  In reality, most surveys employ several detection criteria. The most robust method of determining the efficiency $\epsilon_j$ for each target is to inject planet signals into the data according to the distribution function $d^n N_{\rm pl}/d \vbeta$, and then attempt to recover these signals using the same set of  detection criteria $\cal{D}$ (e.g., the same pipeline) used to construct the original sample of detected planets. 

In order to determine the overall survey efficiency $\Phi(\vbeta_{\cal I})$, one must compute the efficiency for every target, whether or not the target contains a signal that passes the set of detection criteria $\cal{D}$. The total survey efficiency is then
\begin{equation}
    \Phi(\vbeta_{\cal I})= \sum_{j=1}^{N_{\rm tar}} \epsilon_j(\vbeta_{\cal I}).
    \label{eqn:Phi}
\end{equation}
Given the total survey efficiency $\Phi$, it is possible to marginalize over the parameters $\vbeta_{\cal I}$ to estimate the total number of expected planet detections $N_{\rm pl,exp}$ in the survey, given a prior assumption for the distribution function $d^n N_{\rm pl}/d \vbeta_{\cal I}$:
\begin{equation}
N_{\rm pl,exp}=\int_{\beta_{{\cal I},1}} d\beta_{{\cal I},1} \int_{\beta_{{\cal I},2}} d\beta_{{\cal I},2} \cdots \int_{\beta_{{\cal I},n}} d\beta_{{\cal I},n} \frac{d^n N_{\rm pl}}{d\vbeta_{\cal I}}
\Phi(\vbeta_{\cal I}).
\label{eqn:nplexp}
\end{equation}
Of course, it is precisely the distribution function $d^n N_{\rm pl}/d \vbeta_{\cal I}$ that we wish to infer.  Thus, in the usual Bayesian formalism, we must adopt a prior distribution of the parameters of interest, and we then determine if this prior distribution is consistent (in the likelihood sense) with the posterior distribution of these parameters.  If not, then we adjust the prior distribution and repeat.  This is they way one `learns' in the Bayesian formalism.

It is also possible to convert the individual target efficiencies $\epsilon_j(\vbeta_{\cal I})$ as a function of the observable parameters of interest $\vbeta_{\cal I}$ to efficiencies as a function of physical parameters (or more physical parameters) $\valpha$ by simply transforming from $\vbeta_{\cal I}$ to $\valpha$.  In this case, one would replace $\vbeta_{\cal I}\rightarrow \valpha$ in Equation~\ref{eqn:nplexp} after transforming to $\valpha$. In general, such transformations can be relatively straightforward or fairly complicated, depending on the details of what is known about the properties of the target sample, and the variables that are being transformed.  It is also important to note that such transformations can introduce additional sources of uncertainty; for example if one wants to convert from planet/host star mass ratio to planet mass, one must explicitly account for the uncertainty in the host star mass.  Finally, it is important to include the Jacobian of the transformation of adopted prior distribution of the observables $d^n N_{\rm pl}/d^n\vbeta_{\cal{I}}$ to the prior distribution of the physical parameters $d^n N_{\rm pl}/d^n\valpha$, as a given prior distribution on an observable parameter $\beta_{{\cal I},i}$ does not guarantee the desired prior distribution on the physical parameters $\valpha$.

With the individual values of $\epsilon_j$ and total survey $\Phi$ efficiencies in hand, there are two basic approaches that are typically taken to infer the posterior distribution $d^n N_{\rm pl}/d \vbeta_{\cal I}$ (or the distribution of transformed variables $d^n N_{\rm pl}/d\valpha$).\\

\noindent{\bf Binning:} The simplest and most obvious method of inferring the distribution function is simply to define bins in the parameters of interest $\vbeta_{\cal I}$ (or $\valpha$). For definiteness, I assume one is working in the space of the observable parameters of interest $\vbeta_{\cal I}$. One then counts the number of detected planets in each bin $k$ of $\vbeta_{\cal I}$, $N_{{\rm det}, k}$, and then divides by the number of planets expected to be detected in each bin using Equations~\ref{eqn:Phi} and \ref{eqn:nplexp}, where, in this case, the summation and integrals are over the span of each bin.  The estimated frequency of planets (more specifically, the number of planets per star) in each bin $k$ is then
\begin{equation}
   \frac{d^n N_{\rm pl}}{d\vbeta_{\cal I}}\bigg|_{{\rm post},k}\equiv\frac{N_{{\rm det},k}}{N_{{\rm pl,exp},k}},
   \label{eqn:binfreq}
\end{equation}
where the symbol $\big|_{{\rm post},k}$ serves to indicate that the value of $d^n N_{\rm pl}/{d\vbeta_{\cal I}}$ inferred from Equation \ref{eqn:binfreq} is the {\it posterior} distribution inferred for bin $k$.
This gives the frequency of planets in that bin, weighted by the assumed prior distribution function of the bin parameters. In this case, the uncertainty in the inferred planet frequency of each bin is given by Poisson statistics based on the number of detections in each bin, weighted by the prior distribution. Although binning is generally not recommended (for reasons discussed in Section \ref{sec:practice}), it can provide a useful method to visualize the results of the survey. I note that, despite some claims to the contrary, binning is neither a `non-parametric' nor a `prior-free' method of inferring the distribution of planet properties. In reality, all inferences about the distribution of planet properties from survey data are parametric and assume priors (whether those parameters or priors are explicitly stated or not).  This is clear from the discussion above, as the value of $N_{{\rm pl,exp},k}$ is evaluated using Equation \ref{eqn:nplexp}, which itself depends on the assumed {\it prior} distribution of $d^n N_{\rm pl}/d\vbeta_{\cal I}$.

Although this method is conceptually quite straightforward, it is worthwhile to point out that a somewhat different approach has been taken to implement this method in practice, particularly for early results gleaned from the {\it Kepler} data.  This approach, dubbed the inverse detection efficiency method (IDEM) by \cite{Foreman-Mackey:2014}, requires only evaluating the planet detection sensitivity for those stars around which planets have been discovered. Specifically, the planet frequency (again, more specifically the number of planets per star) in a given bin $k$ is given by 
\begin{equation}
   \frac{d^n N_{\rm pl}}{d\vbeta_{\cal I}}\bigg|_{{\rm post},k}\equiv\frac{1}{N_{\rm tar}} \sum_{j=1}^{N_{{\rm det},k}} \frac{1}{\epsilon_j}.
\end{equation}
I again note that, while it may appear that this method is `non-parameteric' or `prior-free', this is not the case, as one must define the boundaries of each bin.

As discussed in detail in \cite{Foreman-Mackey:2014} and \cite{Hsu:2018}, this approach is not optimal, and specifically is likely to yield biased inferences for $N_{\rm pl,exp}$. This issue is particularly acute for transit surveys, for which a small number of detections must be multiplied by large correction factors to estimate the intrinsic frequency of planets.  The fact that this approach is not optimal stems from several reasons, the most important being that the stars with the largest intrinsic sensitivity are those for which it is most likely that planets will be detected.  I note that several analyses of ground-based transit surveys published prior to any results from Kepler (e.g., \cite{Mochejska:2005,Burke:2006, Hartman:2009}) utilized the (more optimal) first method described above (i.e., the method where one estimates and utilizes the sensitivity of all the target stars), rather than the IDEM method.

\noindent{\bf Maximum Likelihood:} A second method to estimate the distribution function $d^n N_{\rm pl}/d \vbeta_{\cal I}$ (or the distribution of transformed variables $d^n N_{\rm pl}/d\valpha$) is to use the maximum likelihood method. The mathematics of this method can be derived by starting with the method of binning described above, and then decreasing the bin size such that there is only one detection in each bin (see, e.g., Appendix A of \cite{Cumming:2008} or Section 3.1 of \cite{Youdin:2011}).  Here one seeks to constrain the parameters of a distribution function $F$, such that \begin{equation}
 \frac{d^n N_{\rm pl}}{d\vbeta_{\cal I}}=F(\vbeta_{\cal I};\Vec{\Theta}),  
\end{equation}
where $\Vec{\Theta}$ are the variables that parameterize the distribution function $F$\footnote{Again, one might instead wish to instead transform $\vbeta_{\cal I} \rightarrow \valpha$ and then constrain $d^n N_{\rm pl}/d\valpha$.  For brevity, I will no longer specifically call out this alternate method.}.
A common form for $F$ when two variables are being considered is a double power-law.  In this case, the vector $\Vec{\Theta}$ would consist of three parameters: a normalization, and one exponent for each of the two parameters. Note this form implicitly assumes a range for each of the two parameters, the minimum and maximum of which can be considered as additional, hidden parameters, the choice of which can affect the maximum likelihood inferences for the other, explicit parameters. 

Given this parameterized form for the distribution function $d^n N_{\rm pl}/d \vbeta_{\cal I}$, then one can determine the likelihood of the observations given the parameters $\Vec{\Theta}$ of the distribution function $F$ (see the references below):
\begin{equation}
{\cal L}(\Vec{\Theta})=\exp^{-N_{\rm pl,exp}}\Pi_{j=1}^{N_{\rm det}} F(\vbeta_{{\cal I},j};\Vec{\Theta})\Phi(\vbeta_{{\cal I},j}),
\label{eqn:likelihood}
\end{equation}
where $N_{\rm det}$ is the number of planets detected in the survey (more specifically, the number of detections that pass all of the detection criteria ${\cal D}$). Equation~\ref{eqn:likelihood} can then be used to estimate the likelihood that the function $F(\vbeta_{\cal I};\Vec{\Theta})$ with a given set of parameters $\Vec{\Theta}$ describe the data.  Using the usual methods of exploring likelihood space (e.g., Markov Chain Monte Carlo), one can determine both the maximum likelihood, as well the confidence intervals of the parameters. These methods are generally well known and have been explored in many publications, and so I will not reiterate them here. 

I note that the above mathematical description of the methodology to estimate the distribution of exoplanet properties (e.g., to determine the "demographics of exoplanets") was exceptionally abstract. This approach was taken intentionally in order to make the discussion as general as possible. Nevertheless, it may be difficult to apply this formalism to specific exoplanet surveys.  I therefore invite the reader to consult the following publications that apply the formalism discussed above (including both binning and maximum likelihood) to specific exoplanet detection methods and surveys.  For readers interested in radial velocity surveys, I suggest starting with \cite{Tabachnik:2002,Cumming:2008,Howard:2010}, for those interested in transit surveys, I suggest starting with \cite{Burke:2006,Youdin:2011,Dong:2013,Dressing:2013,Howard:2012,Burke:2015}, for those interested in microlensing surveys, I suggest starting with \cite{Gaudi:2002, Gould:2010, Suzuki:2016}, and for those interested in direct imaging surveys, I suggest starting with \cite{Bowler:2016, Nielsen:2019}. 

Finally, I note that much more sophisticated statistical methodologies for determining exoplanet demographics have been developed (e.g., hierarchical Bayesian modeling, approximate Bayesian computation), particularly in application to {\it Kepler} (see, e.g., \cite{Foreman-Mackey:2014,Hsu:2018,Hsu:2019}).

\section{Pitfalls to Measuring Exoplanet Demographics in Practice}\label{sec:practice}

In this section, I outline a few of the difficulties in measuring exoplanet demographics in practice. Many of these should be fairly obvious given the discussion in the previous section, but some are much more subtle. I do not claim this to be a comprehensive list of where one might go wrong, but merely a partial list of common pitfalls that one should avoid.
\begin{itemize}
\item {\bf Adopting different detection criteria to estimate the survey efficiency than were used to identify the planet candidates.} This is likely one of the most prevalent source of systematic error when determining exoplanet demographics. An excellent example is the {\it Kepler} survey \cite{Borucki:2010}. The earliest estimates for the intrinsic distribution function of planet parameters assumed simple, ad hoc assumptions for the detection criteria. This approach was required due to the fact that the algorithms used by the {\it Kepler} team to detect the announced planet candidates were not initially publicly available.  As a result, the adopted detection criteria were often significantly different than the true detection criteria, leading to inferences about the planet distribution that were sometimes egregiously incorrect.  This issue is particularly important near the "edges" of the survey sensitivity, where the number of detected planets can be a strong function of the detection criteria (e.g., the signal-to-noise ratio). Later analyses circumvented this issue by developing independent pipelines to identify planet candidates. These same pipelines were then used to determine the detection efficiency by injecting transit signals into the data and recovering them using the same pipeline (e.g., \cite{Dressing:2013,Petigura:2013}). Inferences about the exoplanet distribution made in this way are generally much more robust.  
\item {\bf Use of `by-eye' candidate selection.} Quite often, candidates from exoplanets surveys are subjected to human vetting.  While this can be an excellent method of eliminating false positives, it is obviously difficult to reproduce in an automated way.  While it is possible to inject a large number of signals into the data sets and then have the `artificial' candidates vetted by humans (e.g., \cite{Gould:2006}), this is both labor intensive and can impose biases (e.g., if the humans doing the vetting know that they are searching for injected signals).  It is possible to remove the step of human vetting, as has been done using, e.g., the Robovetter tool developed by the {\it Kepler} team \cite{Thompson:2018}.
\item {\bf Ignoring reliability.}  An implicit assumption that was made in Section \ref{sec:formalism} was that all signals that passed the detection criteria ${\cal D}$ were due to real planets.  This ignores false positives, which can be astrophysical, instrumental, or simple statistical false alarms.  The reliability of a sample of candidate planets is simply the fraction of candidates that are true planets.  Thus, in order to determine the true distribution of planet parameters from a survey, one must not only estimate the efficiency or completeness of the survey (the fraction of true planets that are detected), but one must also asses the reliability (the fraction of detections that are due to true planets).
\item {\bf Overly optimistic detection criteria.}
One of the more straightforward and frequently-used detection criterion is a simple signal-to-noise ratio or $\Delta \chi^2$ cut, or some related statistic that quantifies how much better the data is modelled with a planet signal than the null hypothesis of no planet.  In principle, one should set the threshold value of such detection criteria such that there are few or no statistical false alarms.  One might naively assume that the data have uncertainties that are Gaussian distributed and uncorrelated, in which case the threshold value can generally be estimated analytically (or semi-analytically) by simply computing the probability (say 0.1\%) of a given value of the statistic arising by random fluctuations. In practice, setting the appropriate threshold value is significantly more complicated. First, one requires a detailed knowledge of the noise properties of the data, which are often not Gaussian distributed and are also often correlated on various timescales. Second, one requires an estimate of the number of independent trials used to search for planetary signals to determine the threshold probability. In practice both can be difficult to estimate: the noise properties of the data may be poorly behaved and difficult to characterize, and the number of independent trials often cannot be estimated analytically, and must be estimated via, e.g., Monte Carlo simulations.  A reasonably robust method of estimating the appropriate threshold is to inject simulated planetary signals into the data, and compare the distribution of the desired statistic (e.g., $\Delta\chi^2$) in the data where no planets were injected to the distribution in the data where planets were injected\footnote{In general this method works only if the majority of the targets do not have a planetary signal that is significant compared to the intrinsic noise distribution in the data}. The optimal threshold can therefore be chosen such that the completeness is maximized while minimizing the number of false positives.
\item {\bf The dangers of binning.}
It is often tempting to adopt the procedure outlined above where one collects the detected planets in parameter bins of a given size, and then estimates the intrinsic frequency of planets in that bin by dividing the number of detected planets in that bin by the number of expected detections if every star has a planet in that bin.  Indeed, this is a useful tool that can enable one to visualize the intrinsic distribution function of planet properties, which can then inform the parameterized models to be fitted.  However, it should be noted that binning is not a well-defined procedure.  In particular, one can ask: What is the optimal bin size?  If one makes the bin size too large (in order to, e.g., decrease the Poisson uncertainty), then one risks smoothing over real features in the underlying distribution of planet properties.  On the other hand, if the bin size is too small, the Poisson uncertainties blow up. Indeed, there is {\it no optimal bin size} in this sense, and thus it is generally advised to fit parametric models using the maximum likelihood approach briefly summarized above and derived previously by many authors.
\item {\bf Not accounting for uncertainties in the parameters of the detected planets}
Note that in Equation \ref{eqn:likelihood}, the distribution function $F(\vbeta_{\cal I},\Vec{\Theta})$ and the total survey sensitivity are evaluated at the point estimates of the values of the parameters $\vbeta_{{\cal I},k}$ for each of the $k=[1,N_{\rm det}$] planet detections.  This implicitly assumes that the distributions of these parameters for each detected planet are unimodal with zero uncertainty, neither of which is generically true.  Therefore, one must marginalize the likelihood over the uncertainties in the parameters $\vbeta_{\cal I}$ (properly accounting for covariances between these parameters), and account for multimodal degenerate solutions, as applicable. See \cite{Foreman-Mackey:2014,Hsu:2018} for discussions of the deleterious effects of ignoring the uncertainties on the planet parameters.
\item {\bf Poor knowledge of the properties of the target sample.}
When converting between the observable parameters of detected exoplanet systems to the physical properties of the planets, one must often assume or infer properties of the host star.  When inferring exoplanet demographics as a function of the physical properties of the planets, one must typically assume or infer properties of the entire target sample.  Any systematic errors in these inferences will propagate directly into systematic errors in the inferred exoplanet distributions.  Even if systematic uncertainties are not present, the statistical uncertainties in the host star properties should be taken into account by marginalizing over these uncertainties. As a concrete example, transit surveys are directly sensitive to the transit depth $\delta$, which is just the square of the ratio of the radius of the planet to the radius of the star, $\delta=(R_p/R_*)^2$.  Thus, to infer the demographics of planets as a function of $(R_p,P)$, one must estimate the radii of {\it all} the target stars (not simply the ones that have detected transit signals).  Prior to Gaia \cite{Gaia:2016}, it was generally not possible to robustly estimate radii for most stars; rather these had to be estimated using other observed properties of the star (e.g., $T_{\rm eff},\log{g},{\rm [Fe/H]}$), combined with theoretical evolutionary tracks. This led to systematic differences in the inferred radii of subsamples of the {\it Kepler} targets, which in turn led to discrepancies in the inferred radii of the detected planets, and the inferred planet radius distribution \cite{Mann:2012,Pinsonneault:2012,Dressing:2013,Huber:2014,Gaidos:2016}
\footnote{Fortunately, with the availability of near-UV to near-IR absolute broadband photometry, combined with Gaia parallaxes and stellar atmosphere models, it is possible to measure the radii of most bright stars nearly purely empirically to relatively high precision \cite{Stassun:2017,Stevens:2017}.}. 
\item {\bf Properly distinguishing between the fraction of stars with planets and the average number of planets per star.}
A commonly overlooked subtlety in determining exoplanet demographics from surveys is the distinction between estimates of the average number of planets per star (NPPS) and estimates of the fraction of stars with planets (FSWP).  Depending on the distribution of exoplanet multiplicities, these two quantities can be significantly different \cite{Youdin:2011,Brakensiek:2016}.  Generally, studies that consider all the planets detected in a survey (including multiplanet systems) can be thought of as measuring the NPPS. \cite{Youdin:2011,Brakensiek:2016,Hsu:2018}.
\item {\bf Including post-detection data or detections from other surveys.} When inferred planet occurrence rates from a survey, it is critical that the survey be 'blind'.  In other words, the data that were used to detect the planets and characterize the survey completeness must not have been influenced by the presence (or absence) of significant planetary signals.  For true surveys with predetermined and fixed observation strategies (such as {\it Kepler}), this is generally not an issue.  However, for targeted surveys, is it common to acquire more data on targets that show tentative evidence of a signal in order to bolster the significance of the signal (or determine that is not real).  It is also common to take additional data once a robust detection is made, in order to better characterize the parameters of the system.  Using this data may result in biased estimates of the occurrence rate of planets.  Similarly, it is often the case that targets will stop being observed or will be observed at a lower cadence after a certain time if there no evidence of a signal.  Again, such a strategy can lead to biased estimates of the occurrence rates.  For similar reasons, one must not use detections from other surveys or data from other surveys to confirm marginal detections, as this may also lead to biased estimates of the occurrence rates. 
\item {\bf The dangers of extrapolation.} No single survey or detection method, and indeed even the totality of the surveys that have been conducted with all the detection methods at our disposal, have yielded a "complete" statistical census of exoplanets (see Section~\ref{sec:methods} and the discussion therein). It is therefore tempting to extrapolate from regions of parameter space where exoplanet demographics have been relatively well-measured to more poorly-studied regions of parameter space.  Aside from theoretical arguments why such extrapolations may not be well-motivated, the dangers of such extrapolations have already been empirically demonstrated.  For example, initial extrapolations of power-law fits to the occurrence rate of giant planets with semimajor axes of $\lesssim 3~{\rm au}$ as estimated from RV surveys \cite{Cumming:2008} out to very large separations implied that ground-based direct imaging surveys for young, giant planets should yield a large number of detections.  This prediction has not been borne out, and it is now known that very wide-separation giant planet are relatively rare (see Section~\ref{sec:disurveys}). Perhaps more strikingly,
\cite{Dulz:2020} found that, by extrapolating the double power-law fit to the distribution of planets detected by {\it Kepler} found by \cite{Kopparapu:2018} with periods of $\lesssim 600~{\rm days}$ to longer periods, many (and depending on the adopted uncertainties in the fit by \cite{Kopparapu:2018}, even the majority) of the systems were dynamically unstable.  
\end{itemize}

\section{Methods of Detecting Exoplanets: Inherent Sensitivities and Biases}\label{sec:methods}

\begin{figure}[t]
\includegraphics[width=11cm]{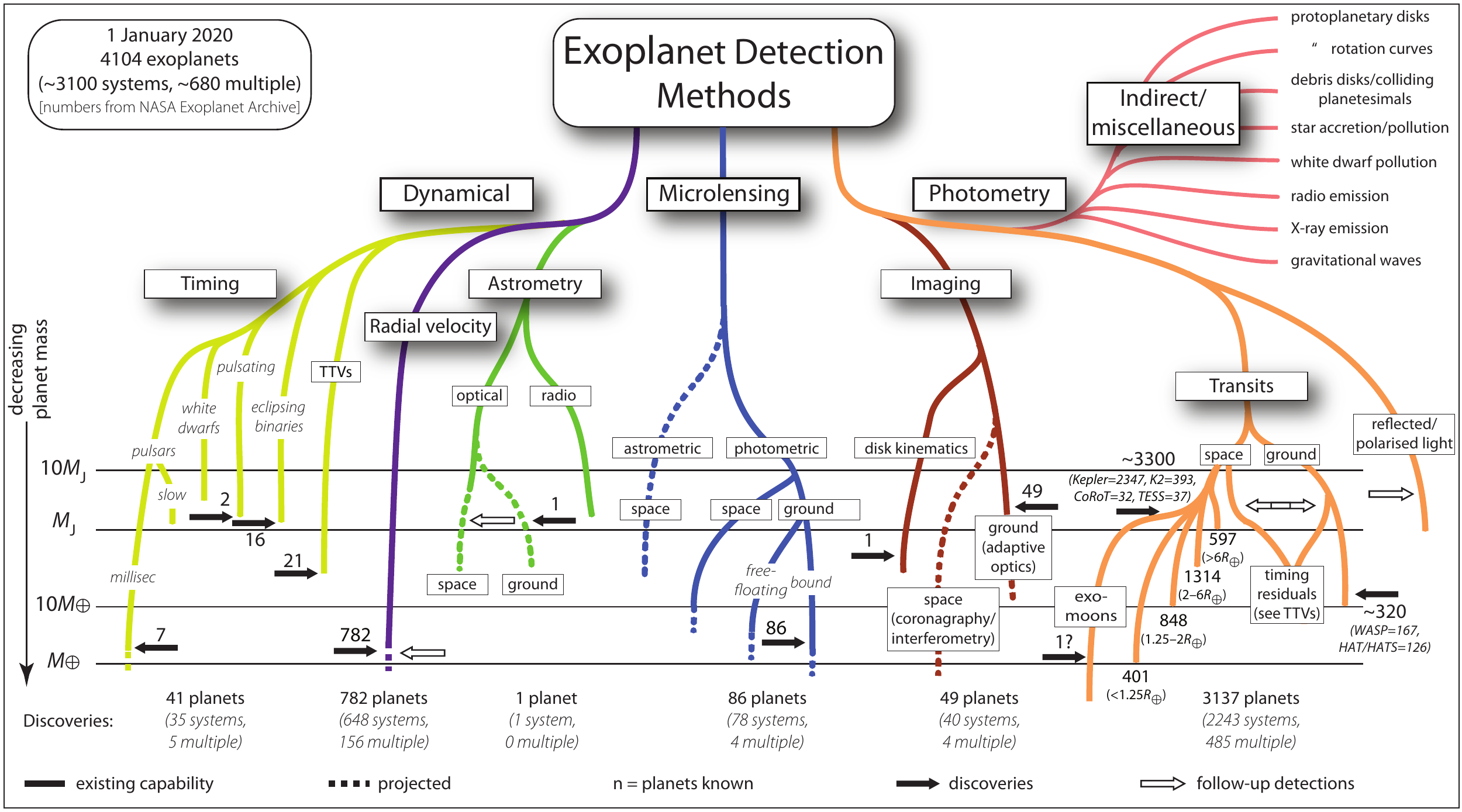}
\caption{A graphical summary of exoplanet detection methods, including the number of discoveries by each method as of January 1, 2020.  Courtesy of Michael Perryman \cite{Perryman:2018} and reproduced with permission.}
\label{fig:perryman}
\end{figure}

There are four primary methods that have been used to detect exoplanets: radial velocities, transits, direct imaging, and microlensing. Another well-known method of detecting exoplanets is astrometry.  However, to date there have been no confirmed detections of planetary companions using astrometry, although there are some candidates that have yet to be confirmed.  Nevertheless, astrometry is an extremely promising method for detecting planets, particularly wide-orbit planets. In particular, as we discuss in Section \ref{sec:future}, Gaia \cite{Gaia:2016} is expected to detect tens of thousands of giant planets on relatively wide orbits \cite{Casertano:2008,Perryman:2014}.  Therefore, I will also discuss astrometry in this section.  There are, of course, many other methods of detecting exoplanets (see Fig.~\ref{fig:perryman}).  However, none of these methods have yielded large samples of planets to date, and so I will not be discussing these in this review.  

\begin{figure}[t]
\includegraphics[width=1.0\textwidth]{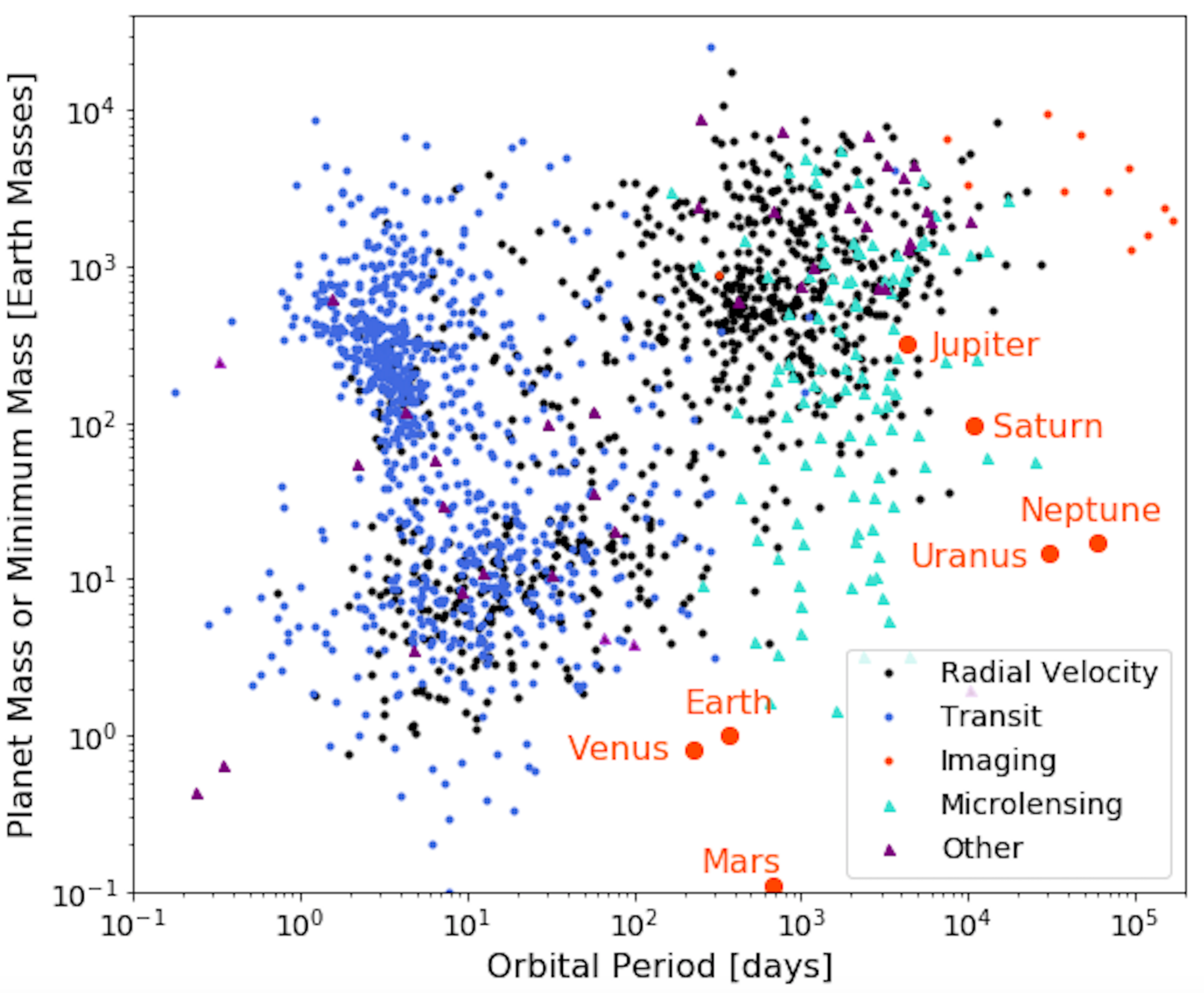}
\caption{The distributions of the $\sim$4300 confirmed low mass companions to stars as a function of their mass (or minimum mass in the case of RV detections) and period. For planets detected by microlensing and direct imaging, the projected semi-major axis has been converted to period using Newton's version of Kepler's Third Law. The color coding denotes the method by which planets were detected.  I note that $\sim25$ planets detected by direct imaging are not shown in this plot because they have periods that are greater than $\gtrsim 10^6$~days.   This figure is based on data from the NASA Exoplanet Archive: https://exoplanetarchive.ipac.caltech.edu/. Courtesy of Jesse Christiansen with assistance from Radek Poleski, reproduced with permission.}
\label{fig:exoplanets-pm}
\end{figure}

\begin{figure}[t]
\includegraphics[width=1.0\textwidth]{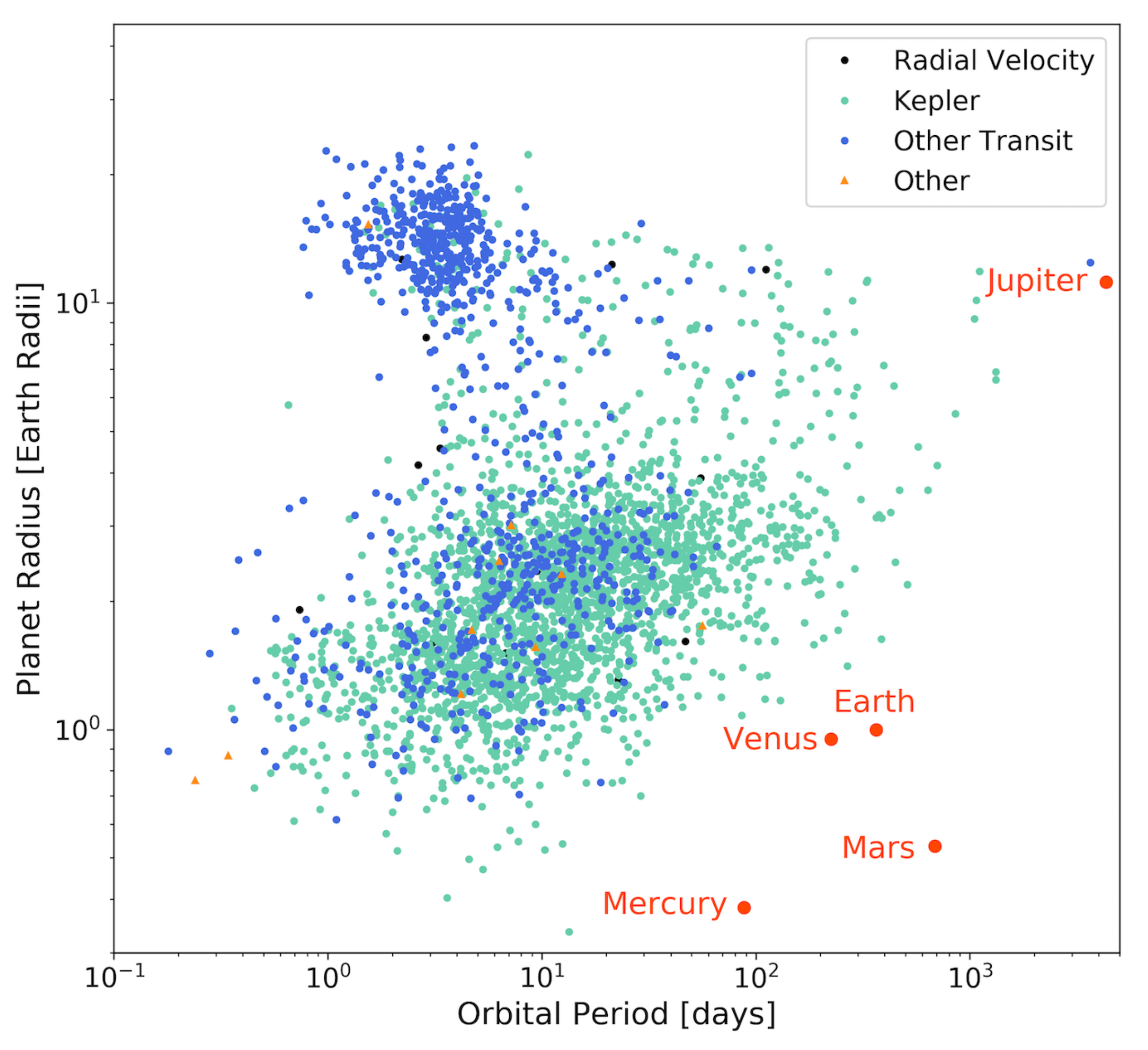}
\caption{Same as Figure~\ref{fig:exoplanets-pm}, except showing the known planets in radius and period space. As with Figure \ref{fig:exoplanets-pm}, the color coding denotes the method by which planets were detected.  This figure is based on data from the NASA Exoplanet Archive: https://exoplanetarchive.ipac.caltech.edu/. Courtesy of Jesse Christiansen, reproduced with permission.}
\label{fig:exoplanets-pr}
\end{figure}

Figures \ref{fig:exoplanets-pm} and \ref{fig:exoplanets-pr} show the distribution of confirmed exoplanets as a function of mass (or minimum mass in the case of RV detections) and period (Fig.~\ref{fig:exoplanets-pm}), or radius and period (Fig.~\ref{fig:exoplanets-pr}).  There are several noteworthy features:
\begin{itemize}
    \item Most of the planets with (minimum) mass measurements and orbital periods greater than $\sim 100$ days were discovered via RV, and thus do not have radius measurements.  Conversely, the majority of planets with mass measurements and periods of less than $\sim 100$ days were discovered by ground based transit surveys (e.g., "hot Jupiters"). This is because, although {\it Kepler} discovered far more transiting planets than ground-based transit surveys, most of the host stars were too faint for RV follow-up, and only a relatively small subset exhibited transit timing variations that allow for a measurement of the planet mass. 
    \item The number of hot Jupiters detected via RV is much smaller than the number of hot Jupiters detected by ground-based transit surveys, and much smaller than the number of Jovian planets (mostly discovered with RV) with periods $\gtrsim 100$ days.  This has several implications. From RV surveys, it is known that hot Jupiters are relatively rare compared to the population of Jupiters with longer periods \cite{Wright:2012}.  On the other hand, when hot Jupiters detected by transits are also considered, the fact that the two populations appear comparable in number is due to the strong selection bias of transit surveys toward short-period planets (e.g., \cite{Gaudi:2005,Gaidos:2013}). 
    \item The vast majority of planets with measured radii do not have mass measurements.  As mentioned above, this is due to the fact that the {\it Kepler} host stars are generally too faint for RV follow-up, and a minority of systems exhibit transiting timing variations.  The majority of planets that have both mass and radius measurements were discovered by ground-based transit surveys.
    \item The fact that the spread in the radii of hot Jupiters is considerably smaller than the spread in mass is a consequence of the fact that the radii of objects with masses that span roughly the mass of Saturn up to the hydrogen burning limit are all approximately constant, with radii of $\sim R_{\rm J}$.
    \item Finally, and most relevant to this review, there is paucity of planets in the lower right corner of both figures (small masses/radii and long periods).  This is purely a selection effect, as I discuss below.  It is also worth noting that this is the area that is spanned by the planets in our solar system.  In particular, essentially no detection method is currently sensitive to analogs of any of the planets in our solar system except for Jupiter (and, in the case of microlensing, Saturn and possibly analogs of the ice giants).  
\end{itemize} 

\subsection{A Note About Stars}\label{sec:stars}
Before discussing the sensitivities and biases of each detection method, I will first make a few comments about stars.  Although initial RV surveys focused primarily on solar type (FGK) stars, to date a wide range of stellar host types have been surveyed for exoplanetary companions.  These hosts span a broad range of properties that can be relevant for exoplanet detection, including radius, mass, effective temperature, projected rotation velocity, luminosity, activity, and local number density, to name a few.  Thus, as is well appreciated, in order to understand the impact of the sensitivities and biases of a given detection method on the population of planets that can be detected, it is essential to understand how these depend on the properties of the host stars (``Know thy star, know thy planet").  Consequently, it is also important to have a reasonably in-depth understanding of stellar properties and how they vary with stellar type, as well as have a detailed accounting of the distribution of stellar properties of any given exoplanet survey. 

For the purposes of this review, I focus on relatively unevolved solar-type FGKM main sequence stars.  For such stars, the mass-luminosity and mass-radius relation can very roughly be approximated by \cite{Torres:2010}
\begin{eqnarray}
L_*&=&L_\odot \left(\frac{M_*}{M_\odot}\right)^{4}\label{eqn:ml}\\
R_*&=&R_\odot \left(\frac{M_*}{M_\odot}\right)\\
\label{eqn:mr}
T_{\rm eff}&=&T_{{\rm eff},\odot}\left(\frac{M_*}{M_\odot}\right)^{1/2}.
\label{eqn:mteff}
\end{eqnarray}
The mass-radius relation holds from roughly the hydrogen burning limit to $\sim 2~M_\odot$ for stars near the zero age main sequence, whereas the luminosity-mass relation holds from roughly the fully convective limit of $\sim 0.35~M_\odot$ up to $\sim 10~M_\odot$.  Below the fully convective limit, the mass-luminosity relation is shallower than Equation \ref{eqn:ml}, and similarly, in this regime the effective temperature-mass relationship deviates from the scaling relation above. I will use these approximate relations to express the sensitivities of the various detection methods as a function of the host star mass.  

\subsection{A General Framework for Characterizing Detectablity}\label{sec:detectablity}

In the following few sections, I will discuss the scaling of the sensitivity of each of the five methods discussed in this section (RV, transits, direct imaging, microlensing, and astrometry) with planet and host star parameters, highlighting the intrinsic sensitivities and biases of each method.  I will not attempt to provide an in-depth discussion of these detection methods, as this material has been covered in numerous other books and reviews \cite{Seager:2010,Perryman:2018,Wright:2013,Fischer:2014}.  

As outlined in \cite{Wright:2013}, although the criteria to detect a planet depends on the details of the planet signal, the data properties (e.g., cadence, uncertainties), and the precise quantitative definition of a detection, one can often estimate the scaling of the signal-to-noise ratio with the planet and host star properties by decomposing the signal into two contributions.  These are the overall magnitude of the signal, and the detailed form of signal itself.
The magnitude of the signal typically depends on the parameters of the system (planet and host), and largely dictates the detectability of the planet.  The detailed form of the signal typically depends on geometrical or nuisance parameters, but generally has an order unity impact on the magnitude of the signal itself.  These two contributions can often be relatively cleanly separated, such that one only needs to consider the magnitude of the signal to gain intuition about the scaling of the
signal-to-noise ratio of a given method with the stellar and planetary parameters, and thus detectability with these parameters. In the language of Section \ref{sec:formalism}, the magnitude of the signal generally depends on the set of the (more) physical parameters of interest $\vbeta_{\cal I}$, whereas the form of the signal depends on set of parameters $\vbeta_{\cal N}$, although I stress that this separability does not strictly apply over all the detection methods. 

Assuming this separability, the approximate signal-to-noise ratio of a planet signal can be written in terms of the magnitude or amplitude of the signal $A$, the number of observations $N_{\rm obs}$, and the typical measurement uncertainty $\sigma$, such that
\begin{equation}
 \snr \simeq A(\vbeta_{\cal I})
 \frac{N_{\rm obs}^{1/2}}{\sigma}g(\vbeta_{\cal N}),
\end{equation}
where $g(\vbeta_{\cal N})$ is a function of $\vbeta_{\cal N}$ whose value depends on the details of the signal, but is typically of order unity.  Thus, to roughly determine the scaling of the $\snr$ with the physical parameters of the planet and star (for an arbitrary survey), one must simply consider the scaling of the amplitude $A$ on the physical parameters. 

\subsection{Radial Velocity}\label{sec:snrrv}
The set of parameters that can be measured with radial velocities are $\vbeta=\left\{K,P,e,\omega,T_p,\gamma\right\}$.  Assuming uniform and dense
sampling of the RV curve over a time span that is long compared to the period $P$, the signal amplitude is $A=K(P,M_*,M_p,e,i)$, and the total signal-to-noise ratio scales as
\begin{equation}
    \snr_{\rm RV} \propto A \propto M_p P^{-1/3} M_*^{-2/3} \propto M_p a^{-1/2} M_*^{-1/2}.
\end{equation}
with a relatively weak dependence on eccentricity for $e \lesssim 0.5$. 

Thus radial velocity surveys are more sensitive to more massive planets, with the minimum detectable mass\footnote{Formally, the minimum detectable $M_p\sin{i}$.} at fixed $\snr$ scales as $P^{1/3}$ for planet periods of $P<T$, where $T$ is the duration of the survey.  

For $P>T$, it becomes increasingly difficult to characterize each of the parameters $\vbeta$ individually.  For $P\gg T$, the signal of the planet is an approximately constant acceleration ${\cal A}_*$, with a magnitude that is ${\cal A}_*=(2\pi K/P)f(\phi,\omega,e)$, where $f(x)$ is a function that depends on $\omega$, $e$, and the phase of the orbit $\phi$. Thus, a measurement of the acceleration can constrain the combination $M_p/P^{4/3}$.  By combining a measurement of ${\cal A}_*$ with the direct detection of the companion causing the acceleration, one can derive a lower limit on $M_p$ \cite{Torres:1999,Crepp:2012}.

\subsection{Transits}\label{sec:snrtr}
The set of parameters than can be measured with transits are $\vbeta=\left\{P,\delta,T,\tau,T_0,F_0\right\}$, where $T$ is the full-width half-maximum duration of the transit, $\tau$ is the ingress/egress duration\footnote{I note that a common alternative parameterization is to use $T_{14}$, the time between first and fourth contact, and $T_{23}$, the time between second and third contact (also referred to as $T_{\rm full}$ and $T_{\rm flat}$).  I strongly advise against adopting this parameterization, for several reasons.  First, the algebra required to transform from this parameterization to the physical parameters is significantly more complicated (compare \cite{Seager:2003} and \cite{Carter:2008}).  Second, $T_{14}$ and $T_{23}$ are generally much more highly correlated than $T$ and $\tau$, making the analytical interpretation of fits using the former parameterization much more difficult that using the latter parameters.  Finally, the timescale estimated from the Boxcar Least Squares algorithm \cite{Kovacs:2002} is much more well approximated by $T$ than $T_{14}$ or $T_{23}$.},  $T_0$ is a fiducial reference time, and $F_0$ is the out-of-transit baseline flux. If the radius of the host star can be estimated, than the radius of the planet can also be inferred via $R_p=\delta^{1/2}R_*$.  If the radial velocity of the host star can also be measured, it is then possible to determine the orbital eccentricity and planet mass $M_p$, and thus the density of the planet $\rho_p$. Of course, with additional follow-up observations, it is also possible to study the atmospheres for some transiting planets (e.g., \cite{Seager:2010}). 

Assuming uniform sampling over a time span that is long compared to the transit period $P$, we have that the signal amplitude is $A=N_{\rm tr}^{1/2}(\delta/\sigma)$, where $N_{\rm tr}= (N_{\rm tot}/\pi)(R_*/a)$ is the number of data points in transit and $N_{\rm tot}$ is the total number of data points. Therefore, $A=f(R_p,R_*,M_*,P)$, or $A=f(R_p,R_*,M_*,a)$. The signal-to-noise ratio of the transit
when folded about the correct planet period scales as \cite{Wright:2013}
\begin{equation}
    \snr_{\rm TR} \propto A \propto R_p^2 P^{-1/3} M_*^{-5/3} \propto M_p a^{-1/2} M_*^{-3/2}.
\end{equation}
Furthermore, the transit probability scales as as 
\begin{equation}
    P_{\rm tr} \propto \frac{R_*}{a} \propto P^{-2/3}M_*^{2/3} \propto a^{-1}M_*.
\end{equation}
Of course, the planet must transit {\it and} and the signal-to-noise ratio requirement must be met to detect the planet.  Finally, transit surveys require at least two transits to estimate the period of the planet, and often require at least three to aid in the elimination of false positives.  Thus the final requirement to detect a planet via transits is $P\leq T/3$.

Thus transit surveys are more sensitive to large planets, with the minimum detectable radius at fixed signal-to-noise ratio scaling as $P^{1/6}$ up until $P=T/3$.  Planets with periods longer than this are essentially undetectable with traditional transit selection cuts (but see Section \ref{sec:trsurveys}).  Furthermore, the transit probability decreases as $P^{-2/3}$.  Because of these two effects, transit surveys are very "front loaded" for reasonable planet distributions that do not rise sharply with increasing period, meaning that, e.g., doubling the duration of the survey will generally not double the yield of planets. 

\subsection{Microlensing}\label{sec:snrmicro}

Unlike most of the other detection methods discussed in this chapter, the detectability of planets via microlensing does not lend itself as well to the simple analytic description described above.  In particular, it is not possible to write down a simple scaling of the signal-to-noise ratio with the planet and/or host star properties. I will therefore simply review the essentials of microlensing and the parameter space of planets and stars to which it is most sensitive.  For more detail, I refer the reader to the following review and references therein \cite{Gaudi:2012}.

Briefly, a microlensing event occurs whenever a foreground compact object (the lens, which could be e.g., a planet, brown dwarf, star, or stellar remnant) passes very close to an unrelated background source star.  In general, for a detectable microlensing event to occur, the lens must pass within an angle of roughly the angular Einstein ring radius $\thetae$ of the lens,
\begin{equation}
\thetae \equiv \left( \kappa M \pi_{\rm rel} \right)^{1/2},
\label{eqn:thetae} 
\end{equation}  
where $\pi_{\rm rel} = \pi_l-\pi_s = {\rm au}/\drel$ is the relative lens ($\pi_l$)-source($\pi_s)$ parallax, $\drel^{-1} \equiv D_l^{-1} -D_s^{-1}$, $D_l$ and $D_s$ are the distances to the lens and source, respectively, and  
$\kappa \equiv 4G/(c^2{\rm au}) = 8.14~{\rm mas}~M_\odot^{-1}$ is constant.  For a typical stellar mass of $M_*=0.5M_\odot$, a source in the Galactic center with a distance of $D_s=8~{\rm kpc}$, and a lens half way to the Galactic center with $D_l=4~{\rm kpc}$, $\thetae\simeq 700~\mu{\rm as}$. The minimum lens-source alignment must be exquisite for a detectable microlensing event to occur, and given the typical number density of lenses along the line of sight and typical lens-source relative proper motions $\mu_{\rm rel}$, microlensing events are exceedingly rare.  Thus most microlensing surveys focus on crowded fields toward the Galactic center, where there are many ongoing microlensing events per square degree at any given time.

When a microlensing event occurs, the lens creates two images of the source, whose separations are of order $2\thetae$ during the event, and are thus generally unresolved (c.f. \cite{Dong:2019}).  However, the background source flux is significantly magnified if the lens passes within a few $\thetae$ of the source, resulting in a transient brightening of the source: a microlensing event.
The characteristic timescale of a microlensing event is the Einstein ring crossing time,
\begin{equation}
t_{\rm E} \equiv \frac{\thetae}{\mu_{\rm rel}},
\label{eqn:tedef}
\end{equation}
where $\mu_{\rm rel}$ is the relative lens-source proper motion. 
The typical timescale of observed microlensing events toward the Galactic bulge is $t_{\rm E}\sim20~$days \cite{Mroz:2019}, but can range from a few days to hundreds of days.  

A bound planetary companion to the lens can be detected during a microlensing event if it happens to have a projected separation and orientation that is close to the paths that the two images create by the host lens trace on the sky.  The planet will then further perturb the light rays from the source, causing a short duration deviation to the otherwise smooth, symmetric microlensing event due to the more massive host \cite{Mao:2011,Gould:1992,Bennett:1996}.  These deviations are also of order hours to days for terrestrial to gas giant masses.  Since the planets must be located close to the paths of one of the two images to create a significant perturbation, and the images are always close to the Einstein during the primary microlensing events, the sensitivity of microlensing is maximized for planets with angular separations of $\sim \thetae$.  At the distance of the lens, $\theta_{\rm E}$ corresponds to a linear Einstein ring radius of 
\begin{eqnarray}
r_{\rm E} &\equiv& \theta_{\rm E}D_l\\
 &=&2.85{\rm au}\left(\frac{M}{0.5 M_\odot}\right)^{1/2}
\left(\frac{D_s}{8~{\rm kpc}}\right)^{1/2}
\left[\frac{x(1-x)}{0.25}\right]^{1/2},
\label{eqn:requant}
\end{eqnarray}
where $x\equiv D_l/D_s$.  Thus the sensitivity of microlensing surveys for bound exoplanets peaks for planets with semimajor axes of $\sim 3~{\rm au} (M_*/0.5~M_\odot)^{1/2}$, which corresponds to orbital separations relative to the snow line (as defined in Eq.~\ref{eqn:snowline}) of $\sim 2~(M_*/0.5~M_\odot)^{-1/2}$.  Planets with significantly smaller semimajor axes become difficult to detect due to the fact that they perturb faint images.  Planets with significantly larger semimajor axis can be detected if the lens-source trajectory is aligned with the host star-planet projected binary axis.  However, the probability of having the requisite alignment decreases with increasing semimajor axis.  Eventually, the probability of detecting the magnification due to both the host and planet becomes exceedingly small.  Thus very wide separation planets are generally only detected as isolated, short-timescale microlensing events, which produce light curves that are essentially indistinguishable from events due to free-floating planets \cite{Han:2005}\footnote{I note that, in some cases, it is possible to disambiguate free-floating planets from bound planets by detecting (or excluding) light from the host.}.  Thus microlensing is, in principle, sensitive to planets with separations out to infinity, e.g., including free-floating planets \cite{Mroz:2018,Mroz:2019b,Mroz:2020,Kim:2020,Ryu:2020}.  The timescales of microlensing events caused by free-floating or widely-bound planets with terrestrial to gas giant masses also range from hours to days (since $t_{\rm E} \propto \thetae \propto M_p^{1/2}$).  

Microlensing events are rare and unpredictable.  Furthermore, planetary perturbations to these events are brief, unpredictable and generally uncommon.  Typical detection probabilities given the existence of a planet located with a factor of $\sim$2 of $\thetae$ are a few percent to tens of percent for terrestrial to gas giant planets \cite{Gould:1992,Bennett:1996}.
As with free-floating planets, the duration of planetary perturbations caused by bound planets scales as $\sim M_p^{1/2}$, and thus with range hours to days.
Therefore, microlensing surveys for exoplanets must observe many square degrees of the Galactic bulge on timescales of tens of minutes to both detect the primary microlensing events and monitor them with the cadence needed to detect the shortest planetary perturbations.

The magnitude of planetary perturbations are essentially independent of the mass of the planet, for planets with angular Einstein ring radii that are larger than the angular size of the source, i.e., when $\theta_{\rm E} \gtrsim \theta_*$.  For a source with $R_* \sim R_\odot$ at a distance of $D_s \sim 8~{\rm kpc}$, $\theta_* \sim 1~{\mu}$as.  The angular Einstein ring radius for an Earth-mass lens at a distance of 4~kpc and source distance of 8~kpc is $\thetae \sim 1~$mas.  Thus, for planets with mass $\lesssim M_\oplus$, finite source effects will begin to dominate, as the planet is only magnifying a fraction of the source at any given time. 

In summary, while the magnitude of the planetary perturbations to microlensing events, as well as the peak magnification of microlensing events due to widely-separated and free-floating planets, is essentially independent of planet mass for planets with mass $\gtrsim M_\oplus$, these signals become rarer and briefer with decreasing planet mass.  Nevertheless, planets with $M_p\gtrsim M_\oplus$ can be detected from current ground-based surveys \cite{Suzuki:2018}, and planets with $M_p\gtrsim 0.01M_\oplus$ (or roughly the mass of the moon) can be detected with a space-based microlensing survey \cite{Bennett:2002,Penny:2019}.  

The parameters that can be measured from a microlensing event with a well-sampled planetary perturbation are $\vbeta=\left\{t_0,t_{\rm E}, u_0, F_s, F_b, \alpha_{\mu{\rm L}}, q,s\right\}$. Here $t_0$ is the time of the peak of primary microlensing event, $u_0$ is the minimum lens-source angular separation in units of $\thetae$ (which occurs at a time $t_0$), $F_s$ is the flux of the source, $F_b$ is the flux of any light blended with the source but not magnified, which can include light from the lens, companions to the lens and/or source, and unrelated stars, $\alpha_{\mu{\rm L}}$ is the angle of the source trajectory relative to the projected planet-star axis\footnote{I note that in the microlensing liturature, this variable is typically simply refereed to as $\alpha$.  Since I have already defined $\alpha$ above, I adopt the form $\alpha_{\mu{\rm L}}$ to avoid confusion.}, $q\equiv M_p/M_*$ is the planet-star mass ratio, and $s$ is the instantaneous projected separation in units of $\thetae$.  Note that $F_s$ and $F_b$ can be (and usually are) measured in several bandpasses. The primary physical parameters of interest are $F_s$, $F_b$, $t_{\rm E}, q$, and $s$.  For most planetary perturbations, the effect of the finite size of the source on the detailed shape of the perturbation can also be detected, which allows one to also infer $\rho=\theta_*/\thetae$.  Since $\theta_*$ can be estimated from the color and flux of the source, $\thetae$ can generally be inferred, leaving a one-parameter degeneracy between the lens mass and distance (assuming $D_s$ can be estimated, as is usually the case).  This degeneracy can be broken in a number of ways (see \cite{Gaudi:2002} for a detailed discussion), but the most common method is to measure the flux of the lens \cite{Bennett:2007}.  This requires isolating light from the lens in the blend flux $F_b$, and thus typically requires high angular resolution from the ground using adaptive optics, or from space, in order to resolve light from unrelated stars from the lens and source flux. For space-based microlensing surveys, it is expected that the lens flux will be detectable for nearly all luminous lenses \cite{Bennett:2007}, and thus it will be possible to estimate $M_*$, $M_p$, $D_l$, and the instantaneous projected separation between the star and planet in physical units, for the majority of planet detections.  Since microlensing events can be caused by stars (or planets) all along the line of sight toward the Galactic bulge, it will be possible determine the frequency of bound and free-floating planets as a function of Galactocentric distance, and in particular determine if the planet population is different in the Galactic disk and bulge \cite{Penny:2016}.

\subsection{Direct Imaging}\label{sec:snrdi}
It is generally more complicated to characterize direct imaging observations. In part this is because there are three general cases of emission from the planet that must be considered.  The first, and the most relevant to current ground-based direct imaging surveys, is the case where the luminosity of the planet is dominated by the residual heat from formation, and is thus decoupled from the luminosity of, and distance from, its host star.  In this case, it is possible to measure the angular separation of the planet from its host star $\theta$.  The amplitude of the signal is simply the flux of the planet in a specific band $A=F_{p,\lambda}$, and thus $\vbeta=\left\{F_{p,\lambda},\theta\right\}$.  With a distance to the star, it is possible to determine the monochromatic luminosity of the planet as well as the instantaneous physical separation projected on the sky.  Using detailed cooling models of exoplanets as well as an estimate of the age of the system, it is possible to use the former to obtain a (model-dependent) estimate of the mass of the planet. With multiple epochs spanning a significant fraction of the orbit of the planet, it is possible to measure its Keplerian orbital elements (up to a two-fold degeneracy in the longitude of the ascending node). 

The second and third cases correspond to when the energy output from the residual heat from formation is negligible compared to the energy input due to the irradiation from the host star (e.g., when the planet is in thermal equilibrium with the host star irradiation).  In this case the spectrum of the planet has two components: reflected starlight and the thermal emission from starlight that is absorbed and then re-radiated as thermal emission. In the case of detection by reflected starlight, the basic parameters that can be measured are again the instantaneous angular separation from the host star and the amplitude of the signal which is simply given by $A=F_{p,\lambda}$. Thus $\vbeta=\left\{F_{p,\lambda},\theta\right\}$.  As before, the orbital elements can be inferred via a distance to the system and multiple epochs of astrometic observations of the planet.  In reflected light, the flux of the planet at any given epoch depends on the radius of the planet, its albedo, and its phase function.  The phase function can be estimated from measurement of the planet flux over multiple epochs of its orbit, leaving a degeneracy between the planet's radius and albedo.  The albedo can be roughly estimated from a spectrum of the planet, with a precision that depends on the particular properties of the system.  For a detection in thermal emission, the directly-measured parameters are (assuming that a spectrum can be obtained) $\vbeta=\left\{\theta,F_{p,\lambda}, T_{\rm eq}\right\}$, where $T_{\rm eq}$ is the equilibrium temperature of the planet.  As with detection in reflected light, the orbital elements can be determined with multiple epochs and a distance to the system.  In this case, the amplitude of the signal is also given by $A=F_{p,\lambda}$. By combining the detection of a planet in both reflected light and thermal emission, it is possible to determine $R_p$, $T_{\rm eq}$ and the albedo.  

In addition to these three kinds of emission, there are also multiple sources of noise in direct imaging surveys.  These include, but are not limited to: Poisson noise from the planet itself, noise from imperfectly removed light from the host, noise from the local zodiacal light, noise from the zodiacal light in the target system (the "exozodi"), and any other sources of background noise (e.g., read noise and dark current).  Which of these dominate (if any one dominate) depends on many factors, including, e.g., the planet/star flux ratio, the limiting contrast floor, the amount of exozodiacal dust in the target system, and the angular resolution of the telescope.

Rather than repeat the discussion of the parameters that can be measured in each case, or how the various physical parameters interplay to affect the detectability of directly-imaged planets, I will simply refer the reader to the discussion in \cite{Wright:2013}. However, in contrast with \cite{Wright:2013},  I will not assume any specific source of noise, and thus will not assume any specific relation between the noise and the properties of the host star.  Therefore, the scalings of the signal-to-noise ratio below reflect only the contributions due to the planet flux signal, and {\bf do not} include any contributions to the noise from the planet or host star, and thus do not include any scalings of the properties of the planet and/or host star with their potential contribution to the noise.  

With these caveats in mind, we have that amplitude of the signal is given by the planet flux $A=F_{p,\lambda}$.  Considering the three different cases discussed above, we have that, for planets that are not necessarily in thermal equilibrium with their host star, $F_{p,\lambda}=R_p^2T_p$, where $T_p$ is the temperature of the planet and I have assumed observations in the Rayleigh-Jeans tail (where the flux of a blackbody is linearly proportional to its temperature). For planets in thermal equilibrium with their host stars, we have that the planet flux is $F_{p,\lambda} \propto L_*R_p^2a^{-2}$ in reflected light, and $F_{p,\lambda} \propto R_p^2 T_p \propto R_p^2 L_*^{1/4}a^{-1/2}$ in the Rayleigh-Jeans tail.  Thus the scalings are 
\begin{eqnarray}
\snr_{\rm dir} &\propto& A \propto R_p^2 a^{-2} M_*^4 \propto R_p^2 P^{-4/3} M^{10/3}\quad  {\rm (Reflected~Light, Equilibrium)}\\
\snr_{\rm dir} &\propto& A \propto R_p^2 a^{-1/2} M_* \propto R_p^2 P^{-1/3} M_*^{5/6}\quad {\rm (Thermal,~Equilibrium,~RJ)}\\
\snr_{\rm dir} &\propto& A \propto R_p^2 T_p \hskip1.375in {\rm (Thermal,~RJ)},
\end{eqnarray}
where the last two expressions are only valid for the Raleigh-Jeans tail.

In addition, the planet must have a maximum angular separation that is outside the inner working angle $\theta_{\rm IWA}$ of the direct imaging survey, which is essentially the minimum angular separation from the star that the planet can be detected. This leads to the requirement that
\begin{equation}
    a \gtrsim \theta_{\rm IWA}d,
\end{equation}
where $d$ is the distance to the the star.
The inner working angle is generally tied to the wavelength of light $\lambda$ of the direct imaging survey, and the effective diameter $D$ of the telescope, or for interferometers, the distance between the individual apertures.  Specifically
\begin{equation}
    \theta_{\rm IWA} \sim N \frac{\lambda}{D},
\end{equation}
where $N$ is a dimensionless number that is typically between $2-4$ that primarily depends on the detailed properties of the starlight suppression system (coronagraph, starshade, or interferometer).  

In general, the above scalings imply that ground-based direct imaging surveys are generally most sensitive to massive, young giant planets at projected seprations of $\gtrsim 10~{\rm au}$ from their host star.  Direct imaging surveys for planets in thermal equilibrium have a more complicated selection function.  In terms of semimajor axis, they are typically most sensitive to planets on semimajor axis that are just outside of the inner working angle.  At fixed semimajor axis and distance, they are more generally sensitive to planets orbiting more massive stars.  However, more massive stars are generally rare and thus more distant (and thus are less likely to meet the inner working angle requirement). Furthermore, the contrast between the planet and star is larger for more massive stars (at fixed planet radius and semimajor axis).  Thus if residual stellar flux is the dominant source of noise, lower mass stars are preferred. The net result of these various considerations is that space-based direct imaging surveys in reflected light are generally most sensitive to solar-type stars, particularly for planets in the habitable zone (e.g., \cite{Agol:2007}).  

\subsection{Astrometry}\label{sec:snrastro}

Assuming a large number of astrometric measurements that cover a time span that is significantly longer than the period of the planet, the parameters than can be measured from the astrometric perturbation of star due to the orbiting planet are $\vbeta=\left\{\theta_{\rm ast},P,e,\omega,T_p,i,\Omega\right\}$, where $\Omega$ is the longitude of the ascending node\footnote{Note that, with only astrometric observations, there is a two-fold ambiguity in $\Omega$.}, $d$ is the distance to star, and the amplitude of the astrometric signal due to the planet is $A=\theta_{\rm ast}$, where
\begin{equation}
    \theta_{\rm ast}\equiv \frac{a}{d}\frac{M_p}{M_*}.
    \label{eqn:thetaast}
\end{equation}
Note that Equation \ref{eqn:thetaast} assumes $M_p \ll M_*$. 

In practice, for planetary-mass companions, the magnitude of the stellar proper motion $\mu_*$ and parallax $\pi$ is significantly larger than $\theta_{\rm ast}$, and thus if the astrometric perturbation from the planet can be detected, so can $\mu_*$ and $d$. The mass of the star can be estimated using the usual methods, and thus $M_p$ and $a$ can be inferred, along with the Keplerian orbital elements and orientation of the orbit on the sky (up to the two-fold degeneracy $\Omega$).  

Again, assuming a large number of astrometric measurements that cover a time span that is significantly longer than the period of the planet, the signal-to-noise ratio scales as
\begin{equation}
    {\rm (S/N)_{AST}} \propto A \propto M_p a M_*^{-1} d^{-1} \propto M_p P^{2/3} M_*^{-2/3} d^{-1}.
\end{equation}
Thus astrometry is more sensitive to more massive planets, as well as planets on longer period orbits.  However, unlike the RV method, the sensitivity of astrometry declines precipitously for planets periods $P$ greater than the duration of the survey $T$.  This is because such planets only produce an approximately linear astrometric deviation of the host star (particularly for $P\gg T$), which is then absorbed when fitting for the (much larger) stellar proper motion \cite{Casertano:2008,Gould:2008}.  

Thus the detection and characterization of exoplanets via astrometry has an additional criterion, namely that $P\lesssim T$.  The net result is that sensitivity function of astrometry in $M_p-P$ space has a 'wedge-shaped' appearance, with the minimum detectable planet mass a given (S/N) decreasing as $P^{-2/3}$ until $P\sim T$, and then increasing precipitously for $P>T$.

\subsection{Summary of the Sensitivities of Exoplanet Detection Methods}\label{sec:snrsummary}

Considering the five primary methods of detecting exoplanets: radial velocities, transits, microlensing, direct imaging, and astrometry, it is clear that all methods are (not surprisingly) more sensitive to more massive or larger planets.  

Radial velocity and transit surveys are generally more sensitive to shorter-period planets.  The sensitivity of the radial velocity method (in the sense of the minimum detectable planet mass) declines as $P^{1/3}$, and maintains this sensitivity scaling up to the survey duration $T$.  Thus, long-period or wide-separations planets can be detected only in radial velocity surveys with sufficiently long baselines.  For planets with $P>T$, planets produce accelerations or "trends", which can constrain a combination of the planet mass and period.  The sensitivity of transit surveys (in the sense of the minimum detectable planet radius) decline as $P^{1/6}$ up until roughly $T/3$, assuming three transits are required for a robust detection.  Planets with periods longer than this are undetectable under this criterion.  Thus radial velocity and transit surveys are generally less well-suited to constraining the demographics of long-period or wide-separation exoplanets. 

Astrometric surveys are more sensitive to longer-period planets, with their sensitivity (in the sense of the minimum detectable planet radius) increasing as $P^{2/3}$, up to $P\sim T$.  For planets with $P>T$, the sensitivity of astrometric surveys drops precipitously, e.g., it is very difficult to detect and characterize planets with $P>T$.

The sensitivity functions for microlensing and direct imaging surveys are generally more complicated.  Microlensing surveys are most sensitive to planets with semimajor axes that are within roughly a factor of two of the Einstein ring radius, which is $r_{\rm E}\sim 3~{\rm au}(M_*/M_\odot)^{1/2}$ for typical lens and source distances.  Microlensing is relatively insensitive to planets with semimajor significantly smaller than $r_{\rm E}$, but maintains some sensitivity to planets with separation significantly larger than $r_{\rm E}$, although the detection probability drops as $\sim a^{-1}$.  Microlensing is the only method capable of detecting old free-floating planets with masses significantly less than $\sim M_J$. The sensitivity functions of direct imaging surveys are similarly complicated.  Current ground-based surveys, which typically detect young planets via thermal radiation from their residual heat from formation are typically sensitive to planets that are relatively widely separated from (the glare of) their host stars.  Future space-based direct imaging surveys designed to detect mature planets in thermal equilibrium with their host stars in reflected light are generally only sensitive to planets with semimajor axes greater than $\sim~{\rm au}$ (due to the requirement that the planet angular semimajor axis is outside the inner working angle, which is typically a few $\lambda/D$), and their sensitivity declines as $R_p \propto a$.  Space-based direct imaging mission surveys designed to detect mature planets in thermal equilibrium with their host stars via their thermal emission are also generally only sensitive to planets with semimajor axis $\gtrsim 1{\rm au}$, and their sensitivity declines as $R_p \propto a^{-1/4}$.  

Thus, the strongest constraints on the demographics of long-period or wide-separation planets are expected to come from long-running radial velocity surveys, microlensing surveys, astrometic surveys, and direct imaging surveys.  Relatively short-duration radial velocity surveys and transit surveys in general do not constrain the demographics of long-period planets. 

Finally, I will make a few comments about the sensitivity of the various methods to the mass of host star.  All of the methods discussed here are more sensitive to planets at fixed mass $M_p$ or radius $R_p$ and period $P$ that orbit lower-mass stars, with the exception of direct imaging surveys for planets in thermal equilibrium.  For radial velocity surveys, the scaling of the (S/N) is $\propto M_*^{-2/3}$.  For transits, the scaling of the (S/N) is $\propto M_*^{-5/3}$.  For microlensing, the scaling is approximately $\propto M_*^{-1/2}$.  For direct imaging surveys, the scaling ranges from being independent of the host star mass (for young planets detected in thermal emission) to $\propto M_*^4$ (for planets in thermal equilibrium detected in reflected light).  For astrometric surveys, the (S/N) scales as $\propto M_*^{-2/3}$.

\section{The Demographics of Wide-Separation Planets}\label{sec:demo}

I now turn to summarizing some of the most important extant constraints on the demographics of long-period or wide separation planets.  I will first summarize results from each of the individual detection techniques that have placed at least some constraints on wide-separation planets, and then discuss efforts to synthesize results from multiple surveys using the same technique, and multiple surveys using different techniques.  There have been very few attempts to synthesize results from multiple surveys (regardless of whether or not they use the same detection technique) in general, for many reasons, some of which are discussed in Section \ref{sec:practice}.  However, because each technique is more or less sensitive to a particular region of planet ($M_p$ or $R_p$ and $P$ or $a$) and host star parameter space, such a synthesis is needed to assemble as complete a picture of exoplanet demographics as possible. Such a broad synthesis of exoplanet demographics, when performed correctly, provides the empirical ground truth to which all theories of planet formation and evolution must match. Thus synthesizing results from multiple surveys remains a fruitful avenue of future research.  

I note that, while I will attempt to highlight the most relevant results from the literature, it is simply not possible to provide a comprehensive and complete summary of all exoplanet demographics surveys in the limited amount of space available here.  In particular, I will generally not discuss results regarding the demographics of short-period planets, as this topic will be covered in another chapter in this book.  Specifically, I will not be discussing the vast literature on inferences about the demographics of relatively short-period planets from {\it Kepler}.

\subsection{Results from Radial Velocity Surveys}\label{sec:rvsurveys}

Radial velocity surveys for exoplanets have been ongoing since the late 1980s \cite{Campbell:1988,Latham:1989,Mayor:1992,Marcy:1992}, with the first widely-accepted discovery of an exoplanet using the radial velocity technique announced in 1995 with the discovery of 51 Pegasi b \cite{Mayor:1995}. A few of these radial velocity surveys have been ongoing since the early 1990s, and thus have accrued a baseline of roughly 30 years (corresponding to a semimajor axis of $\sim 10~{\rm au}$ for a solar-type star, or roughly the semimajor axis of Saturn).  The minimum achievable precision of these surveys have generally decreased with time, but the most relevant precision for detecting long-period planets is roughly the worst precision, which for the surveys mentioned above was in the range of a few m/s.  For a Jupiter-sun analog, $K\sim 13~{\rm m/s}$, $P\sim 12~{\rm yr}$ whereas for a Saturn-sun analog, $K\sim 3~{\rm m/s}$, $P\sim 30~{\rm yr}$.  Thus these long-term surveys are readily sensitive to Jupiter analogs orbiting sunlike stars, but Saturn analogs orbiting sunlike stars are just at the edge of their sensitivity \cite{Bryan:2016}.  Thus, extant radial velocity surveys can only constrain the population of giant planets ($M_p \gtrsim 0.15~M_{\rm Jup}$ beyond the snow line of sunlike stars. 

Surveys for planets orbiting M dwarfs generally have shorter baselines, but cover the complete orbits of planets with semimajor axes out to $3~{\rm au}$ \cite{Johnson:2010,Bonfils:2013}.  Giant planets with orbits longer than the baseline of these surveys can be detected via their trends, or, for sufficiently massive planets and sufficiently precise radial velocity observations, the planet properties can be characterized using partial orbits \cite{Bonfils:2013,Montet:2014}. However, as with surveys for planets orbiting solar-like stars, these surveys can generally only constrain the demographics of relatively massive planets beyond the snow line of M dwarfs.

\begin{figure}[t]
\includegraphics[width=6.5cm]{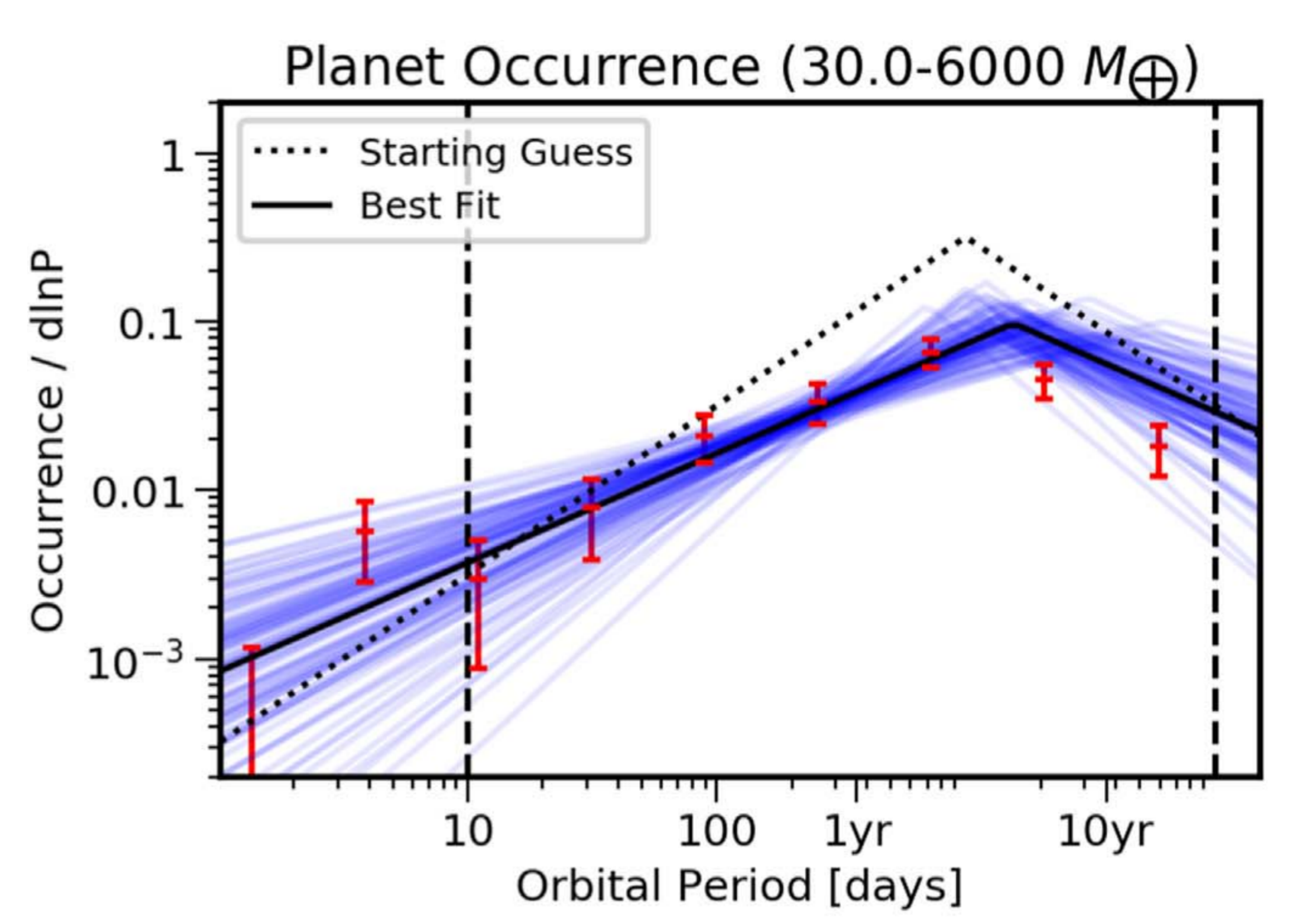}
\includegraphics[width=5cm]{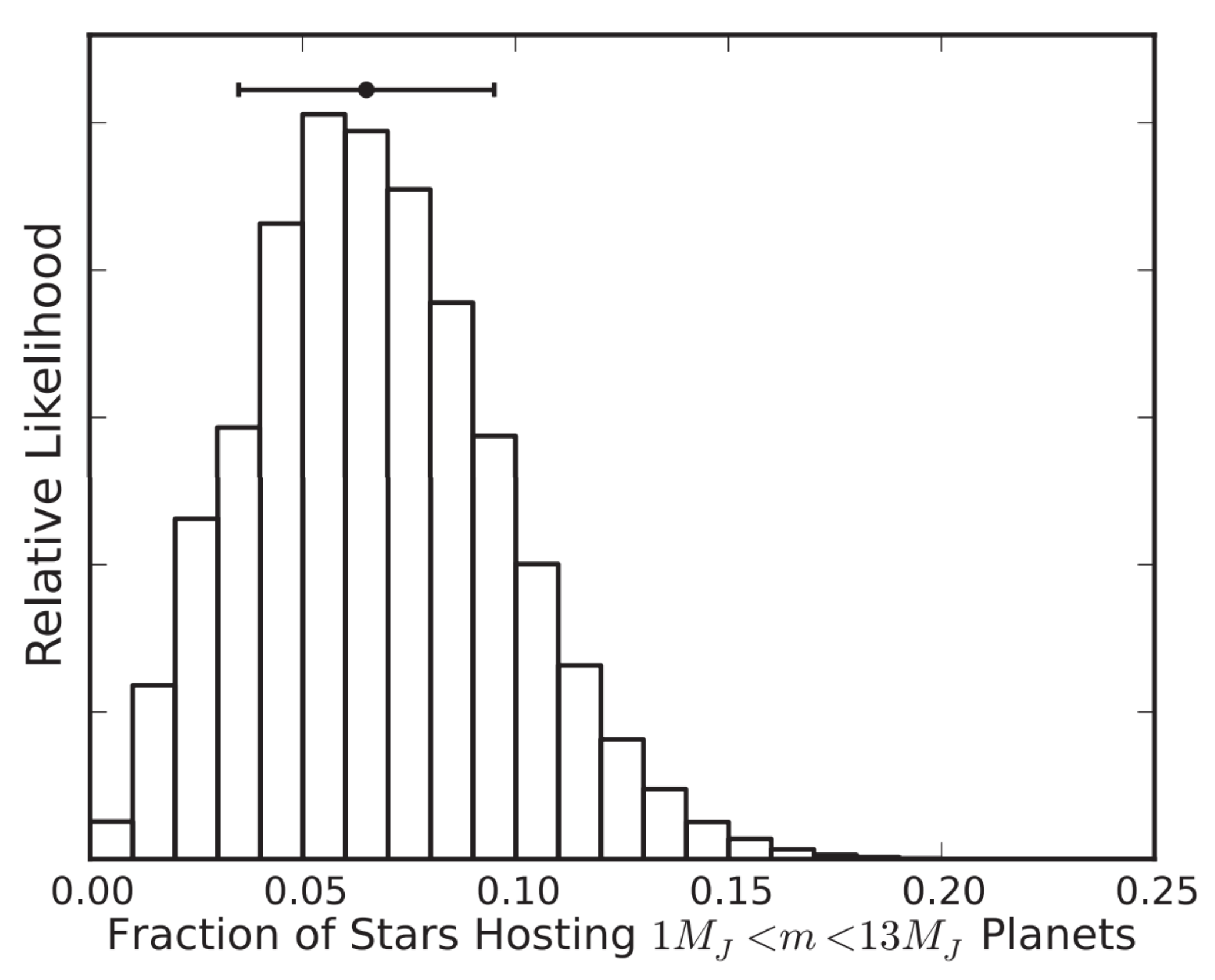}
\caption{(Left) Occurrence rate per $\ln{P}$ as a function of orbital period $P$ based on the analysis of the HARPS/CORALIE survey for exoplanets \cite{Mayor:2011}.
The red points show the binned occurrence rate, whereas the black broken power law shows the maximum likelihood fit to the unbinned data.  The blue broken power laws show samples from the $1\sigma$ range of fits.  The dashed line show the initial starting value of the fit.  Finally, the vertical lines show the range of periods considered in the fit. From \cite{Fernandes:2019}. \copyright AAS. Reproduced with permission. (Right) The distribution of the giant planet occurrence rates inferred from the California-Carnegie radial velocity survey of 111 M dwarfs. The median and $68\%$ confidence interval of the distribution implies that $6.5\% \pm 3.0\%$ of M dwarfs host a planet with mass between $M_{\rm Jup}-13M_{\rm Jup}$, with separations $<20~{\rm au}$.  From \cite{Montet:2014}. \copyright AAS. Reproduced with permission.}
\label{fig:rvconst}
\end{figure}

Using the sample of planets discovered in the HARPS and CORALIE radial velocity surveys of planets orbiting primarily sunlike stars from \cite{Mayor:2011}, and accounting for the survey completeness, \cite{Fernandes:2019} infer that the frequency of giant planets with masses in the range of $0.1-20~M_{\rm Jup}$ rises with increasing period (in agreement with previous results, e.g., \cite{Cumming:2008}) up to roughly $2-3~{\rm au}$ (e.g., roughly the snow line for sunlike stars), at which point the frequency of such planets begins to decrease with increasing period (see Figure \ref{fig:rvconst}). \cite{Wittenmyer:2016} analyzed 17 years of data from the Anglo-Australian Planet Search radial velocity survey of sunlike stars.   They estimated the frequency of "Jupiter analogs", which they define as planets with semimajor axes between $a=3-7~{\rm au}$, mass of $M_p>0.3~M_{\rm Jup}$, and eccentricity $e < 0.3$, to be $6.2^{+2.8}_{-1.6}\%$.  This is consistent with the result from \cite{Fernandes:2019}, implying that true Jupiter analogs are relatively rare.  Furthermore, extrapolating the declining frequency of giant planets beyond the snow line as inferred by \cite{Fernandes:2019} results in a frequency of giant planets accessible to direct imaging surveys that is consistent with detection rates from those surveys (with some caveats), which is not the case if one simply extrapolates the increasing frequency of giant planets inferred by \cite{Cumming:2008} from the analysis of the shorter-baseline California-Carnegie planet survey out to separations where direct imaging surveys would be sensitive to (young analogs) of them.  

Using a database of 123 known exoplanetary systems (primarily hot and warm Jupiters) monitored by Keck for nearly 20 years, combined with NIRC2 K-band adaptive optics (AO) imaging, \cite{Bryan:2016} estimated the frequency of giant planets with masses between $M_p=1-20~M_{\rm Jup}$ and $a=5-20~{\rm au}$ to be $\sim52\pm 5\%$.  This frequency is quite high, and in particular higher than the extrapolation of the results from \cite{Cumming:2008}. This suggests that the existence of hot/warm Jupiters and cold Jupiters may be correlated.  In other words, the conditional probability of a particular star hosting a cold Jupiter, given the existence of a hot/warm Jupiter, is not random.   

The constraints on the population of wide-separation planets orbiting low-mass stars are generally weaker.  Surveys for giant planets orbiting low-mass M stars ($M_*\lesssim 0.5~M_\odot$) have generally found a paucity of planets on relatively short orbital periods interior to the snow line \cite{Johnson:2010}.  This is seemingly a victory for the core-accretion model of planet formation, which generally predicts that giant planets should be rare around low-mass stars, due to their lower-mass disks and longer dynamical timescales at the distances where giant planets are thought to have formed \cite{Laughlin:2004,Johnson:2010b}.  However, these constraints were generally only applicable for planets with semimajor axes less than a few au, implying that giant planets could still form efficiently around low-mass stars, but perhaps do not migrate to the relatively close-in orbits where they can be detected via the relatively short time baseline radial velocity surveys for exoplanets orbiting low-mass stars.  Indeed, a closer inspection of the results from the HARPS \cite{Bonfils:2013} and California-Carnegie \cite{Montet:2014} surveys for planets orbiting M dwarfs hint at a significant population of giant planets at orbits with periods roughly equal to the survey duration.  See Figure \ref{fig:rvconst}.

\subsection{Results from Transit Surveys}\label{sec:trsurveys}

As discussed in Section \ref{sec:snrtr}, transit surveys are generally not sensitive to planets with periods longer than $T/3$ of the survey duration $T$.  This is because such surveys impose an arbitrary (but reasonable) criterion that at least three transits must be detected.  A robust detection of at least two transits is generally required to infer the period of the planet, whereas the detection of a third transit largely eliminates most false positives.  

However, as first pointed out by \cite{Yee:2008}, based on the arguments from \cite{Seager:2003}, the robust detection of a single transit can be used to estimate the period of the planet, given an estimate of the density of the star $\rho_*$ and assuming zero eccentricity.  Since single transits from long-period planets typically have long durations, they also typically have quite large signal-to-noise ratios, which are sufficient to allow an estimate of the planet period, under the assumptions above \cite{Yee:2008}. Indeed, \cite{Yee:2008} used this fact to argue that the follow-up of single transit events could be used to extend the period sensitivity of transit surveys, and in particular that of {\it Kepler}.  As \cite{Yee:2008} argued, not only does this require an estimate of $\rho_*$, but it also generally requires nearly immediate radial velocity observations of the target star to measure the acceleration of the star due to the planetary companion. 

\begin{figure}[t]
\includegraphics[width=12cm]{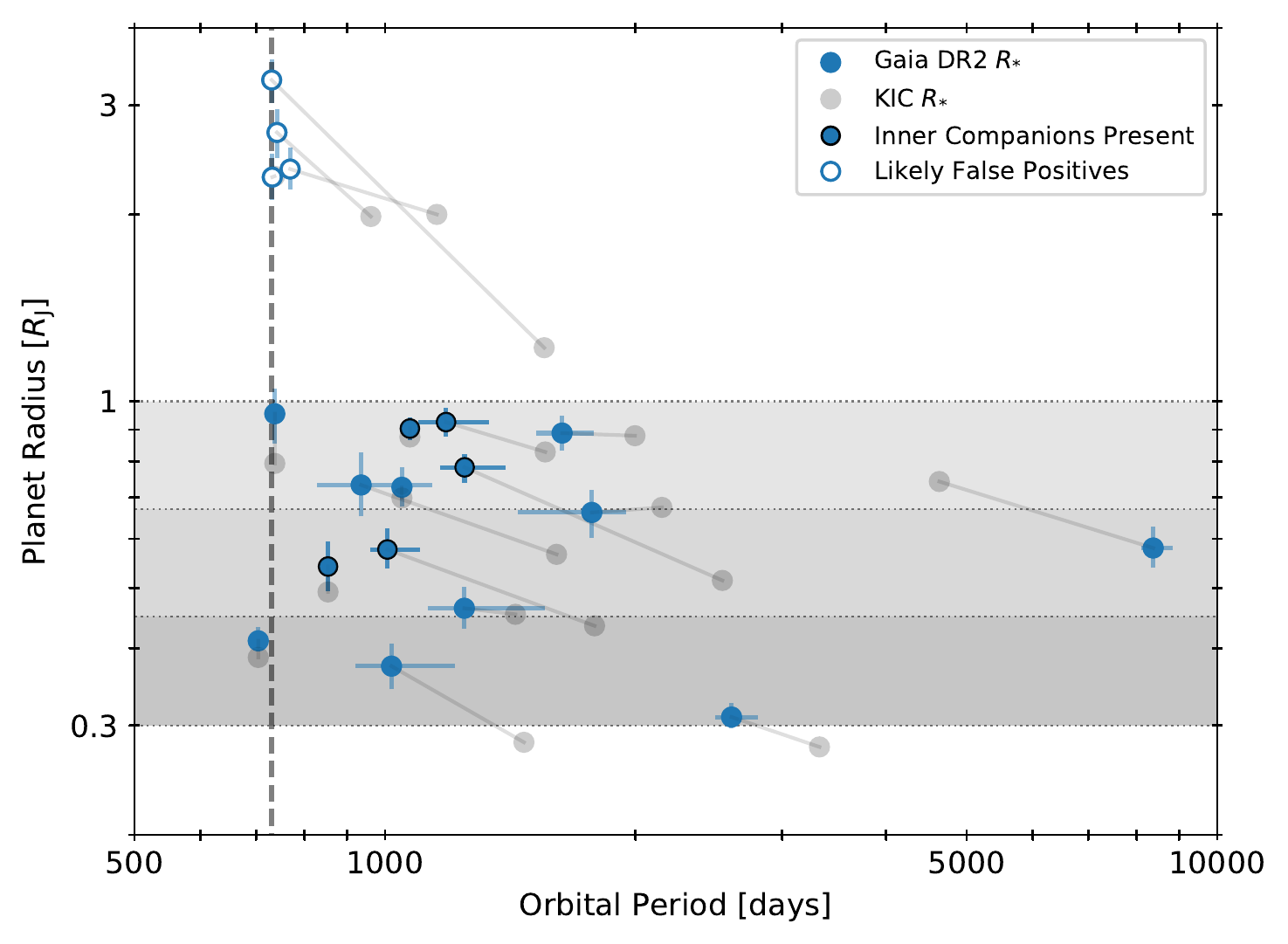}
\caption{The results of the single-transit candidate search from \cite{Herman:2019}. The parameters of the candidates using revised stellar parameters are plotted in blue, while gray points
indicate the parameters inferred using the older {\it Kepler} input catalog values. The vertical dashed line denotes the maximum possible period to exhibit at least three transits during the {\it Kepler} primary mission. From \cite{Herman:2019}. \copyright AAS. Reproduced with permission.}
\label{fig:trconst}
\end{figure}

The suggestion of \cite{Yee:2008} was largely ignored during the primary {\it Kepler} mission.  Likely this was because the yield of such single-transit events was expected to be small.  Nevertheless, two groups \cite{Foreman-Mackey:2016,Herman:2019} endeavored to estimate the planet occurrence rate for planets with periods beyond the nominal range of the {\it Kepler} primary mission using the methods outlined in \cite{Yee:2008}. In particular, \cite{Herman:2019} used updated information about the planet host stars from the Gaia DR2 release\cite{Gaia:2018} to improve the purity of the long-period single-transit sample (see Figure \ref{fig:trconst}). They inferred a frequency of cold giant planets (radii between $0.3-1~R_{\rm Jup}$ and periods of $\sim 2-10~{\rm yr}$ of   $0.70^{+0.40}_{-0.20}$.  This rate is consistent with the results of \cite{Cumming:2008}, albeit with larger uncertainties.  They also infer a radius distribution of planets beyond the snow line of 
\begin{equation}
    \frac{d N}{d \log{R_p}}=R_p^{-1.6^{+1.0}_{-0.9}},
\end{equation}
consistent with the conclusion from microlensing surveys that Neptunes are more common than Jovian planets beyond the snow line \cite{Gould:2006b}.  Finally, they note that 5 of the 13 long-period planet candidates they identify have confirmed inner transiting planets, indicating that there is a strong correlation between the presence of cold
planets with warm/hot inner planets, and that the mutual inclinations between the inner and outer planets must be small.  The results of \cite{Herman:2019} demonstrate the importance of a holistic picture of exoplanet demographics.

\subsection{Results from Direct Imaging Surveys}\label{sec:disurveys}

A relatively recent, cogent, and exceptionally comprehensive review of direct imaging surveys has been provided by Bowler \cite{Bowler:2016}. Rather than attempt to reproduce or replicate the contents of that exceptional review, I will simply highlight the most important conclusions from direct imaging surveys as summarized therein. 

\begin{figure}[t]
\includegraphics[width=12cm]{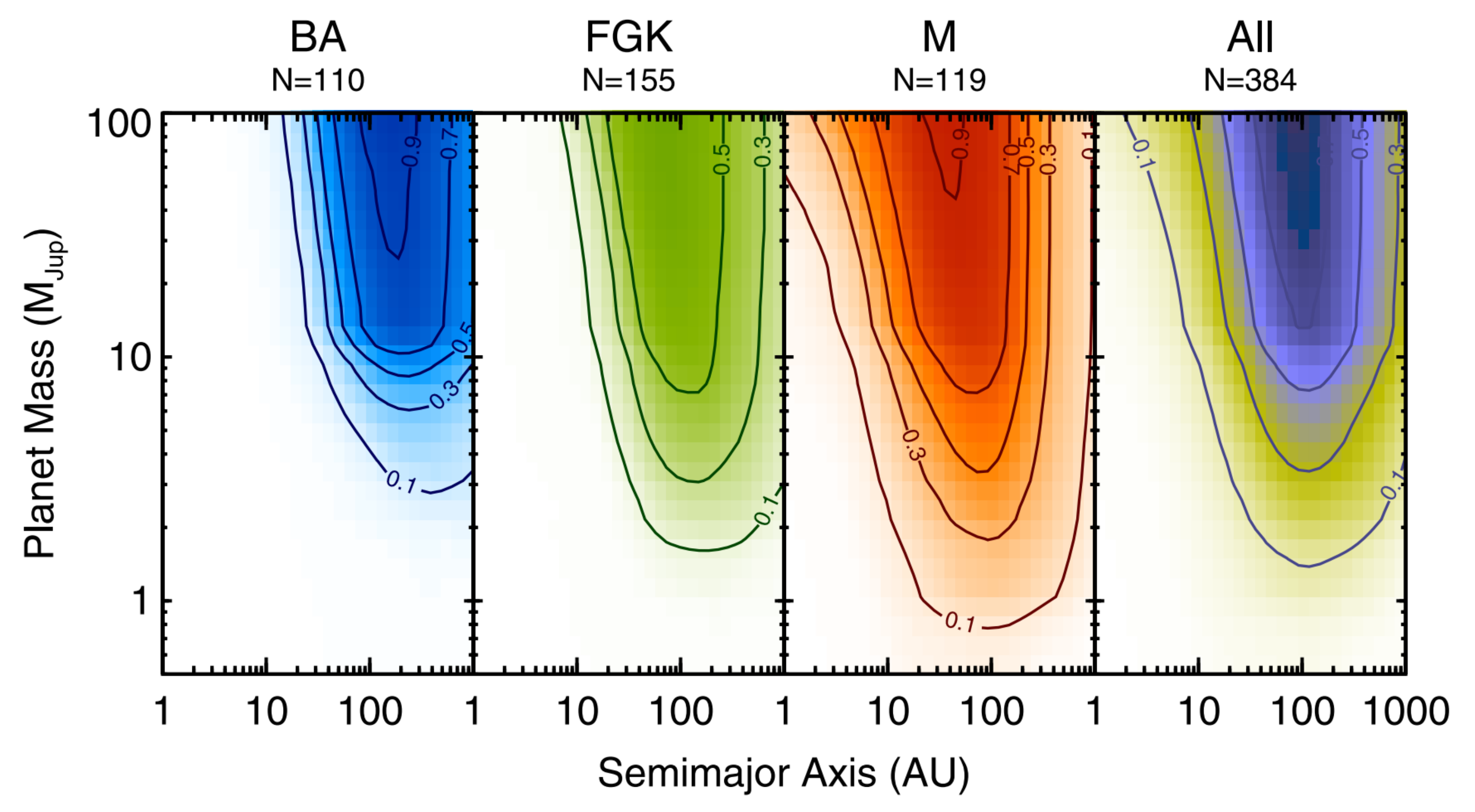}
\caption{Completeness contours from an ensemble analysis of $\sim380$ unique stars with published high-contrast imaging observations. The contours denote
10\%, 30\%, 50\%, 70\%, and 90\% completeness limits.  Note that low-mass stars provide stronger constraints for planets at a fixed mass and semimajor axis. From \cite{Bowler:2016} \copyright ~Publications and the Astronomical Society of the Pacific. Reproduced with permission.
}
\label{fig:diconst}
\end{figure}

As shown in Figure \ref{fig:diconst}, in the current state of the art direct imaging surveys for exoplanets \cite{Hinkley:2011,Macintosh:2014,Beuzit:2019,Skemer:2014} are primarily sensitive to planets with masses of $\gtrsim M_J$ and orbits of $\gtrsim 10~{\rm au}$. 

These surveys have searched for planetary companions orbiting roughly $\sim 400$, relatively young ($<300~{\rm Myr}$) stars, with spectral types ranging from low-mass M stars to B stars. Roughly 8 planetary ($\lesssim 13~M_{\rm Jup}$) candidates with semimajor axes of $\lesssim 100~{\rm au}$ have been found orbiting main-sequence stars \cite{Lagrange:2010,Macintosh:2015,Rameau:2013,Marois:2008,Marios:2010,Kraus:2012}.  

The majority of these candidates are very dissimilar to the giant planets in our solar system: they are typically more massive and have much larger separations.  This has led to the speculation that these planets formed via a different mechanism than the core accretion mechanism \cite{Pollack:1996} that is typically invoked to explain the formation of shorter-period giant planets detected by other methods.  Indeed, the massive, long-period planets detected by direct imaging have been suggested to be evidence of planet formation via gravitational collapse \cite{Boss:1997,Dodson-Robinson:2009}. However, it has been argued that there are serious theoretical difficulties with this formation mechanism \cite{Rafikov:2005}.  If gravitational instability can create long-period giant planets, the theoretical prediction is that there should be a larger population of brown dwarfs \cite{Kratter:2010}.  Indeed, this has been observed \cite{Nielsen:2019}. There is some evidence that the orbits of directly-imaged planets are distinct from those found via the radial velocity method, perhaps suggesting that there are indeed two mechanisms for giant planet formation \cite{Bowler:2020}.  

Even if there are two channels of forming giant planets, the population of planets formed by these two different channels may not occupy distinct regions of parameter space.  For example, the directly-imaged planet candidate 51 Eridani b \cite{Macintosh:2015}, which has an estimated mass of $\sim 2~M_{\rm Jup}$ and a semimajor axis of $\sim 13~{\rm au}$, has properties that are quite similar to the giant planets detected via radial velocity surveys, as well as the giant planets in our solar system.

\begin{figure}[t]
\includegraphics[width=12cm]{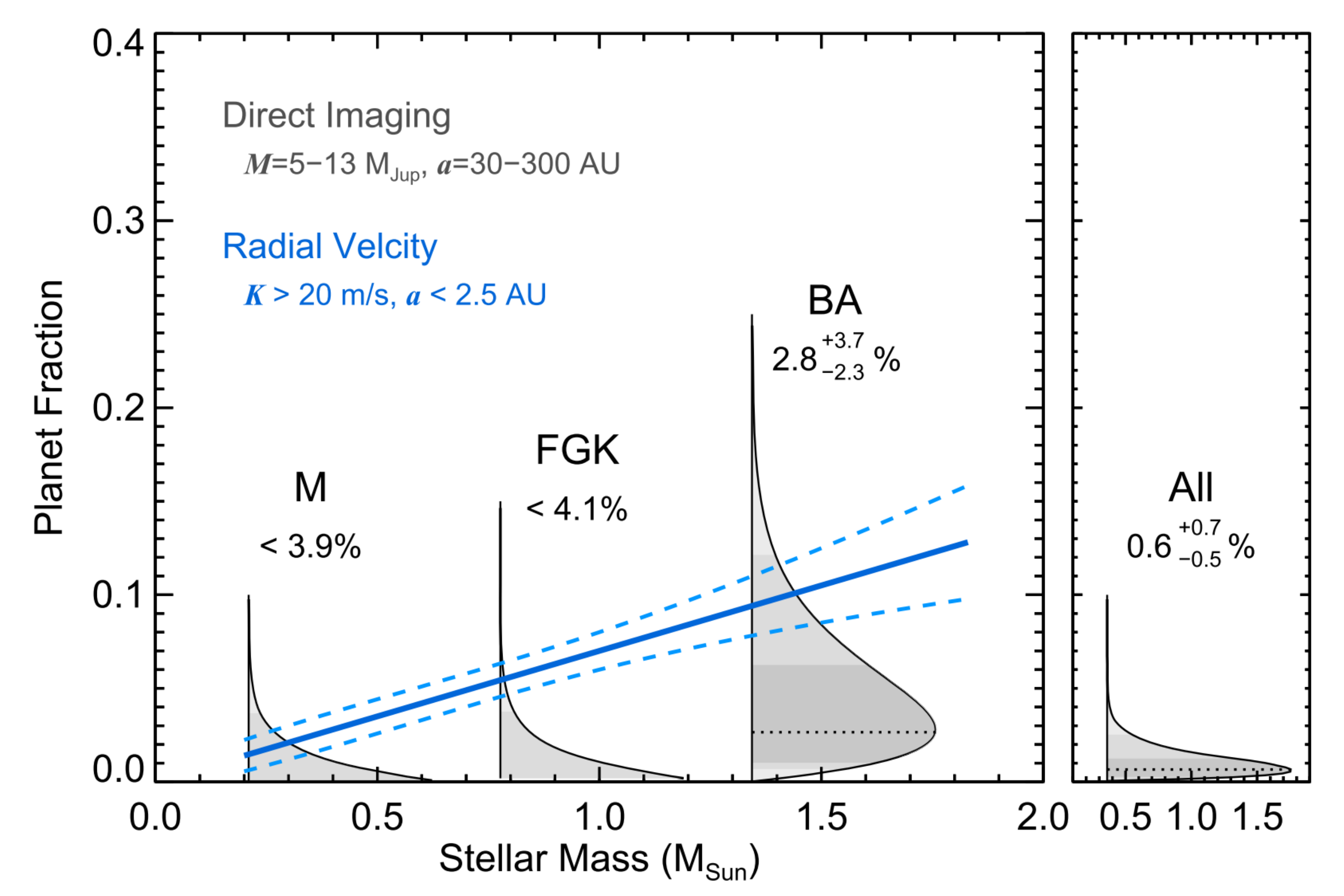}
\caption{
Distributions of the occurrence rate of giant planets from an ensemble analysis of direct imaging surveys in the literature, as compared to the results from radial velocity surveys for giant plants at small separations ($\lesssim 2.5~{\rm au}$) \cite{Johnson:2010b}.
There is no statistically significant evidence for a correlation between the giant planet occurrence and host star spectral type for planets at large separations probed by direct imaging surveys.  Determining whether or not this is consistent with the significant trend inferred by \cite{Johnson:2010b} will require a larger sample of directly-imaged planets. From \cite{Bowler:2016}. \copyright~Publications of the Astronomical Society of the Pacific. Reproduced with permission.}
\label{fig:dioccur}
\end{figure}

Finally, it is worth noting that there is no statistically significant evidence for an increase in the frequency of long-period giant planets with host star spectral type (a proxy for host star mass) from direct imaging surveys, as shown in Figure \ref{fig:dioccur}.  However, given the relatively small number of giant planet candidates detected by direct imaging surveys, this result is not statistically discrepant with the result from radial velocity surveys that the frequency of shorter-period giant plants increases with increasing host mass \cite{Johnson:2010b}. 

\subsection{Results from Microlensing Surveys}\label{sec:mlsurveys}

\begin{figure}[t]
\includegraphics[width=12cm]{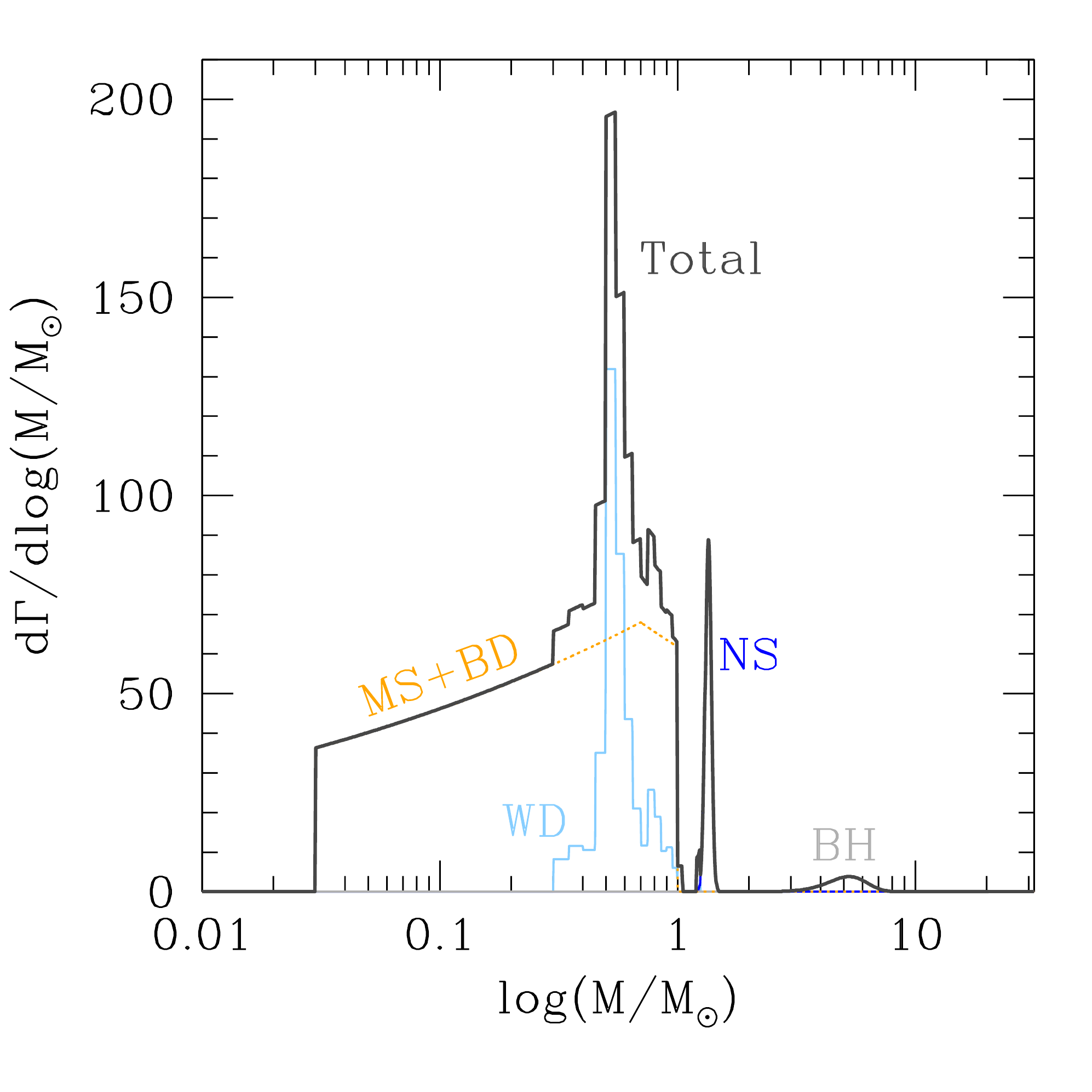}
\caption{
The predicted distribution of host masses for microlensing surveys for exoplanets.  Brown dwarfs BD), main-sequence (MS) stars, and remnants [including white dwarfs (WD), neutron stars (NS), and black holes (BH)] are included.  Considering only MS stars, the median host star mass surveyed by microlensing is $\sim 0.4~M_\odot$.  Note that MS stars with masses above the bulge turn-off have been suppressed; in reality the distribution of host star masses will include a small contribution from more massive MS stars in the Galactic disk.  From \cite{Gaudi:2002}, adapted from \cite{Gould:2000}. © ARAA. Reproduced with permission.}
\label{fig:micromass}
\end{figure}

Microlensing surveys for exoplanets cannot choose their target stars - rather, the sample of hosts around which microlensing surveys can constrain the properties of planetary systems is dictated by the number density of compact objects (brown dwarfs, stars, and remnants) weighted by the event rate, which scales as $\thetae \propto M_*^{1/2}$.  Figure \ref{fig:micromass} shows the predicted distribution of host masses probed by microlensing \cite{Henderson:2014,Gould:2000}.  Ignoring brown dwarfs and remnants, the median mass of main-sequence (MS) stars probed by microlensing surveys is $\sim 0.4 M_\odot$.  Thus, although a little over half the MS host stars will be M dwarfs, microlensing surveys still probe the frequency of planets beyond the snow line orbiting G and K type stars, a fact that is not widely appreciated.  

The first constraints on the frequency of planets from microlensing surveys was by \cite{Gaudi:2002}, based on five years of data from the Probing Lensing Anomalies with a world-wide NETwork (PLANET) collaboration \cite{Albrow:1998}.  Although they did not detect any planets, they were able to place a robust upper limit on the frequency of massive companions.  They concluded that $<33\%$ of hosts have $\sim M_{\rm Jup}$ companions with separations between 1.5-4~au, and less than $<45\%$ of hosts have $\sim 3 M_{\rm Jup}$ with separations between 1-7~au.  As the majority of these hosts were M dwarfs, this result provided the first significant limits on planetary companions to M dwarfs.

The first conclusive discovery of an exoplanet by microlensing was in 2004 \cite{Bond:2004}. Additional detections followed soon after \cite{Udalski:2005,Beaulieu:2006,Gould:2006b}.  Of the first four planets detected via microlensing, two had Jovian mass-ratios, whereas the other two has super-Earth/Neptune mass ratios.  Given the decreasing sensitivity of microlensing for smaller mass ratios (See \ref{sec:snrmicro}), this immediately implied that cold low-mass (super-Earth to Neptune) mass planets were much more common than cold Jovian planets \cite{Gould:2006b}. 

\begin{figure}[t]
\includegraphics[width=12cm]{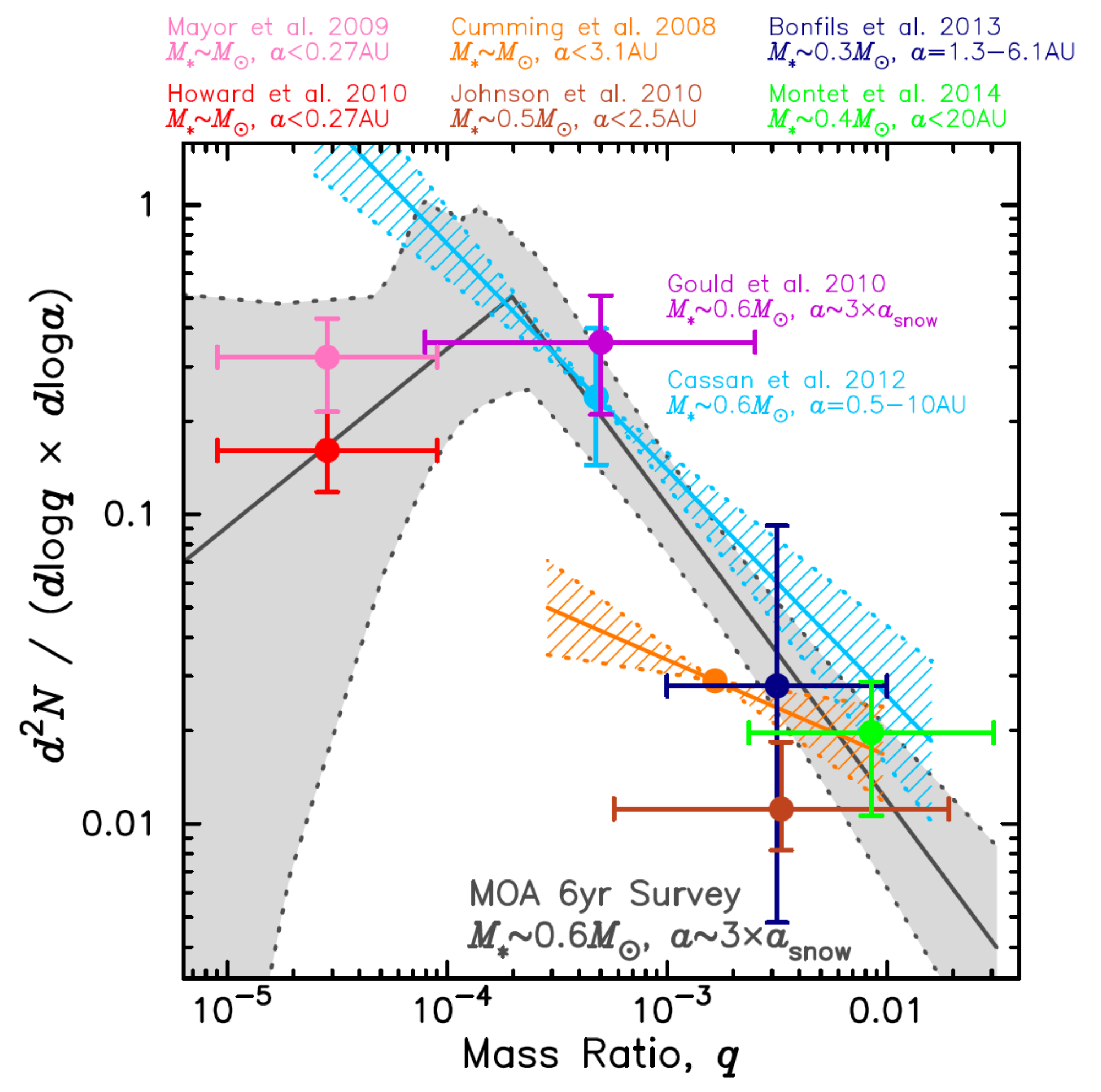}
\caption{
The distribution of planet/star mass ratios inferrred from \cite{Suzuki:2016}. The black line shows best-fit
broken power-law mass-ratio function (see Equation~\ref{eqn:suzuki}), whereas the gray shaded region show the uncertainties about this best-fit model.  This result is compared to compendium of demographic constraints from other radial velocity and microlensing surveys \cite{Mayor:2011,Howard:2010,Cumming:2008,Johnson:2010,Bonfils:2013,Montet:2014}. Note that the typical primary host mass and semimajor axis range vary amongst the various results.  From \cite{Suzuki:2016}, adapted and updated from \cite{Gould:2000} and \cite{Gaudi:2002}. \copyright AAS. Reproduced with permission.
}
\label{fig:suzuki}
\end{figure}

To date, over 100 planets have been detected by microlensing\footnote{See https://exoplanetarchive.ipac.caltech.edu/}. Individual surveys have detected a sufficiently large sample of planets that they have been able to place robust constraints on the population of cold planets orbiting main-sequence stars \cite{Sumi:2010,Gould:2010,Cassan:2012,Shvartzvald:2016}, see Figure \ref{fig:suzuki} for a graphical representation of the constraints derived by \cite{Gould:2010,Cassan:2012}.  

The most recent and thorough analysis with the largest sample of planets is that by \cite{Suzuki:2016}, who analyzed six years of data from the second generation Microlensing Observations in Astrophysics (MOA-II) collaboration \cite{Sumi:2013}. The analysis included 23 planets detected from 1474 alerted microlensing events.  They find that the distribution of mass ratios $q$ and projected separations $s$ is well-described by a broken power law, with the form:
\begin{equation}
   \frac{dN}{d\log{q}~d\log{s}}= 
   0.61^{+0.21}_{-0.16}\left[\left(\frac{q}{q_{\rm br}}\right)^{-0.93\pm 0.13}{\cal H}(q-q_{\rm br}) + \left(\frac{q}{q_{\rm br}}\right)^{0.6^{+0.5}_{-0.4}}{\cal H}(q_{\rm br}-q)\right]s^{0.49^{+0.47}_{-0.49}},
   \label{eqn:suzuki}
\end{equation}
where, as before, ${\cal H}(x)$ is the Heaviside step function.
Thus, \cite{Suzuki:2016} find that the mass ratio function of cold exoplanets with mass ratios above that of $q_{\rm br}\sim 1.7\times 10^{-4}$ (roughly the mass ratio of Neptune to the sun) is steeper than that mass function for shorter-period giant planets found by \cite{Cumming:2008}.  See Figure \ref{fig:suzuki}.  In addition, the distribution of orbits for planets beyond the snow line is consistent with a log-uniform distribution, and planets with Neptune/sun mass ratios are likely the most common planets beyond the snow line.  \cite{Suzuki:2016} also synthesized their MOA-II constraints with those of \cite{Cassan:2012}, which also included constraints from \cite{Sumi:2010, Gould:2010}.  The combined results from these four surveys (\cite{Sumi:2010,Gould:2010,Cassan:2012,Suzuki:2016}) are shown in Figure \ref{fig:suzuki2018}. 

Using a sample of seven planets detected by microlensing with well-measured mass ratios of $q<10^{-4}$ and unique solutions, \cite{Udalski:2018} confirmed the result from \cite{Suzuki:2016} that the mass ratio function for planets with mass ratio below $q_{\rm br}$ declines with decreasing mass ratio.  They inferred a power-law index in this regime of $1.05^{+0.78}_{-0.68}$, as compared to the value of $0.6^{+0.5}_{-0.4}$ found by \cite{Suzuki:2016}.  By combining their results with those of \cite{Suzuki:2016} and \cite{Udalski:2018}, they refine the power-law index in this regime to
$0.73^{+0.42}_{-0.34}$.

However, using a sample of 15 planets with well-measured mass ratios of $q<3\times 10^{-4}$, \cite{Jung:2019} arrived at a somewhat different conclusion than \cite{Suzuki:2016} and \cite{Udalski:2018}.  Again assuming the same double power-law form as Equation~\ref{eqn:suzuki}, they find a smaller value for the break in the mass function of $q_{\rm br}=0.55 \times 10^{-4}$, e.g., a factor of $\sim 3$ times smaller than found by \cite{Suzuki:2016}. Furthermore, they inferred a very large difference in the slope of the mass function for mass ratios above and below $q_{\rm br}$, with a best fit of 5.5 ($>3.3$ at $1\sigma$). Assuming the power-law index of $-0.93$ for $q>q_{\rm br}$ found by \cite{Suzuki:2016}, which they did not constrain, this implies a very steep power law index for $q<q_{\rm br}$ of $4.6$ (with an upper limit of $2.4$ at $1\sigma$), as compared to the value of $0.6^{+0.5}_{-0.4}$ and $0.73^{+0.42}_{-0.34}$ found by \cite{Suzuki:2016} and \cite{Udalski:2018}, respectively.  \cite{Jung:2019} also note an apparent `pile-up' of planets with mass ratio similar to that of Neptune to the sun.  Specifically, four of their 15 planets have mass ratios between $0.55\times 10^{-4}$ and $5.9\times10{-4}$, a span of only $\Delta \log q = 0.030$, as compared to the full range spanned by their sample of $\Delta \log q = 0.875$.  However, they were unable determine conclusively whether this `pile-up' is real or due to a statistical fluctuation. 

As discussed in Section \ref{sec:snrmicro}, microlensing is uniquely sensitive to widely-bound and free-floating planets.  These are detectable as isolated, very short timescale (hours-to-days) microlensing events.  The first constraints on the frequency of free-floating or widely bound planets from microlensing was by \cite{Sumi:2011}. Based on an excess of short timescale ($\sim 1~{\rm day}$) events, the argued for the existence of a population of free-floating or widely-separation planets with masses of $\sim M_{\rm Jup}$, with a frequency of roughly twice that of stars in the Milky Way.  By comparing to the frequency of giant planets found by direct imaging surveys, \cite{Clanton:2017} were able to demonstrate that $\gtrsim 70\%$ of these events must be due to free-floating planets.  There are significant difficulties with producing such a large population of free-floating planets, which I will not expound upon here (but see \cite{Veras:2012, Ma:2016}).  This is because a subsequent analysis of the Optical Gravitational Lensing Experiment (OGLE, \cite{Udalski:2015}) data conclusively demonstrated that the purported excess of short-timescale microlensing events by MOA-II was spurious \cite{Mroz:2017}.

Curiously, \cite{Mroz:2017} do find an excess of very short timescale ($\lesssim 0.5~{\rm days}$) candidate microlensing events, which may be an indication of population of free-floating or widely-bound planets with planets of mass of $\sim 1-10~M_\oplus$.  Some theories of planet formation predict the ejection of a significant number of such low-mass planets during the chaotic phase of planet formation.

Since the \cite{Mroz:2017} result, a total of seven robust free-floating planet or wide-orbit planets have been discovered via microlensing \cite{Mroz:2018,Mroz:2019b,Mroz:2020,Kim:2020,Ryu:2020,Mroz:2020b} primarily using data from the OGLE and Korea Microlensing Telescope Network (KMTNet) collaborations \cite{Henderson:2014,Kim:2016}.  

\begin{figure}[t]
\includegraphics[width=12cm]{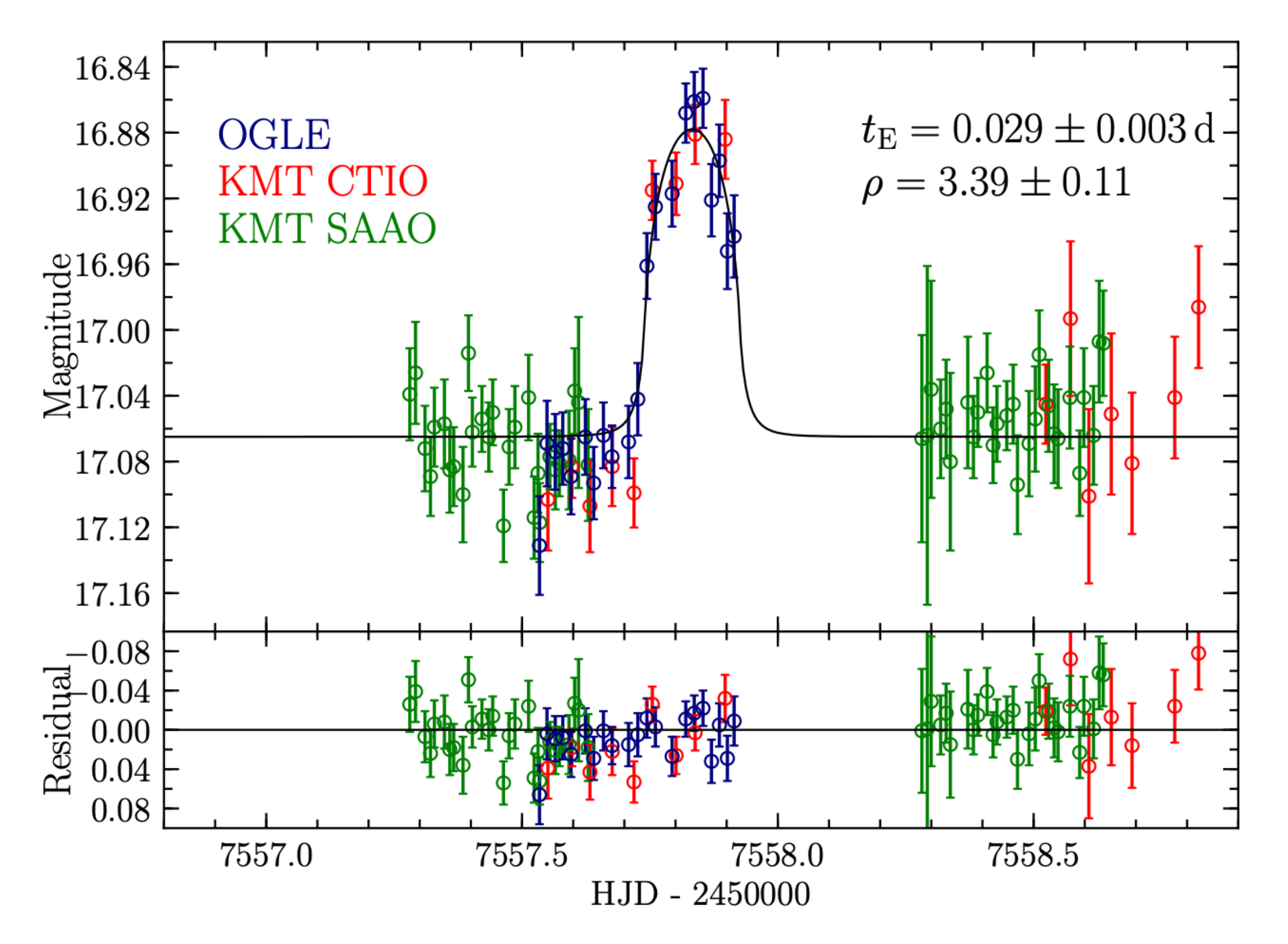}
\caption{
Data for microlensing event OGLE-2016-BLG-1928, which has the shortest Einstein timescale of any microlensing event detected to date. It is likely the lowest-mass free-floating or widely-bound ($\gtrsim 10~{\rm au}$) planet detected to date, with an estimated mass between Mars and the Earth.  From \cite{Mroz:2020}. \copyright~AAS. Reproduced with permission.  
}
\label{fig:mroz}
\end{figure}

Figure \ref{fig:mroz} shows the data for the shortest-timescale microlensing event detected to date, with an Einstein timescale of only $t_{\rm E} \sim 40~{\rm minutes}$ \cite{Mroz:2020}.  The lens likely has a mass in the Mars- to Earth-mass regime, with lower masses being favored.  If the planet is bound to a host star, it must have a projected separation of $\gtrsim 10~{\rm au}$. 

\subsection{Synthesizing Wide-separation Exoplanet Demographics}\label{sec:synthesis}

Ultimately, because of the intrinsic sensitivities and biases of all exoplanet detection methods (as discussed in Section \ref{sec:methods}), it is not possible for any single method or survey to provide the broad constraints on exoplanet demographics that are needed to properly constrain and refine planet formation theories.  Thus, multiple surveys using multiple detection methods must be "stitched together" to provide the needed empirical constraints.  

We are fortunate that the various detection methods at our disposal are largely complementary, and can, in principle, constrain exoplanet demographics over nearly the full range of planet and host-star properties needed to fully test planet formation theories. However, combining the results from various surveys and methods is not trivial. Often, exoplanet researchers have the relevant expertise in only one or perhaps two detection methods.  Surveys often do not report the details needed to combine their results with other surveys, such as providing the appropriate information about their target sample, or providing the individual detection sensitivities for each of their targets (including those for which no planetary candidates were detected).

For these reasons and others, the obstacles to synthesizing the demographics of exoplanets are significant, which is likely the reason why very little progress in this area has been made.  Nevertheless, exoplanet surveys now have significant overlap in terms of the parameter space of the planet and host star properties, and thus there is an opportunity to make significant progress in constructing a broad statistical census of exoplanet demographics, using results that are already in hand. 

In this section, I will highlight a few notable attempts to synthesize the results of wide-orbit demographics from multiple surveys using multiple methods. However, I emphasize that much more work needs to be done in this area.

Two of the first rigorous attempts to compare the frequency of giant planets as constrained by radial velocity and microlensing surveys were performed by \cite{Montet:2014} and \cite{Clanton:2014a,Clanton:2014b}.  Both groups focused on low-mass stellar hosts, and used the results from trends found in radial velocity surveys of relatively low-mass hosts (e.g., evidence for companions with periods longer than the duration of the survey) to constrain the frequency of long-period giant planets.  In particular, \cite{Montet:2014} used AO imaging to constrain the mass of companions causing such trends to be in the planetary regime.  They then used their constraints to determine if the population of long-period giant planets was consistent with that found by microlensing.  In contrast, \cite{Clanton:2014a,Clanton:2014b} took a different approach, and mapped the distribution of planets orbiting low-mass stars as inferred from microlensing to that expected from RV surveys of low-mass stars. Both groups concluded that the demographics of long-period giant planets as determined by radial velocity and microlensing surveys were consistent.  In particular, both groups found that there exists a significant population of Jovian planets at relatively long periods. Specifically, \cite{Clanton:2014b} found that the frequency of Jupiters and super-Jupiters ($1 < M_p\sin{i}/M_{\rm Jup} < 13$) with periods $1 < P/{\rm days} < 10^4$ is $0.029^{+0.013}_{-0.015}$. This is a median factor of $4.3$ smaller than the inferred frequency of such planets around FGK stars of $0.11\pm 0.02$ \cite{Cumming:2008}.  Thus, although low-mass stars do indeed host giant planets, they are less common than giant planets orbiting sunlike stars, and tend to be at larger separations (compared to the snow line \cite{Kennedy:2008}).

\begin{figure}[t]
\includegraphics[width=12cm]{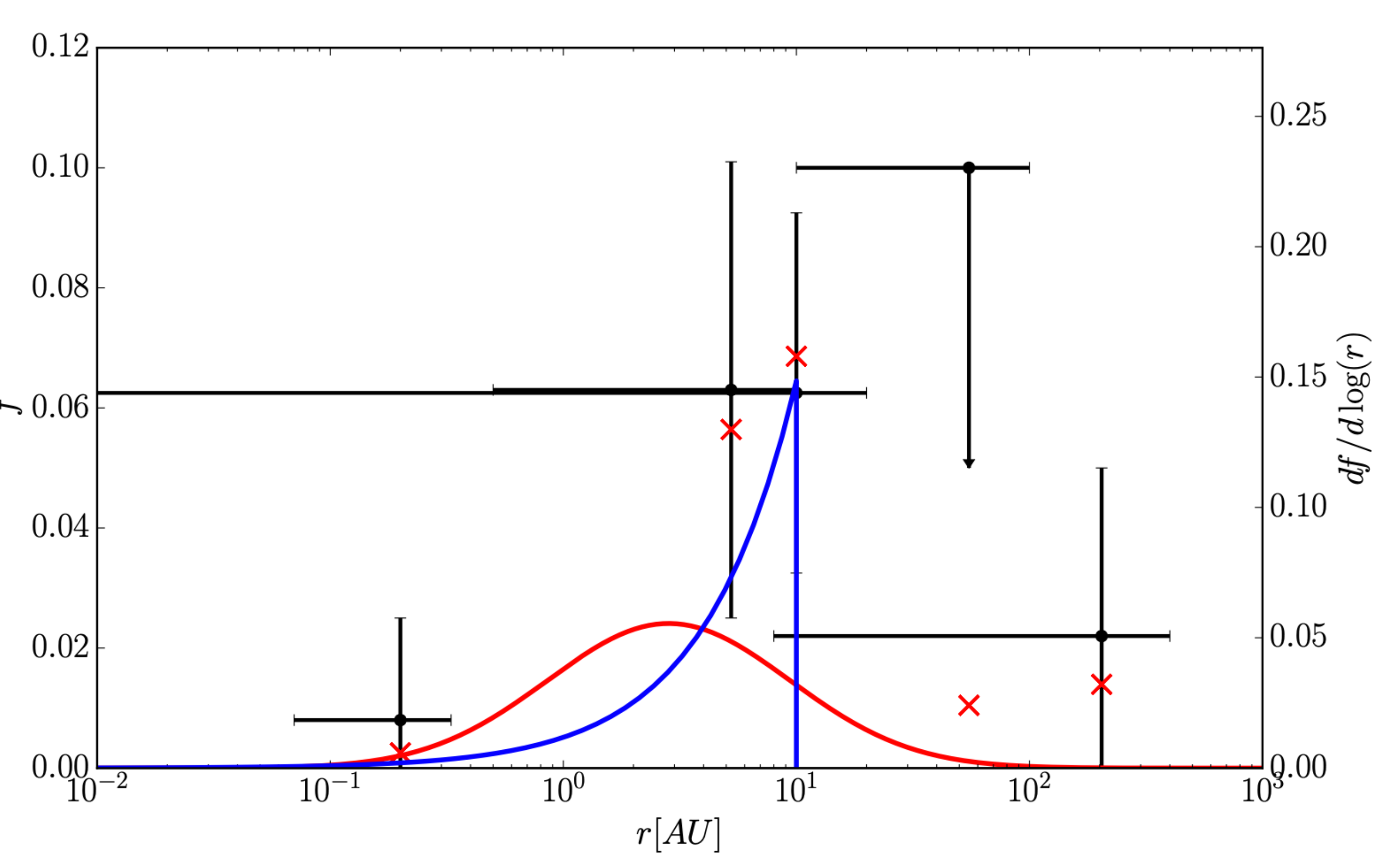}
\caption{
Semimajor axis distributions of roughly Jovian-mass companions to M dwarfs. The vertical axis shows point estimates of the semimajor axis distribution from several surveys (red "X"s and uncertainties and/or upper limits shown as black error bars), as well as parametric fits to these distributions from \cite{Clanton:2016} (blue curve) and \cite{Meyer:2018} (red curve). From \cite{Meyer:2018}. Reproduced with permission \copyright ESO.  
}
\label{fig:meyer}
\end{figure}

Several authors have combined results from radial velocity, microlensing, and direct imaging surveys to constrain the population of giant planets with large semimajor axes orbiting low-mass stars.  In particular, \cite{Clanton:2016} synthesized constraints on the population of long-period planets from five different exoplanet surveys using three independent detection methods: microlensing, radial velocity, and direct imaging.  Adopting a power-law form for properties of long-period ($>2~{\rm au})$ planets, they found
\begin{equation}
   \frac{d^2N}{d\log{M_p}~d\log{a}}= 
   0.21^{+0.20}_{-0.15}\left(\frac{M_p}{M_{\rm Sat}}\right)^{-0.86^{+0.21}_{-0.19}}\left(\frac{a}{2.5~{\rm au}}\right)^{1.1^{+1.9}_{-1.4}},
   \label{eqn:clanton}
\end{equation}
with an outer cutoff of $a_{\rm out} = 10^{+26}_{-4.7} {\rm au}$.  This result was for "hot-start" models, but the results for "cold-start" models are very similar.  This is because the typical host stars are quite old, and as such the luminosity at fixed mass of planets assuming "hot start" and "cold start" models have largely converged.

A similar analysis was performed by \cite{Meyer:2018}, although they fit a (likely) more well-motivated log-normal model for the semimajor axis distribution of giant planets orbiting M dwarfs.  Their results are shown in Figure \ref{fig:meyer}. Their conclusions are broadly consistent with those of \cite{Clanton:2017}: generally speaking, the frequency of giant planets orbiting M dwarfs increases with increasing semimajor axis up to a few au, and then declines for semimajor axes beyond $\sim 10$~au. An interesting but unanswered question is how the distribution of giant planets found by \cite{Clanton:2017} and \cite{Meyer:2018} differ from that found by radial velocity surveys of giant planets for solar-type (FGK) stars, and whether or not they are consistent with the expectations of ab initio planet formation theories.

I note that the analyses of both \cite{Clanton:2017} and \cite{Meyer:2018} did not include the most recent and comprehensive microlensing constraints on the demographics of planets from \cite{Suzuki:2016}.  Therefore, there is a clear opportunity for improving and updating the syntheses provided in \cite{Clanton:2017} and \cite{Meyer:2018}.

There have been several other studies that have attempted to synthesize the demographics of exoplanets determined by various methods. 
\begin{itemize}
\item \cite{Howard:2012} compared early demographic constraints from {\it Kepler} for planets with periods of $\lesssim 50~{\rm days}$ with the constraints from the Keck/HIRES Eta-Earth RV survey for planets with periods in the same range \cite{Howard:2010}.  By adopting a deterministic density-radius relation and restricting the analysis to planets with masses $\gtrsim 3~M_\oplus$ and radii $\gtrsim 2~R_\oplus$ (where the Eta-Earth and [then available] {\it Kepler} results were mostly complete), they were able to map the radius distribution inferred from {\it Kepler} to the $M_p\sin{i}$ distribution from the Eta-Earth survey.  They found good agreement, particularly when they assumed that the density of planets increased with decreasing radii.
\item Both \cite{Gould:2000} and \cite{Suzuki:2016} compared constraints on the frequency of planets inferred from microlensing surveys to several estimates of the frequency of shorter-period planets found by several RV surveys of both solar-type FGK stars as well as M stars (see, e.g., Figure \ref{fig:suzuki}). 
\item As discussed in more detail in Section \ref{sec:disurveys}, \cite{Bowler:2016} compared the frequency distribution giant planets as a function host star spectral type found by RV surveys \cite{Johnson:2010b} with the distribution found by direct imaging surveys (see Figure~\ref{fig:dioccur}).  
\item Using the DR25 {\it Kepler} catalog of planets between $\sim 1-6~R_\oplus$ and $P<100~{\rm days}$, and converting planet radii to planet masses using the \cite{Chen:2017} planet mass-radius relation, \cite{Pascucci:2018} found a break in the mass ratio function of planets at $q\sim 0.3 \times 10^{-4}$, independent of host star mass.  This break is at a mass ratio that is $\sim 3-10$ times lower than the break in the mass-ratio function for longer-period planets found by microlensing as estimated by \cite{Suzuki:2016,Udalski:2018} (Figure \ref{fig:suzuki}), but is similar (a factor of $\sim 2$ smaller) than that inferred by \cite{Jung:2019}. Assuming the latter result holds, this implies that Neptune mass-ratio planets are the most common planet both interior and exterior to the snow line, as least down to the mass ratios that have been probed by transits and microlensing of $\gtrsim 0.5\times 10^{-6}$, or slightly larger than the Earth/sun mass ratio. 
\item Using results from both RV surveys and {\it Kepler}, \cite{Zhu:2018} studied the relationship between the population of small-separation ($a\lesssim1{\rm au}$) super-Earths ($M_p$ roughly between that of the Earth and Neptune) and 'cold' Jupiters ($a>1{\rm au}$ and $M_p>0.3~M_{\rm Jup}$) orbiting sunlike stars.  They found that the conditional probability of a system with a super-Earth hosting a cold Jupiter was $\sim 30\%$, three times higher than the frequency of cold Jupiters orbiting typical sunlike field stars (e.g., \cite{Cumming:2008}).  Given the prevalence of super-Earths found from {\it Kepler}, this implies that nearly every star that hosts a cold Jupiter also hosts an inner super Earth.  Since our solar system has a cold Jupiter but does not host an inner super-Earth, the corollary to this result is that solar systems with architectures like ours are rare, $\sim1\%$. 
\item Using a meta-analysis to combine the results of many demographics studies of transiting planets detected by {\it Kepler}, the NASA Exoplanet Exploration Program Analysis Group (ExoPAG) Study Analysis Group (SAG) 13 determined a consensus estimate of the occurrence rates for planets with relatively short periods of $P=10-640~{\rm days}$, and radii from roughly that of the Earth to that of Jupiter.  This estimate, and the process by which it was determined, is described in detail in \cite{Kopparapu:2018}. The double power-law fit to the SAG 13 occurrence rate was extrapolated to longer periods by \cite{Dulz:2020}, who also synthesized these results with the frequency of relatively long-period gas giants planets as determined by \cite{Cumming:2008}, \cite{Bryan:2016}, and \cite{Fernandes:2019}.  Taking care to eliminate systems that were dynamically unstable, \cite{Dulz:2020} provide a comprehensive synthesis of the demographics of planets with masses of $0.1-10^3~M_\oplus$ and semimajor axes of $0.1-30~{\rm au}$, albeit with significant reliance on extrapolation.
\end{itemize}
These and other results begin the process of "stitching together" the demographics constraint of multiple surveys using multiple detection methods. However, the studies itemized above, as well as the majority of other similar studies, have primarily focused on comparing the demographics of relatively short-period planets detected by different surveys, or by comparing the demographics of close and wide-orbit companions.  Thus, while very important, a comprehensive review of such studies is beyond the scope of this chapter. 

\begin{figure}[t]
\includegraphics[width=12cm]{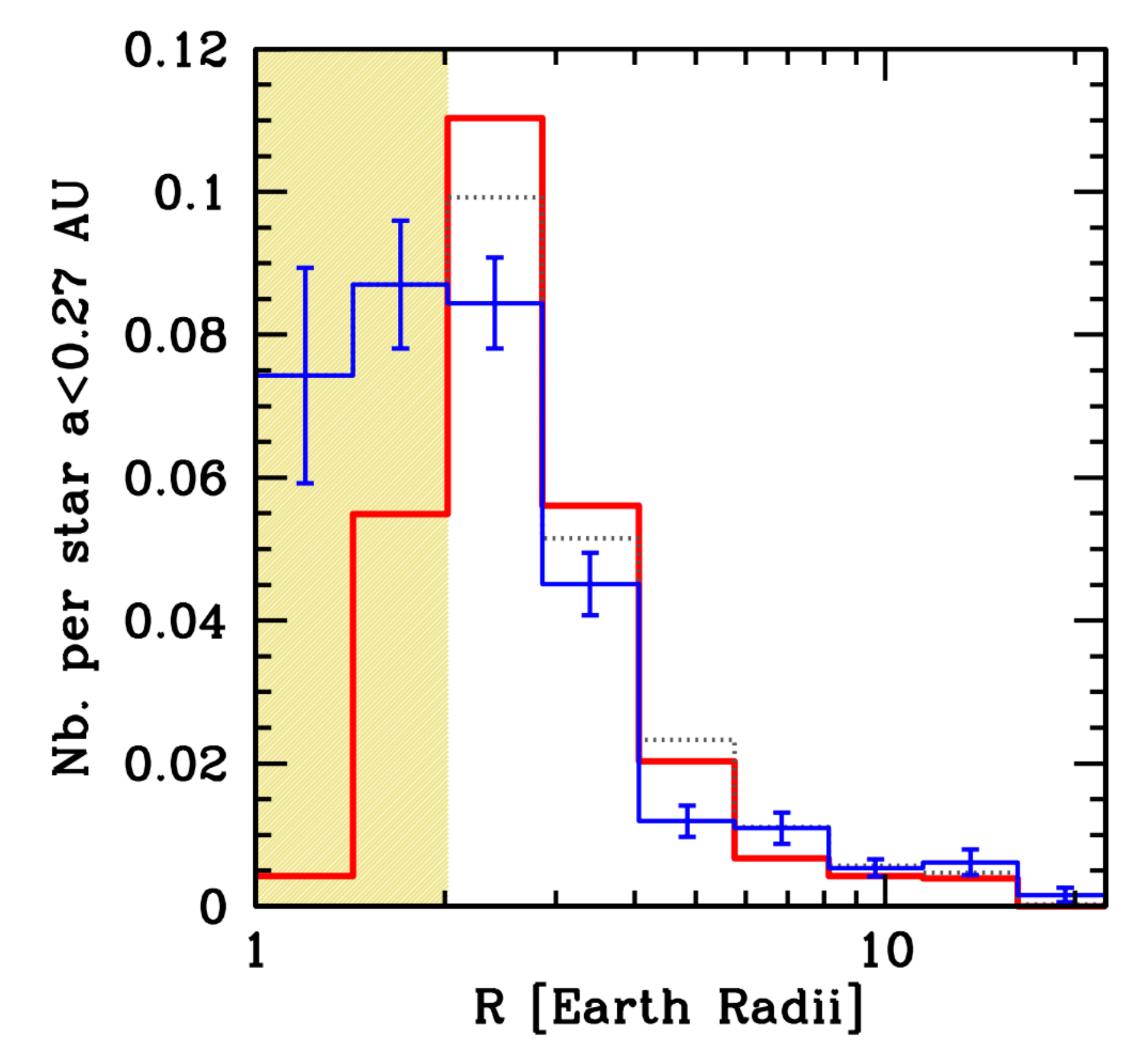}
\caption{Number of planets per star in a given radius bin for semimajor axes of $a < 0.27$~au. The thick red histogram
shows the predictions from the ab initio population synthesis models of \cite{Mordasini:2012}. The blue histogram with error bars shows the empirical distribution inferred by \cite{Howard:2010} from an early release of {\it Kepler} data. The black dotted histogram is a preliminary analysis from the initial {\it Kepler} data.  The {\it Kepler} data from the yellow shaded region ($R_p < 2 R_\oplus$) was substantially incomplete at the time of this study. From \cite{Mordasini:2012}.
Reproduced with permission \copyright ESO. }
\label{fig:mordasini}
\end{figure}

\section{Comparisons Between Theory and Observations}\label{sec:compare}

Remarkably, despite the large number of exoplanets that have been confirmed to date ($\sim$4300), there have been relatively few studies that rigorously compare the predictions of ab initio exoplanet population synthesis models to empirical exoplanet demographic data.  The most important reason for this is that the majority of the observational results on exoplanet demographics do not supply the data needed to compare these theories with the empirical predictions. Such data include (but are not limited to) the details of the target samples, the detection efficiencies (or completeness) of all of the target sample (not just those that have detected planets), and the quantification of the false positive rate (or reliability).  

One example of an attempt to compare the prediction of ab initio models of planet formation with empirical constraints is provided by \cite{Mordasini:2012}.  They compare an estimate of the radius distribution of planets from some of the first results from {\it Kepler} \cite{Howard:2012} with the predictions from their population synthesis models for planets with semimajor axis of $a\lesssim 0.3$au.  They find reasonable agreement between the predictions and the empirical results (see Figure \ref{fig:mordasini}).  
I note that this result applies to relatively short-period planets, and is thus formally out of the scope or this review. Nevertheless, it does provide an important example of the quantitative comparison of empirical results of the demographics of exoplanets with ab initio
theories. 

\begin{figure}[t]
\includegraphics[width=12cm]{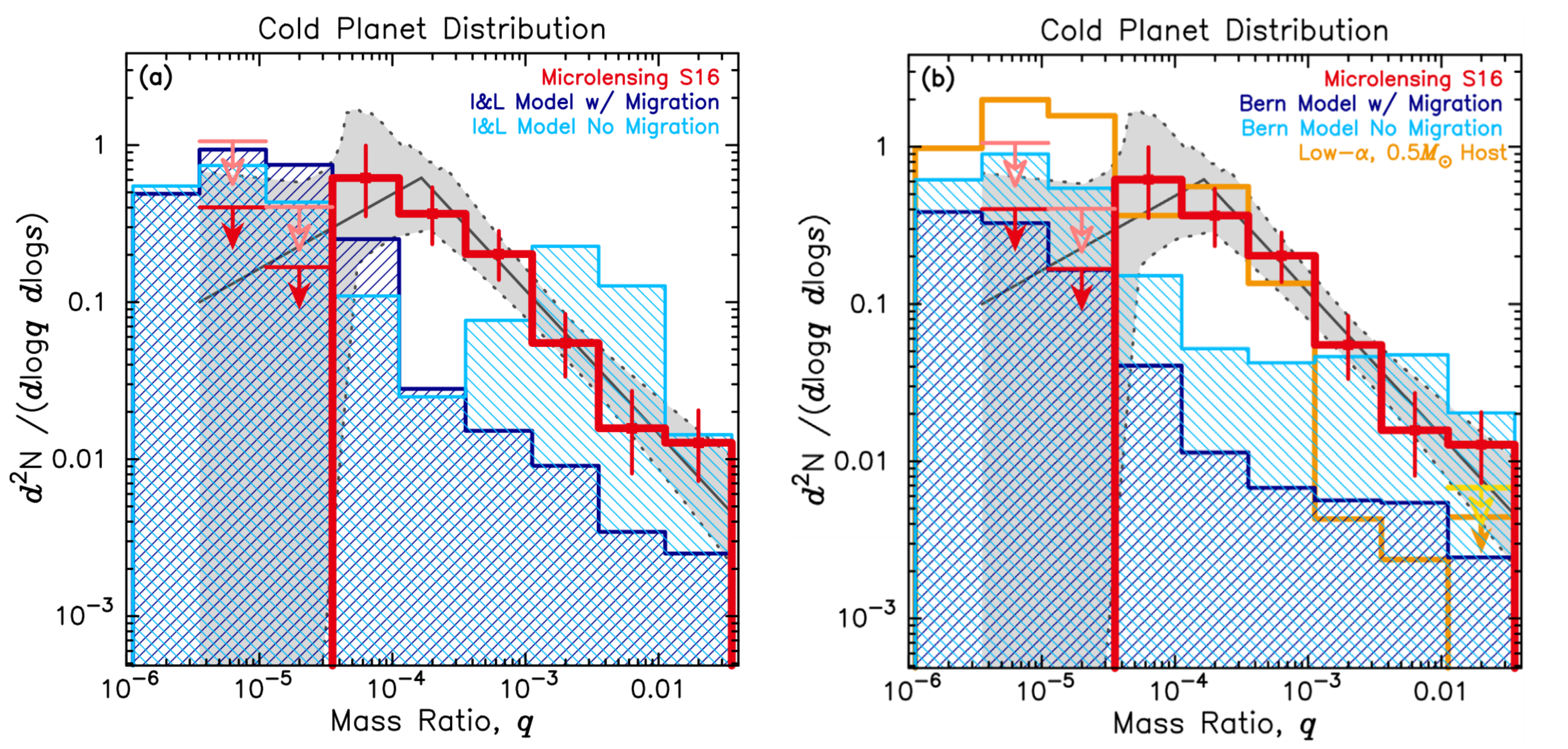}
\caption{The planet/host star mass ratio distribution as measured by microlensing surveys for exoplanets \cite{Gould:2000,Cassan:2012,Suzuki:2016}, compared to the predicted mass ratio distribution function from ab initio models of planet formation (see \cite{Ida:2013} and \cite{Mordasini:2015} and references therein). The red histogram shows the measured mass-ratio distribution from microlensing, along with the best-fit broken power-law model and 1$\sigma$ uncertainty indicated by the solid black line and
gray shaded regions. The dark and light blue histograms
show the predicted mass-ratio functions from the population synthesis models with migration, and the alternative migration-free models. (left) Comparison to models from Ida \& Lin (e.g, \cite{Ida:2013} and references therein). (right) Comparison to models from the Bern group (e.g., \cite{Mordasini:2015} and references therein).  The gold histogram shows the results for a lower-viscosity disk model and $0.5~M_\odot$ host stars. From \cite{Suzuki:2018} \copyright AAS. Reproduced with permission.} 
\label{fig:suzuki2018}
\end{figure}

A direct comparison between the ab initio predictions of multiple independent planet formation theories (see \cite{Ida:2013} and \cite{Mordasini:2015} and references therein) and the demographics of cold exoplanets as constrained by microlensing was explored in \cite{Suzuki:2016}.  As shown in Figure \ref{fig:suzuki2018}, they find that the generic prediction of the core accretion (or nucleated instability) model of giant formation predicts a paucity of planets with masses roughly between that of Neptune and Jupiter (e.g., \cite{Pollack:1996}). This prediction is not confirmed by the results from microlensing surveys.  This result appears to be fairly robust against some of the model assumptions, including the treatment of migration and the viscosity of the protostellar disk.  Some possible resolutions to this discrepancy are discussed in \cite{Suzuki:2018}.

One potential complication is that the condition for core-nucleated instability (e.g., runaway gas growth to become a giant planet) may be more sensitive to planet mass than planet mass ratio\footnote{Formally, the condition for runaway growth is that the mass of the gaseous envelop becomes larger than that of the core, leading to a Jeans-like instability and rapid gas accretion (e.g., \cite{Mizuno:1980,Stevenson:1982,Pollack:1996}.  However, the simplest models of protoplanetary disks generally predict core masses of $\sim 10~M_\oplus$, and thus the total critical mass for runaway accretion of $\sim 20~M_\oplus$, e.g., somewhat larger than the masses of the ice giants.}.  If this is the case, then the mass gap predicted by the core-accretion theory may be smoothed out in the microlensing mass-ratio distribution, given the relatively broad range of host masses probed in microlensing surveys (see Figure~\ref{fig:micromass}).  This can be tested by measuring the masses of the host stars of planets detected by microlensing (e.g., \cite{Bennett:2007}).  Indeed, \cite{Bhattacharya:2018} measure the host and planaet star masses of the microlensing planet OGLE-2012-BLG-0950Lb, finding a host star mass of $M_*= 0.58\pm0.04~M_\odot$ and a planet mass of $M_p=39\pm 8~M_\oplus$, placing the planet in the middle of the mass gap predicted by generic models of giant planet formation via core accretion.  

\section{Future Prospects for Completing the Census of Exoplanets}
\label{sec:future}

While substantial progress has been made in determining the demographics of wide-separation planets, it is nevertheless the case that it is this regime of $M_p-P$ and $R_p-P$ parameter space that remains the most incomplete, particularly for planets with masses and radii less than that of Neptune (see Figures~\ref{fig:exoplanets-pm} and \ref{fig:exoplanets-pr}).  Fortunately, there are several planned or candidate surveys on the horizon that will largely fill in this region of parameter space, enabling a nearly complete statistical census of planets with masses/radii greater than that of the Earth, and periods from $<1~{\rm day}$ to essentially infinity, including free-floating planets.  In this section, I will briefly summarize these future prospects.

\begin{figure}
\includegraphics[width=6.3cm]{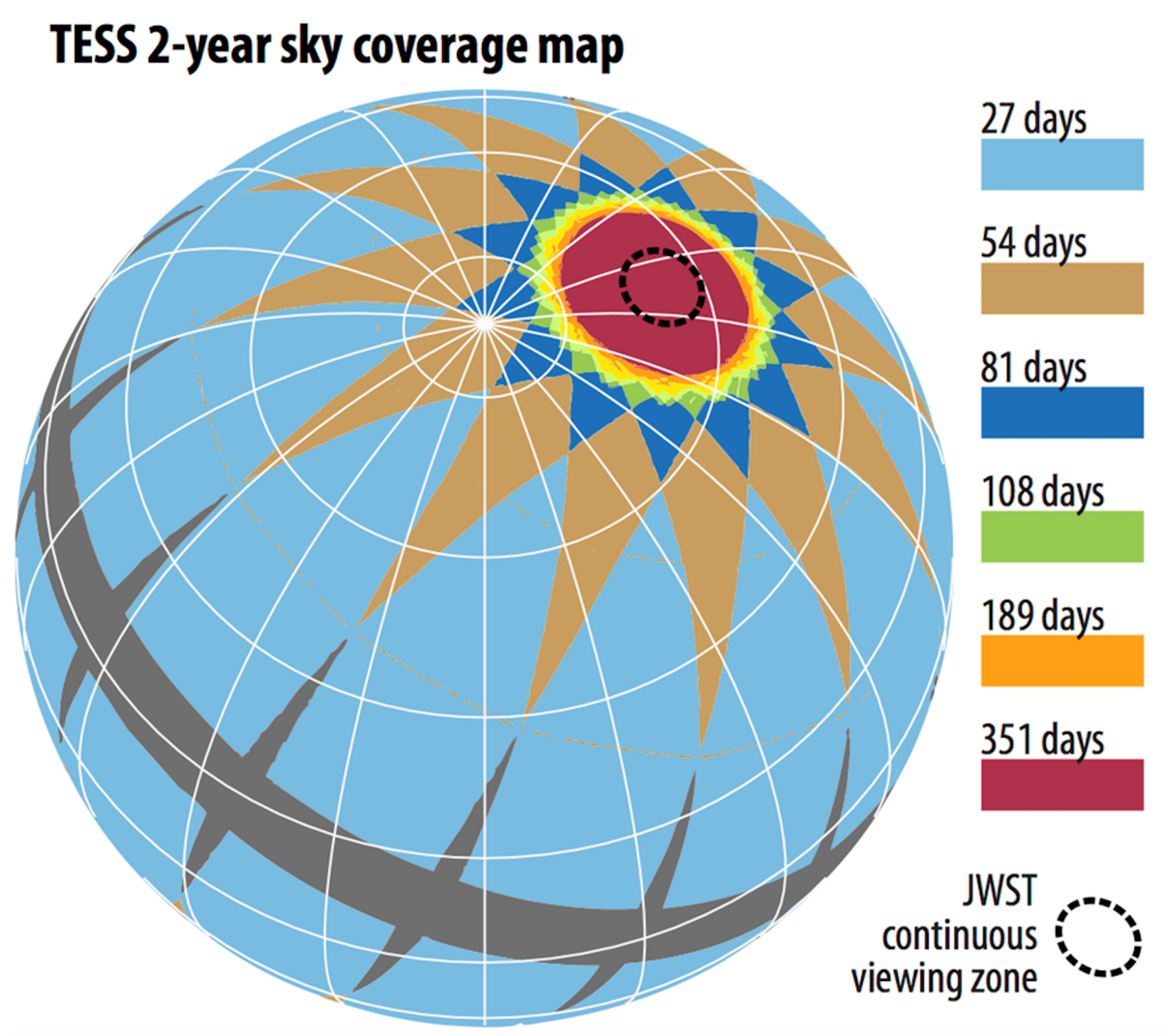}
\includegraphics[width=5.2cm]{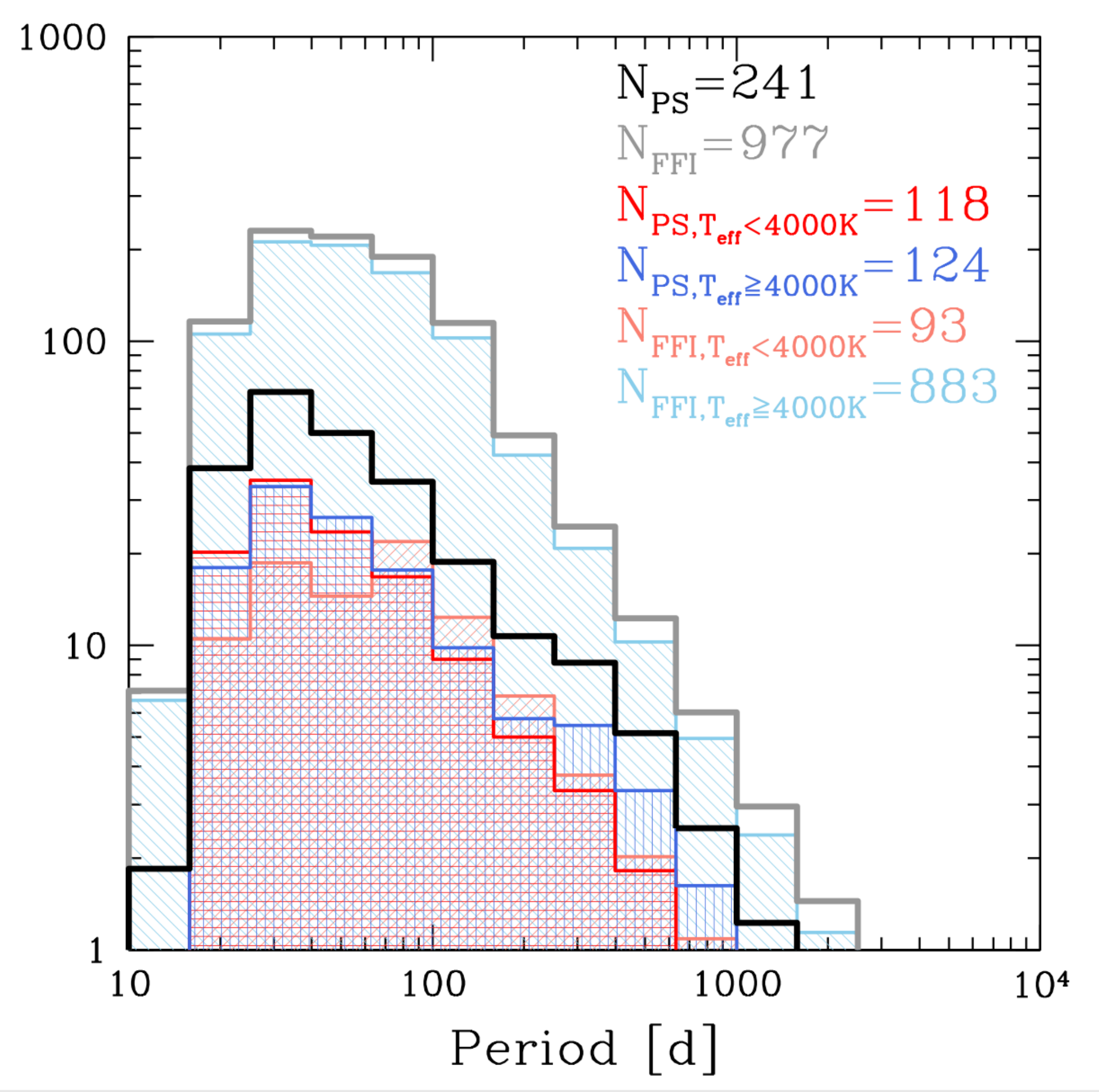}
\caption{(Left)
The original planned Transiting Exoplanet Survey Satellite (TESS) 2-year mission sky coverage, as a function of equatorial coordinates (solid white lines).  TESS originally planned to monitor $\sim 83\%$ of the sky, excluding regions within $\sim 6^\circ$ of the ecliptic equator.  $\sim 74\%$ of its survey area was planned to be monitored for 27 days during the prime mission, whereas a region around the ecliptic poles (including the JWST continuous viewing zone) would be monitored for up to $\sim 350$~days.  Stray light issues forced the TESS team to deviate somewhat from this survey strategy in the northern ecliptic hemisphere.  Courtesy of G. Ricker, reproduced by permission.
(Right) An estimate of the number and period distribution of
single-transit events expected from the TESS primary mission.  Single transit events from the 2-minute cadence postage stamp targets are shown in the black histogram, whereas those from stars in the 30-minute cadence full frame images (FFIs) are shown in grey. Single transit events found for stars with $T_{\rm eff}> 4000$K are shown in blue, whereas those for stars with stars $T_{\rm eff}<4000$K are in red. The darker shades are for the postage-stamp targets, while the lighter shades are for the FFI stars. A total of 241 single-transit events due to planets are predicted in the postage stamp data and at least 977 single-transit events are expected in the FFIs. Reproduced from \cite{Villenueva:2019}. \copyright AAS. Reproduced with permission.}
\label{fig:TESS}
\end{figure}

\begin{itemize}
    \item {\bf Radial Velocity surveys}.  The longest-running RV surveys have been monitoring a sample of bright FGK stars stars for roughly 30 years, with precisions of a few to $\sim 10~{\rm m/s}$.  These surveys are now sensitive to Jupiter analogs with $P\sim 12~{\rm years}$ and $M_p \gtrsim M_{\rm Jup}$ (e.g., \cite{Wittenmyer:2016}). For stars with the longest baselines and RV precisions of a few m/s, Saturn analogs ($P\sim 30~{\rm years}$ and $M_p \gtrsim M_{\rm Sat}$) are barely detectable, with $K\sim 3~{\rm m/s}$. RV surveys are unlikely to be sensitive to analogs of the ice giants in our solar system \cite{Kane:2011} in the foreseeable future, and are similarly unlikely to be sensitive to Neptune-mass planets on orbits beyond $\sim 3~{\rm au}$, which would have RV semiamplitudes of $\sim 1~{\rm m/s}$ and periods of $P\sim 5~{\rm years}$.  Beyond the difficulties with maintaining such RV precisions over such long time spans, it is likely to be the case that the majority of the (already oversubscribed) precision radial velocity resources will be focused on following-up planets detected in current and future transit surveys such as the Transiting Exoplanet Survey Satellite (TESS), \cite{Ricker:2015}) and the PLAnetary Transits and Oscillation of stars mission (PLATO, \cite{Rauer:2014}), as well as search for Earth analogs (e.g., \cite{Burt:2020}). Therefore, I reluctantly conclude that radial velocity surveys are unlikely to contribute to significantly expanding our knowledge of the population of wide-separation planets.\\
    
    \item {\bf The Transiting Exoplanet Survey Satellite}.  TESS is a NASA Medium-Class Explorers (MIDEX) mission \cite{Ricker:2015}, whose goal is to survey nearly the entire sky ($\sim 83\%$) to find systems with of transiting planets orbiting bright host stars, which are the most amenable to detailed characterization of the planet and host star.  In particular, many of the planets detected by TESS will be ideal targets for atmospheric characterization with the James Webb Space Telescope (JWST). See, e.g., \cite{Beichman:2014}.  Because TESS is designed to survey the brightest stars in the sky for transiting planets, its dwell time for the majority ($\sim 74\%$) of its survey area is only 27 days for the prime mission (see Figure~\ref{fig:TESS}).  For stars in the two continuous viewing zones at the ecliptic poles, TESS will obtain nearly continuous observations for $\sim 350~{\rm days}$.  Thus, even considering extended missions, TESS's region of sensitivity in the $M_p-P$ planet will be entirely within that of {\it Kepler}. Simply put, although TESS will (of course) provide important demographic constraints, and will provide them with a higher fidelity than {\it Kepler} (primarily because the brightness of the target stars), in general TESS will not significantly expand our knowledge of exoplanet demographics beyond what has been learned by {\it Kepler}.  Therefore, it will generally not contribute to the demographics of wide-orbit planets.\\
    
    However, one area where TESS may contribute to the demographics of wide-orbit planets is via single transit events.  As discussed in Section \ref{sec:trsurveys}, single transit events can be used to constrain the demographics of planets with periods beyond the survey baseline.  Because TESS will be looking at a much larger number of stars for a shorter period of time than the primary {\it Kepler} survey, its yield of single-transit events is expected to be significantly higher than that of {\it Kepler}.  The yield of single transit events in the 2-year TESS primary mission has been predicted by \cite{Villenueva:2019}.  Their results are shown in Figure \ref{fig:TESS}. They find that over 1000 single-transit events due to planets are expected from the TESS primary mission, with 241 of these coming from the the 2 minute cadence targeted postage-stampe data and a lower limit of 977 coming from stars in the full-frame images (FFIs).\\

\begin{figure}[t]
\includegraphics[width=5.7cm]{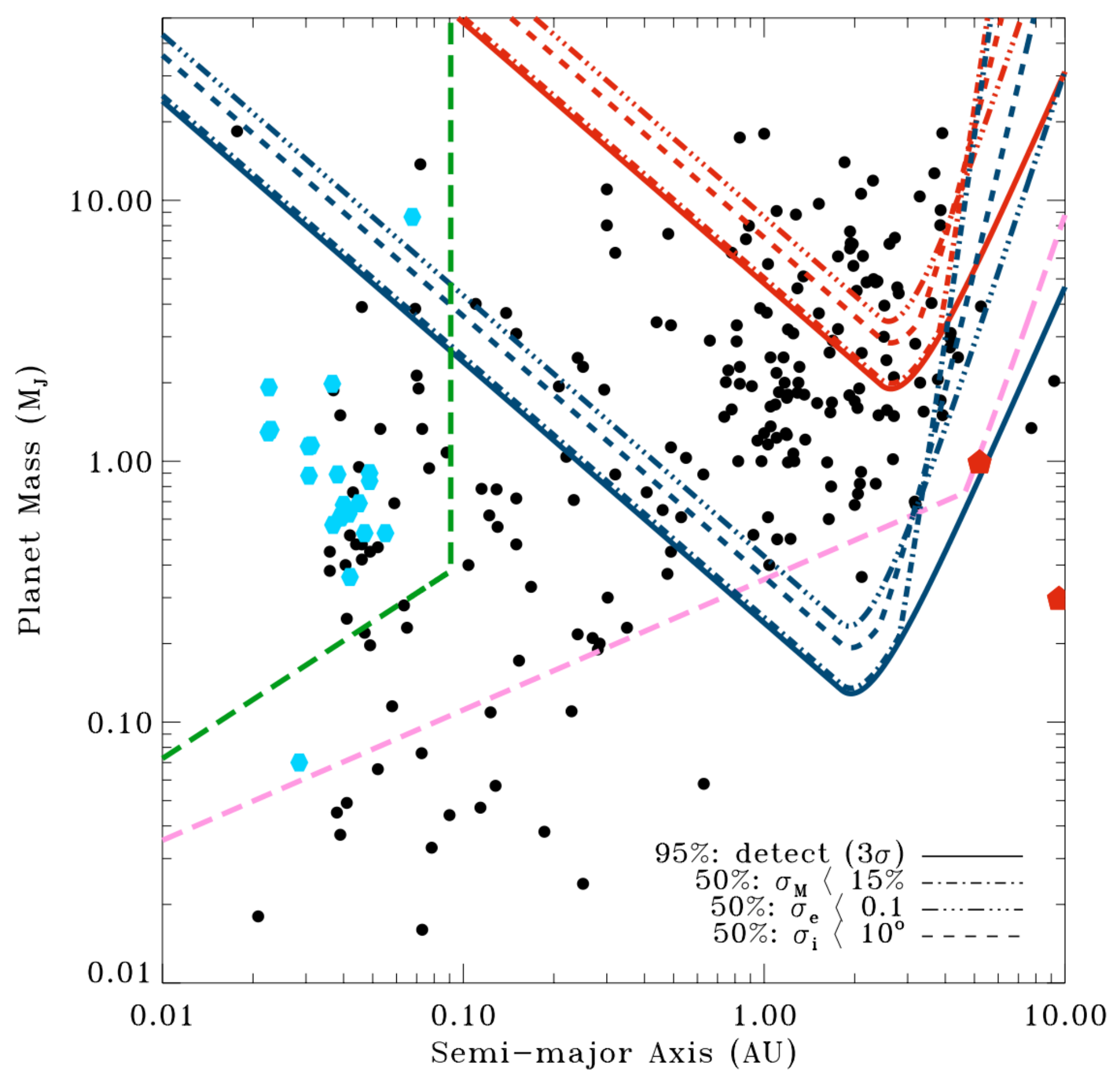}
\includegraphics[width=5.6cm]{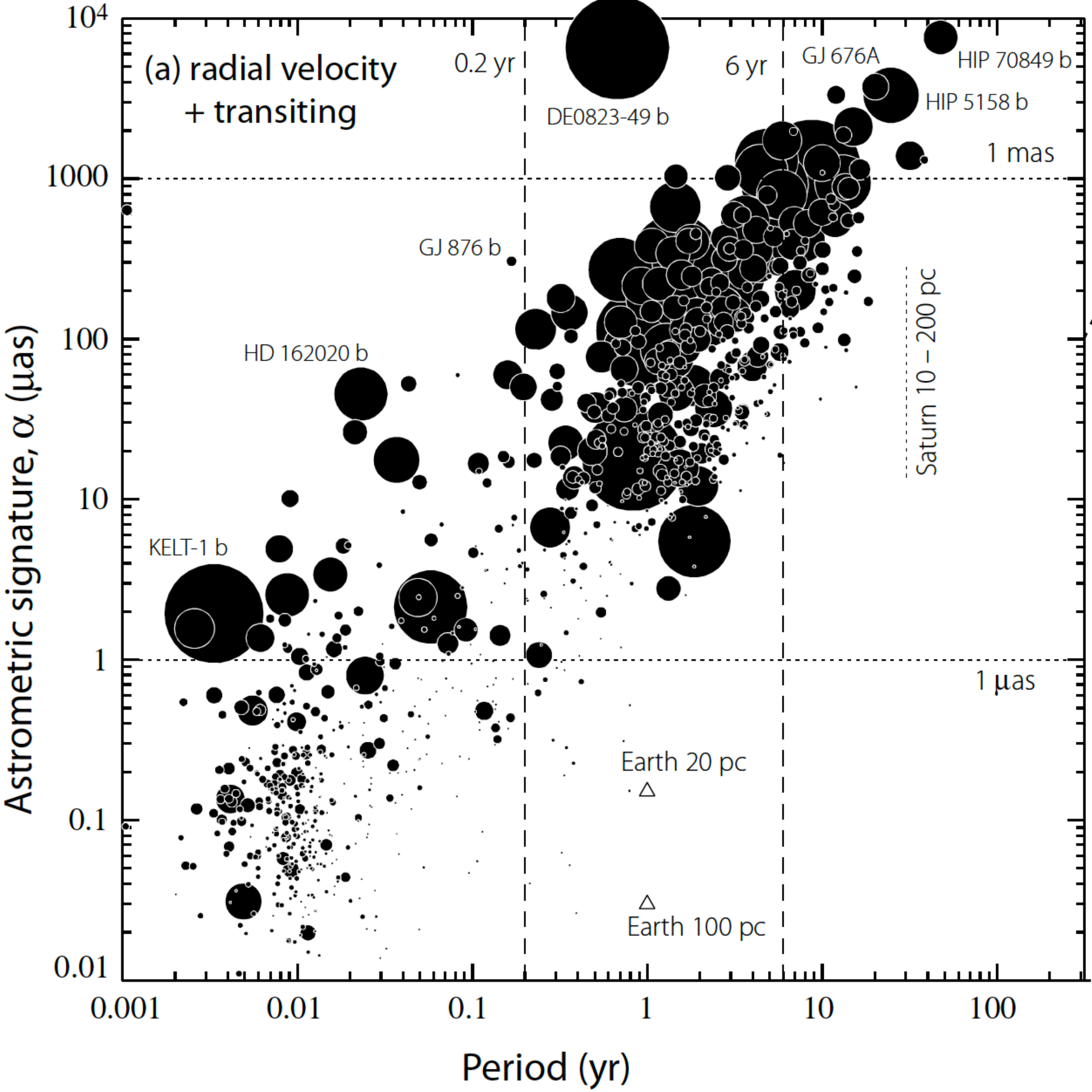}
\caption{(Left)
Gaia sensitivity as a function of planet mass and semimajor axis. The red curves show the estimated Gaia sensitivity for a $\sim M_\odot$ primary at 200~pc. The meaning of the different line styles are indicated in the figure.  The blue curves are the same, except for a $\sim 0.5~M_\odot$ primary at 25~pc. The pink line shows the sensitivity of an RV survey assuming a precision of 3~m/s, a $M_\odot$ primary, and a 10 year survey. The green curve shows the sensitivity of a transit survey assuming $\gtrsim 1000$ data points (approximately uniformly sampled in phase), each with a photometric precision of $\sim 5$~mmag, and a $M_\odot$ and $R_\odot$ primary. Black dots indicate the inventory of
exoplanets as of September 2007. Transiting
systems are shown as light-blue filled pentagons. Jupiter and Saturn are also shown as red
pentagons.  From \cite{Casertano:2008}. Reproduced with permission \copyright ESO.
(Right)
Astrometric signature versus period for the planets listed in the exoplanet.eu archive as of September 1, 2014. The sizes of the circles are proportional to the planet mass. The vertical lines roughly bracket the range in periods in which Gaia will be most sensitive.  The horizontal line delineates an astrometric signature of $1~\mu{\rm as}$. From \cite{Perryman:2018}.  Courtesy of Michael Perryman. Reproduced with permission. 
}
\label{fig:Gaia}
\end{figure}

    \item {\bf Gaia}. The primary goal of European Space Agency's (ESA) Gaia mission is to provide exquisite astrometric measurements of $\sim 10^{10}$ stars down to a magnitude of $V\sim 20$ \cite{Gaia:2016}. These measurements will produce a sample of accurate and precise stellar proper motions and distances that is orders of magnitude larger than is currently available.  A `by-product' of this unprecedented database will be the detection of many thousands of giant planets at intermediate periods via their astrometric perturbations on their host stars.   There have been many studies of expected yield of planets with Gaia; here I will focus on two papers (but see also \cite{Sozzetti:2014}). \\
    
    Using double-blind experiments, \cite{Casertano:2008} demonstrated that planets can be detected if they induce astrometric signatures of roughly $3\sigma_{\rm ast}$, where $\sigma_{\rm ast}$ is the single-measurement precision, provided that the period of the planet is less than the survey duration (5 years for the primary Gaia mission). They also noted that the threshold for reliable inference about the orbital parameters was significantly higher. They determined that at twice the detection limit of $3\sigma_{\rm ast}$, the uncertainties in orbital parameters and planet masses were typically of order 15\% - 20\%.  They also demonstrated that for planets with periods longer than the survey duration, the planet detectability dropped off more slowly with increasing planet period than the precision with which the parameters of the system, and in particular the planet mass, can be measured (see Figure \ref{fig:Gaia} and the discussion in Section \ref{sec:snrastro}). Thus, although the astrometric signal of planets with periods longer than the survey may be detectable, these detections will be significantly compromised as they will have large uncertainties in the orbital elements and, in particular, the planet mass.\\
    
    A more recent estimate of the yield of planets from Gaia was performed by \cite{Perryman:2018}. In particular, these authors used updated planet occurrence rate estimates, as well as updated estimates of the expected Gaia single-measurement astrometric precision.  They consider the detectability of known planetary companions with Gaia (see Figure \ref{fig:Gaia}), as well as projections for as-yet undetected planets.  They find that Gaia should discover $\sim 10^4$ planets with masses in the range of $\sim 1-15~M_{\rm Jup}$, the majority of which will have semimajor axes in the range of $\sim 2-5$~au.  They also estimate that Gaia should find $\sim 25-50$ intermediate-period $P\sim 2-3~{\rm year}$ transiting systems.\\
    
    Thus, although Gaia will likely deliver a very large number of planets, including a large number of planetary systems, it will not be sensitive to a currently unexplored region of $M_p-a$ parameter space.  Nevertheless, Gaia will be sensitive to the mutual inclinations of planets in multi-planet systems, an important property of planetary systems that has been poorly explored to date.  Furthermore, because Gaia should detect a very large number of planets, it will be sensitive to the "tails" of the distribution of planetary properties, e.g., the "oddball" exoplanet systems. Such systems often provide unique insights into the physics of planet formation and evolution.  Finally, the fact that it will uncover temperate transiting giant planets will enable the estimate of the mass-radius relationship for such planets, which will provide important priors on the properties of the giant planets detected by future direct imaging surveys.\\ 
    
\begin{figure}[t]
\includegraphics[width=12cm]{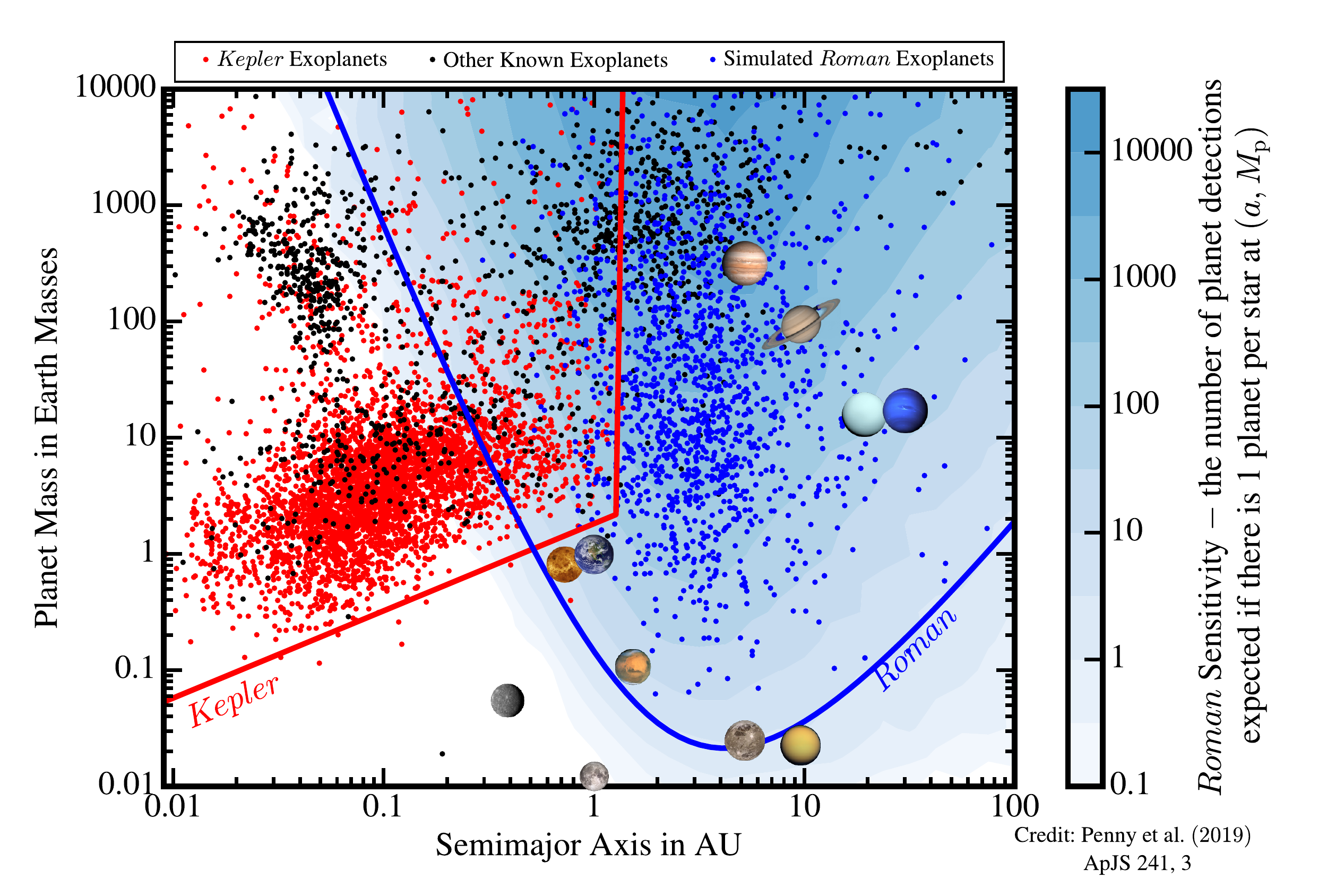}
\caption{(Left) The region of sensitivity in the $M_p-a$ plane of the {\it Kepler} prime mission (red solid line) versus the predicted region of sensitivity of the {\it Roman} (n{\' e}e WFIRST) Galactic Exoplanet Survey (RGES survey, blue solid line). The red dots show {\it Kepler} candidate and confirmed planets, while the black dots show all other known planets extracted from the NASA exoplanet archive as of 2/28/2018. The blue dots show a simulated realization of the planets detected by the RGES survey. Solar system bodies are shown by their images, including the satellites Ganymede, Titan, and the Moon at the semimajor axis of their hosts. Images of the solar system planets courtesy of NASA. Adapted from \cite{Penny:2019}. Courtesy of M. Penny. \copyright AAS. Reproduced with permission. 
}
\label{fig:Roman}
\end{figure}
    
    \item {\bf The Nancy Grace Roman Space Telescope}. The Nancy Grace Roman Space Telescope, or {\it Roman} (n{\' e}e WFIRST), was the highest-priority large space mission from National Academies of Science 2010 Astronomy Decadal Survey \cite{Astro2010}. As outlined by the Astro2010 Decadal Survey, one of {\it Roman}'s primary goals is to "open up a new frontier of exoplanet studies by monitoring a large sample of stars in the central bulge of the Milky Way for changes in brightness due to microlensing by intervening solar systems. This census, combined with that made by the {\it Kepler} mission, will determine how common Earth-like planets are over a wide range of orbital parameters." This application of {\it Roman}, based originally on the concept and simulations by \cite{Bennett:2002}, promises to probe a broad region of planet mass/semimajor axis parameter space that is inaccessible by any other exoplanet detection methods or surveys, including ground-based microlensing surveys.  Initial estimates of the yield of such a space-based microlensing survey can be found in \cite{Bennett:2002,Green:2011,Green:2012,Spergel:2013,Spergel:2015}.  The most up-to-date estimate of the yield of bound planets by microlensing are provided by \cite{Penny:2019}.  In summary, {\it Roman} is expected to detect roughly $\sim 1500$ bound and free-floating planets with masses $\gtrsim M_\oplus$ with separations $>1~{\rm au}$.  At the peak of its sensitivity, {\it Roman} will be sensitive to bound planets with masses as low as $\sim 0.02~M_\oplus$, or roughly twice the mass of the moon and roughly the mass of Ganymede.  {\it Roman} will also be sensitive to free-floating planets, as explored by \cite{Johnson:2020}.  They predict that {\it Roman} could detect $\sim 250$ free-floating planets with masses down to that of Mars, including $\sim 60$ with masses less than that of the Earth.\\
    
    By combining the demographic constraints from {\it Kepler} with those from {\it Roman}, it will be possible to obtain a nearly complete statistical census of exoplanets with mass $\gtrsim M_\oplus$ with arbitrary separations.  See Figure \ref{fig:Roman}.  It is also worth noting that {\it Roman} will also have some sensitivity to potentially habitable planets, and thus may provide constraints on $\eta_\oplus$.  Finally, it will also be sensitive to massive satellites to wide-separation bound planets \cite{Bennett:2002}, thus complementing the sensitivity of transit surveys to massive satellites orbiting shorter-period planets (e.g., \cite{Kipping:2009,Kipping:2009b,Kipping:2009c}).\\

    \item {\bf The PLAnetary Transits and Oscillations of stars mission}. PLATO \cite{Rauer:2014} is an ESA M-class mission that is designed to find transiting planets with periods of $\lesssim 2~{\rm years}$ orbiting relatively bright stars. The PLATO mission can be thought of as an intermediate mission between {\it Kepler} and TESS.  It will survey stars with a longer baseline than TESS, and will survey brighter stars than those surveyed by {\it Kepler}. Because the hosts of the transiting planets discovered by PLATO will be brighter than {\it Kepler}, they will be more amenable to follow-up. Given the current mission architecture and survey design, while PLATO is predicted to expand upon {\it Kepler}'s sensitivity range in the $R_p-P$ plane, it will nevertheless not contribute significantly to our understanding of the demographics of wide-orbit planets, excepting for the possibility of following-up single transit events (e.g., \cite{Yee:2008,Herman:2019, Villenueva:2019}).\\

\begin{figure}[t]
\includegraphics[width=5.8cm]{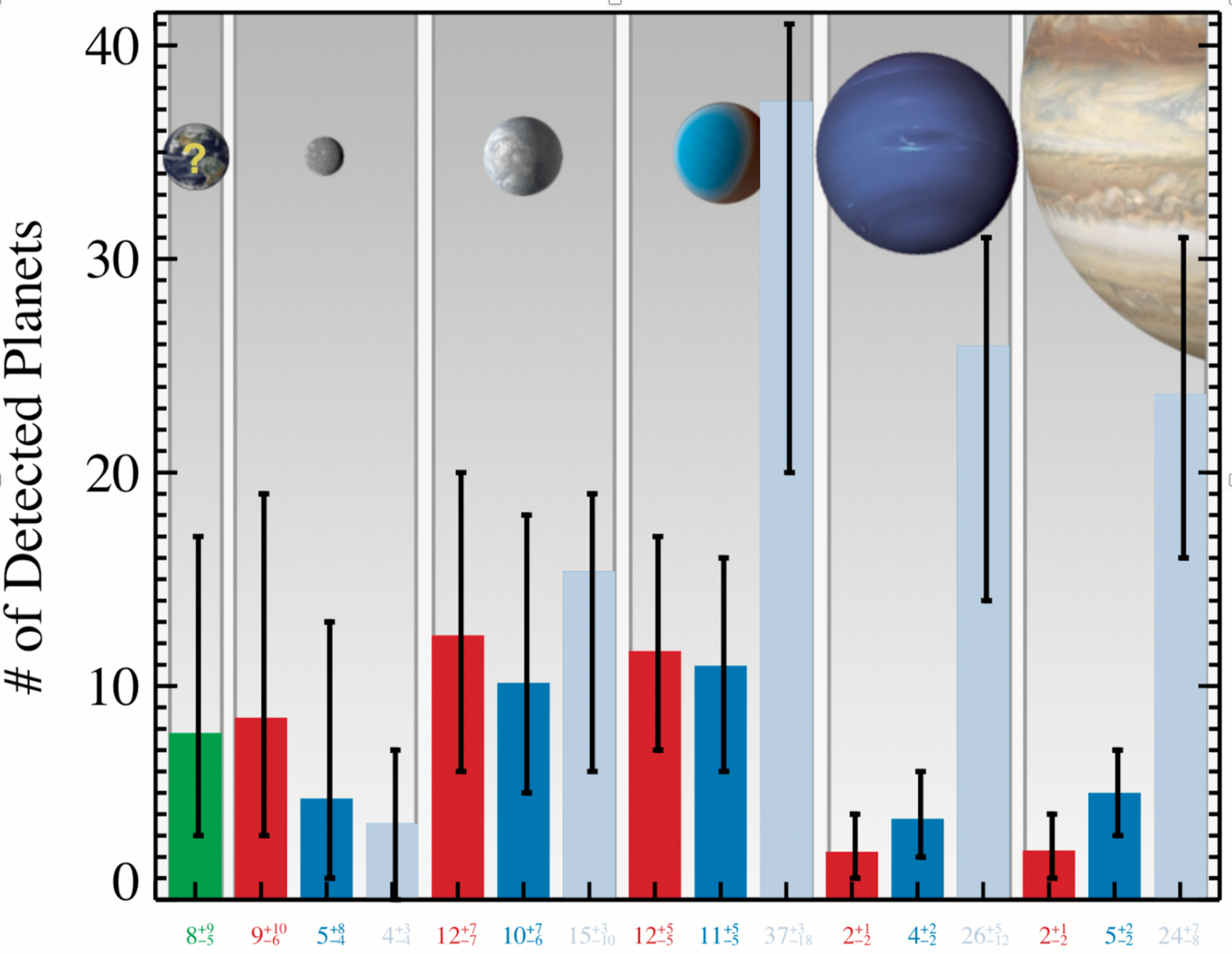}
\includegraphics[width=5.8cm]{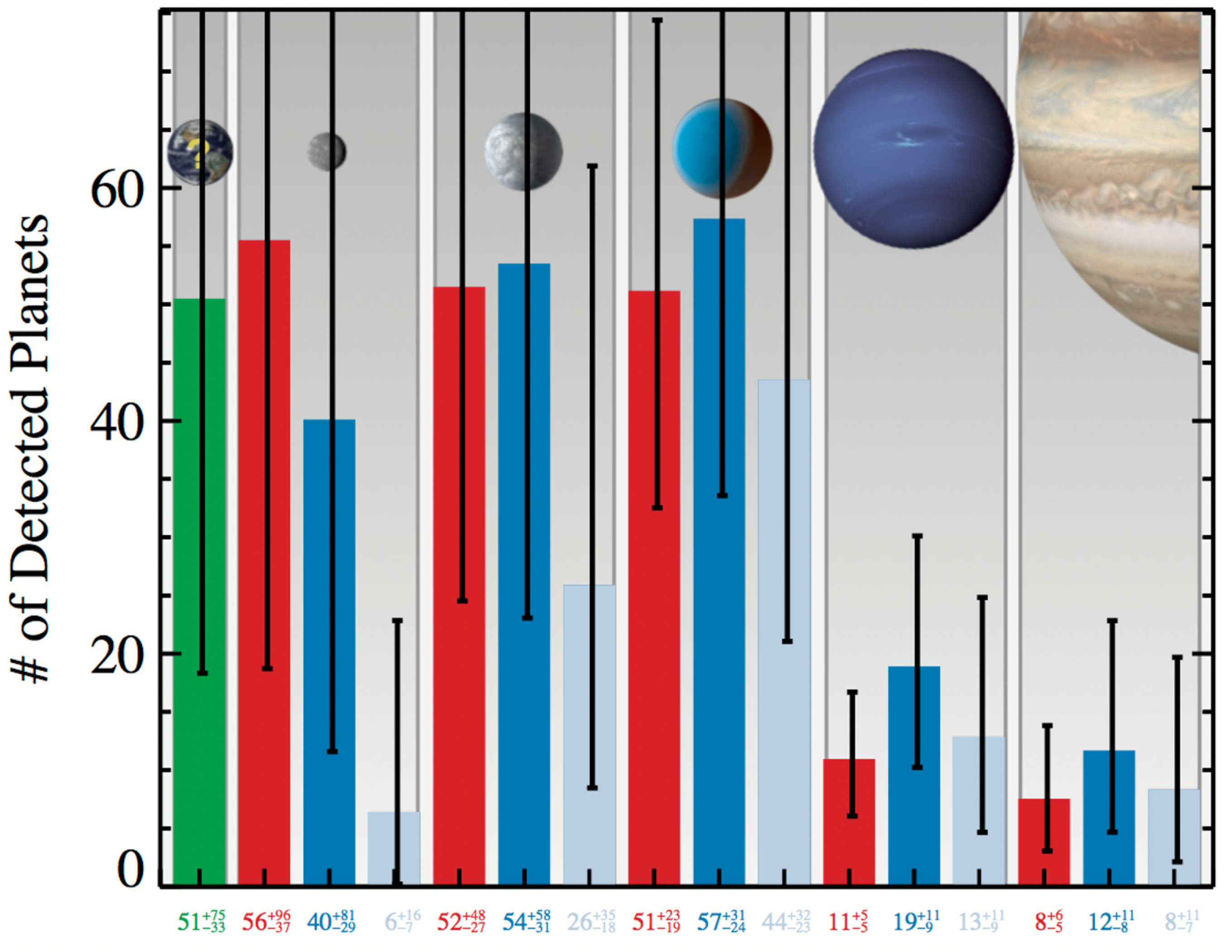}
\caption{Predicted exoplanet detection yields with uncertainties for the HabEx and LUVOIR mission concepts. Planet class types from left to right are: exoEarth candidates (green bar), rocky planets, super-Earths, sub-Neptunes, Neptunes, and Jupiters. Red, blue, and ice blue bars indicate hot, warm, and cold planets, respectively. The predicted yields are indicated under each bar. (Left) Predictions for HabEx. (Right) Predictions for LUVOIR A. Both figures courtesy of Christopher Stark. Reproduced with permission.}
\label{fig:DIyield}
\end{figure}

   \item{\bf Future Direct Imaging Surveys}. 
   A comprehensive review of the potential of future ground and space-based direct imaging surveys to contribute to our knowledge of the demographics of exoplanets is well beyond the scope of this chapter.  Rather, I refer the reader to the review in Chapter 4 of National Academies of Science Exoplanet Science Strategy Report \cite{NASESS:2018}, and in particular Figure 4.3 of that report. I will, however, make a few general comments.  First, current ground-based direct imaging surveys (e.g., \cite{Macintosh:2014,Beuzit:2019}) are expected to continue to improve upon their current sensitivity in terms of inner working angle and contrast, but are unlikely to do so by an order of magnitude.  Second, direct-imaging surveys using JWST will provide modest improvements over ground-based surveys \cite{Beichman:2018}, but are nevertheless not expected to reach the contrast ratios needed to detect mature planets in reflected light or thermal emission.  Third, the next generation of ground-based Giant Segmented Mirror Telescopes (GSMTs) offer opportunities for dramatic improvements in the capabilities of direct-imaging surveys relative to current ground-based facilities.  These opportunities include the potential to directly detect mature "warm Jupiters" in reflected light, as well as to detect temperate planets in thermal emission (e.g, \cite{Wang:2019}).  However, these improvements in capabilities also require dramatic breakthroughs in several key technology areas.  Finally, I will note that there exists an enormous opportunity for a future space-based direct imaging mission in the thermal mid-infrared (e.g. \cite{Quanz:2019}).\\
   
   \item {\bf Future Space-Based Reflected Light Direct Imaging Surveys}. Direct imaging surveys for mature planets orbiting sunlike stars in reflected light generally require space-based missions.  This is because, for planets with radii and separations similar to those in our solar system, the planet/star contrast ratio is $\lesssim 10^{-8}$ and the planet/star angular separation for a system at $\sim 10~{\rm pc}$ is $\lesssim 0.5"$.  For a Jupiter/sun analog at 10~pc, the contrast is $\sim 10^{-9}$ with an angular separation of $\sim 0.5"$, whereas for a Earth/sun analog at 10~pc, the contrast is $\sim 10^{-10}$ at an angular separation of $\sim 0.1"$. Achieving these contrasts at these angular separations is essentially impossible from the ground \cite{Guyon:2005}.\\
   
   Indeed, achieving these contrast ratios at these angular separation is exceptionally challenging, even from space.  Nevertheless, there are two promising techniques for doing so, namely internal coronographs (see, e.g.,  \cite{Guyon:2006}, and references therein) and external occulters, or starshades \cite{Spitzer:1962, Cash:2006}. Both techniques have been extensively studied; and a comprehensive review of these methods is well beyond the scope of this chapter.  Suffice it to say that, after decades of laboratory demonstrations, there does not appear to be any insurmountable obstacles to achieving the required contrasts at the required angular separations using either technique.\\
   
   In preparation for the National Academies of Science 2020 Astronomy Decadal Survey (Astro2020), NASA constituted studies of four large space mission concepts, currently named the Habitable Exoplanet Observatory (HabEx \cite{Gaudi:2020}), the Large UV-Optical-InfraRed Telescope (LUVOIR \cite{LUVOIR:2019}), Lynx (\cite{Lynx:2018}), and Origins \cite{Meixner:2019}).  Two of these mission concepts, namely HabEx and LUVOIR, would be capable of providing important constraints on the demographics of both close and wide-orbit planets orbiting relatively nearby ($\sim 10-20~{\rm pc}$) FGK stars.\\
   
   HabEx and LUVOIR differ primarily in ambition.  The fiducial architecture of the HabEx telescope is a 4m monolithic, off-axis primary that utilizes both internal coronagraphy and an external starshade to directly image and characterize planets orbiting nearby stars.  LUVOIR studied two architectures, but here I focus on the most ambitious architecture for context. The LUVOIR A architecture baselines a 15-m diameter on-axis primary mirror that utilizes coronagraphy to directly image exoplanets. Because of its larger aperture, LUVOIR A would be able to detect a much larger sample of planets than the baseline architecture of HabEx. On the other hand, because HabEx utilizes a (primarily) achromatic starshade, it would be able to better characterize the (smaller number) of exoplanets it would detect.  The estimated yield of both mission concepts is shown in Figure \ref{fig:DIyield}.\\
   
   Regardless of the specific architecture, both HabEx and LUVOIR would be able to determine the demographics of planetary systems orbiting nearby stars over a wide range of planet orbits and sizes {\it for individual systems}. This in contrast to the statistical compendium of different systems that will be enabled by combining the demographic results from, e.g., {\it Kepler} and {\it Roman}.  In particular, both HabEx and LUVOIR would be able to address such fundamental questions as: "What is the conditional probability that, given the detection of a potentially habitable planet, there exists a outer gas giant planet?" The answers to such questions are likely to be essential for understanding the context of habitability. 
\end{itemize}

\section{Conclusion}\label{sec:conclusion}

The demographics of exoplanets, i.e., the distribution of exoplanets as a function of the physical parameters, may encode the physical processes of planet formation and evolution, and thus provides the empirical ground truth that all {\it ab initio} planet formation theories must reproduce.  There exist numerous challenges to determining the demographics of exoplanets over as broad a region of parameter space as possible.  In particular, the exoplanet detection methods at our disposal are sensitive to planets in different regions of (planet and host star) parameter space, as well as being sensitive to planets orbiting different host star parameters.  While this presents a challenge, in that we must construct robust statistical methodologies to combine the results of different surveys and different detection methods, it also provides an opportunity to more completely survey the demographics of exoplanets over the relevant regions of planet and host star parameter space. There exist many exciting future opportunities to expand our understanding of the demographics of exoplanets. 

The future of exoplanet demographics is bright. I expect that, within the next few decades, we will have a nearly complete statistical census of exoplanets with masses/radii greater than roughly than that of the earth, with essentially arbitrary separations (including free-floating planets). As well as providing the empirical ground truth for theories of the formation and evolution of exoplanetary systems, this census will provide the essential context for our detailed characterization of exoplanet properties, and ultimately of our understanding of the conditions of exoplanet habitability. 

\begin{acknowledgement}
I would like to thank the organizers of the $3^{\rm rd}$ Advanced School on Exoplanetary Science: Demographics of Exoplanetary Systems for inviting me to attend a very enjoyable school held at an extraordinarily beautiful venue, and for giving me the opportunity to present on the topic of "Wide-Separation Exoplanets".  I am very much indebted to the editors of the proceedings (L.\ Mancini, K.\ Biazzo, V.\ Bozza, and A.\ Sozzetti) for their exceptional patience as I wrote this chapter. I would like to thank the many people who have shaped my thinking about exoplanet demographics over the past 20+ years, including (but not limited to) Thomas Beatty, Chas Beichman, David Bennett, Gary Blackwood, Brendan Bowler, Chris Burke, Jennifer Burt, Dave Charbonneau, Jesse Christiansen, Christian Clanton, Andrew Cumming, Martin Dominik, Subo Dong, Courtney Dressing, Debra Fischer, Eric Ford, Andrew Gould, Calen Henderson, Andrew Howard, Marshall Johnson, Bruce Macintosh, Eric Mamajek, Michael Meyer, Matthew Penny, Michael Perryman, Peter Plavchan, Radek Poleski, Aki Roberge, Penny Sackett, Sara Seager, Yossi Shvartzvald,  Karl Stapelfeldt, Christopher Stark, Keivan Stassun, Takahiro Sumi, Andrzej Udalski, Steven Villanueva, Jr., Ji Wang, Josh Winn, Jennifer Yee, Andrew Youdin, and Wei Zhu. I would like to thank Wei Zhu and Subo Dong in particular for helpful discussions.  Apologies to those I forgot to include in this list and those I forgot to cite in this review.  Finally, I recognize the support from the Thomas Jefferson Chair for Space Exploration endowment from the Ohio State University, and the Jet Propulsion Laboratory.  This research has made use of the NASA Exoplanet Archive, which is operated by the California Institute of Technology, under contract with the National Aeronautics and Space Administration under the Exoplanet Exploration Program.

\end{acknowledgement}

\clearpage
\newpage

\bibliographystyle{spmpsci}
\bibliography{refs.bib}

\end{document}